\documentclass[12pt,preprint]{aastex}

\singlespace

\shorttitle{\map\ Temperature Analysis}
\shortauthors{Hinshaw et al.}

\newcommand{\map}    {{\sl WMAP}}
\newcommand{\cobe}   {{\sl COBE}}
\newcommand{\n}      {{\bf{n}}}
\newcommand{\EV}[1]  {\langle#1\rangle}
\newcommand{\lmax}   {l_{\rm max}}
\newcommand{\mmax}   {m_{\rm max}}
\newcommand{\uKsq}   {\mbox{$\mu{\rm K}^2$}}
\newcommand{\dg}     {\mbox{$^{\circ}$}}
\newcommand{\lsim}   {\mbox{$_<\atop^{\sim}$}}
\newcommand{\gsim}   {\mbox{$_>\atop^{\sim}$}}
\newcommand{\lt}     {\mbox{$<$}}
\newcommand{\gt}     {\mbox{$>$}} 
\newcommand{\order}  {{\cal O}}
\newcommand{\JJ}     {{\cal J}}

\newcommand{\amin}   {\mbox{$^\prime\ $}}

\newcommand{\ddeg}   {\mbox{${\rlap.}^\circ$}}
\newcommand{\wjjj}[6]{\left(\begin{array}{ccc}{#1}&{#2}&{#3}\\{#4}&{#5}&{#6}\end{array}\right)}
\newcommand{\beq}    {\begin{equation}}
\newcommand{\eeq}    {\end{equation}}
\newcommand{\beqa}   {\begin{eqnarray}}
\newcommand{\eeqa}   {\end{eqnarray}}

\newcommand{\refeqnt}[1]{{equation~(\ref{#1})}}

\newcommand{\reffigt}[1]{{Figure~\ref{#1}}}

\slugcomment{ApJ, in press, January 5, 2007}

\begin{document}

\title{Three-Year Wilkinson Microwave Anisotropy Probe 
(WMAP\altaffilmark{1}) Observations:\\
Temperature Analysis}

\author{G. Hinshaw \altaffilmark{2},
M. R. Nolta   \altaffilmark{3},
C. L. Bennett \altaffilmark{4},
R. Bean	 \altaffilmark{5},
O. Dor\'{e}	 \altaffilmark{3,11},
M. R. Greason \altaffilmark{6},
M. Halpern \altaffilmark{7},
R. S. Hill	 \altaffilmark{6},
N. Jarosik \altaffilmark{8},
A. Kogut   \altaffilmark{2},
E. Komatsu \altaffilmark{9},
M. Limon   \altaffilmark{6},
N. Odegard \altaffilmark{6},
S. S. Meyer   \altaffilmark{10},
L. Page	 \altaffilmark{8},
H. V. Peiris  \altaffilmark{10,15},
D. N. Spergel \altaffilmark{11},
G. S. Tucker  \altaffilmark{12},
L. Verde   \altaffilmark{13},
J. L. Weiland \altaffilmark{6},
E. Wollack \altaffilmark{2},
E. L. Wright  \altaffilmark{14}}

\altaffiltext{1}{\map\ is the result of a partnership between Princeton
University and NASA's Goddard Space Flight Center. Scientific guidance is
provided by the \map\ Science Team.}
\altaffiltext{2}{Code 665, NASA/Goddard Space Flight Center, 
Greenbelt, MD 20771}
\altaffiltext{3}{Canadian Institute for Theoretical Astrophysics, 
60 St. George St, University of Toronto, 
Toronto, ON  Canada M5S 3H8}
\altaffiltext{4}{Dept. of Physics \& Astronomy, 
The Johns Hopkins University, 3400 N. Charles St., 
Baltimore, MD  21218-2686}
\altaffiltext{5}{612 Space Sciences Building, 
Cornell University, Ithaca, NY  14853}
\altaffiltext{6}{Science Systems and Applications, Inc. (SSAI), 
10210 Greenbelt Road, Suite 600 Lanham, Maryland 20706}
\altaffiltext{7}{Dept. of Physics and Astronomy, University of 
British Columbia, Vancouver, BC  Canada V6T 1Z1}
\altaffiltext{8}{Dept. of Physics, Jadwin Hall, 
Princeton University, Princeton, NJ 08544-0708}
\altaffiltext{9}{Univ. of Texas, Austin, Dept. of Astronomy, 
2511 Speedway, RLM 15.306, Austin, TX 78712}
\altaffiltext{10}{Depts. of Astrophysics and Physics, KICP and EFI, 
University of Chicago, Chicago, IL 60637}
\altaffiltext{11}{Dept. of Astrophysical Sciences, 
Peyton Hall, Princeton University, Princeton, NJ 08544-1001}
\altaffiltext{12}{Dept. of Physics, Brown University, 
182 Hope St., Providence, RI 02912-1843}
\altaffiltext{13}{Univ. of Pennsylvania, Dept. of Physics and Astronomy, 
Philadelphia, PA  19104}
\altaffiltext{14}{UCLA Astronomy, PO Box 951562, Los Angeles, CA 90095-1562}
\altaffiltext{15}{Hubble Fellow}

\email{Gary.F.Hinshaw@nasa.gov}

\begin{abstract}

We present new full-sky temperature maps in five frequency bands from 23 to 94
GHz, based on data from the first three years of the \map\ sky survey.  The new
maps are consistent with the first-year maps and are more sensitive.  The
three-year maps incorporate several improvements in data processing made
possible by the additional years of data and by a more complete analysis of the
polarization signal.  These include several new consistency tests as well as
refinements in the gain calibration and beam response models
\citep{jarosik/etal:prep}.

We employ two forms of multi-frequency analysis to separate astrophysical
foreground signals from the CMB, each of which improves on our first-year
analyses.  First, we form an improved ``Internal Linear Combination'' (ILC) map,
based solely on \map\ data, by adding a  bias correction step and by quantifying
residual uncertainties in the resulting map. Second, we fit and subtract new
spatial templates that trace Galactic emission; in particular, we now use
low-frequency \map\ data to trace synchrotron emission instead of the 408 MHz
sky survey.  The \map\ point source catalog is updated to include 115 new
sources whose detection is made possible by the improved sky map sensitivity.

We derive the angular power spectrum of the temperature anisotropy using a
hybrid approach that combines a maximum likelihood estimate at low $l$ (large
angular scales) with a quadratic cross-power estimate for $l>30$.  The resulting
multi-frequency spectra are analyzed for residual point source contamination. 
At 94 GHz the unmasked sources contribute $128 \pm 27$ $\mu$K$^2$ to
$l(l+1)C_l/2\pi$ at $l=1000$. After subtracting this contribution, our best
estimate of the CMB power spectrum is derived by averaging cross-power spectra
from 153 statistically independent channel pairs.  The combined spectrum is
cosmic variance limited to $l=400$, and the signal-to-noise ratio per $l$-mode
exceeds unity up to $l=850$. For bins of width $\Delta l / l = 3$\%, the
signal-to-noise ratio exceeds unity up to $l=1000$. The first two acoustic peaks
are seen at $l=220.8 \pm 0.7$ and $l=530.9 \pm 3.8$, respectively, while the
first two troughs are seen at $l=412.4 \pm 1.9$ and $l=675.2 \pm 11.1$,
respectively.  The rise to the third peak is unambiguous; when the \map\ data
are combined with higher resolution CMB measurements, the existence of a third
acoustic peak is well established.

\citet{spergel/etal:prep} use the three-year temperature and polarization data
to constrain cosmological model parameters.  A simple six parameter $\Lambda$CDM
model continues to fit CMB data and other measures of large scale structure
remarkably well.  The new polarization data \citep{page/etal:prep} produce a
better measurement of the optical depth to re-ionization, $\tau = 0.089 \pm
0.03$.  This new and tighter constraint on $\tau$ helps break a degeneracy with
the scalar spectral index which is now found to be $n_s = 0.958 \pm 0.016$.   If
additional cosmological data sets are included in the analysis, the spectral
index is found to be $n_s = 0.947 \pm 0.015$.

\end{abstract}


\section{INTRODUCTION}
\label{sec:intro}

The Wilkinson Microwave Anisotropy Probe (\map) is a Medium-class Explorer
(MIDEX) mission designed to elucidate cosmology by producing full-sky maps of
the cosmic microwave background (CMB) anisotropy.  Results from the first year
of \map\ observations  were reported in a suite of papers published in the
Astrophysical Journal Supplement Series in September 2003 
\citep{bennett/etal:2003b, jarosik/etal:2003b, page/etal:2003b,
barnes/etal:2003, hinshaw/etal:2003b, bennett/etal:2003c, komatsu/etal:2003,
hinshaw/etal:2003, kogut/etal:2003, spergel/etal:2003, verde/etal:2003,
peiris/etal:2003, page/etal:2003c, bennett/etal:2003, page/etal:2003,
barnes/etal:2002, jarosik/etal:2003, nolta/etal:2003}. The data were made
available to the research community via the Legacy Archive for Microwave
Background Data Analysis (LAMBDA), NASA's CMB Thematic Data Center, and were
described in detail in the \map\ Explanatory Supplement \citep{limon/etal:2003}.

Papers based on the first-year \map\ results cover a wide range of topics,
including: constraints on inflation, the nature of the dark energy, the dark
matter density, implications for supersymmetry, the CMB and \map\ as the premier
baryometer, intriguing features in the large-scale data, the topology of the
universe, deviations from Gaussian statistics, time-variable cosmic parameters,
the Galactic interstellar medium, microwave point sources, the Sunyaev-Zeldovich
effect, and the ionization history of the universe.  The \map\ data has also
been used to establish the calibration of other CMB data sets.

Our analysis of the first three years of \map\ data is now complete and the
results are presented here and in companion papers \citep{jarosik/etal:prep,
page/etal:prep, spergel/etal:prep}. The three-year \map\ results improve upon
the first-year set in many ways, the most important of which are the following.
(1) A thorough analysis of the polarization data has produced full-sky
polarization maps and power spectra, and an improved understanding of many
aspects of the data. (2) Additional data reduces the instrument noise, producing
power spectra that are 3 times more sensitive in the noise limited regime. (3)
Independent years of data enable cross-checks that were not previously possible.
(4) The instrument calibration and beam response have been better characterized.

This paper presents the analysis of the three-year temperature data, focusing on
foreground modeling and removal, evaluation of the angular power spectrum, and
selected topics beyond the power spectrum.  Companion papers present the new 
polarization maps and polarization-specific scientific results
\citep{page/etal:prep}, and discuss the cosmological implications of the
three-year \map\ data \citep{spergel/etal:prep}.  \citet{jarosik/etal:prep}
present our new data processing methods and place systematic error limits on the
maps.

In \S\ref{sec:change} we summarize the major changes we have made to the data
processing since the first-year analysis, and \S\ref{sec:maps} presents a
synopsis of the three-year temperature maps.  In \S\ref{sec_gal_fg} we discuss
Galactic foreground emission and our attempts to separate the emission
components using a Maximum Entropy Method (MEM) analysis. 
\S\ref{sec:gal_remove} illustrates two methods we employ to remove Galactic
foreground emission from the maps in preparation for CMB analysis. 
\S\ref{sec:exfg} updates the \map\ point source catalog and presents a search
for the Sunyaev-Zeldovich effect in the three-year maps.  \S\ref{sec:tt}
evaluates the angular power spectrum and compares it to the previous \map\
spectrum and to other contemporary CMB results.  In \S\ref{sec:phase} we
survey the claims that have been made regarding odd features in the \map\
first-year sky maps, and we offer conclusions in \S\ref{sec:conclude}.

\section{CHANGES IN THE THREE-YEAR DATA ANALYSIS}
\label{sec:change}

The first-year data analysis was described in detail in the suite of first-year
\map\ papers listed above.  In large part, the three-year analysis employs the
same methods, with the following exceptions.

In the first-year analysis we subtracted the COBE dipole from the time-ordered
data to minimize the effect of signal aliasing that arises from pixelizing a
signal with a steep gradient.  Since the \map\ gain calibration procedure uses
the Doppler effect induced by \map's velocity with respect to the Sun to
establish the absolute calibration scale, \map\ data may be used independently
to determine the CMB dipole.  Consequently, we subtract the \map\ first-year
dipole \citep{bennett/etal:2003b} from the time-ordered data in the present
analysis.

A small temperature dependent pointing error ($\sim 1$ arcmin) was found during
the course of the first-year analysis.  The effect is caused by thermal stresses
on the spacecraft structure that induce slight movement of the star tracker with
respect to the instrument.  While the error was small enough to ignore in the
first-year data, it is now corrected with a temperature dependent model of the
relative motion \citep{jarosik/etal:prep}.

The radiometer gain model described by \citet{jarosik/etal:2003} has been
updated to include a dependence on the temperature of the warm-stage (RXB)
amplifiers.  While this term was not required by the first-year data, it is
required now for the model to fit the full three-year data with a single
parameterization.  The new model, and its residual errors, are discussed by
\citet{jarosik/etal:prep}.

The \map\ beam response has now been measured with six independent ``seasons''
of Jupiter observations.  In addition, we have now produced a physical model of
one side of our symmetric optical system, the A-side, based on simultaneous fits
to all 10 A-side beam pattern measurements \citep{jarosik/etal:prep}.  We use
this model to augment the beam response data at very low signal-to-noise ratio,
which in turn allows us to determine better the total solid angle and window
function of each beam.  

The far sidelobe response of the beam was determined from a combination of
ground measurements and in-flight lunar data taken early in the mission
\citep{barnes/etal:2003}.  In the first-year processing we applied a small
far-sidelobe correction to the K-band sky map.  For the current analysis, we
have implemented a new far sidelobe correction and gain re-calibration that
operates on the time-ordered data \citep{jarosik/etal:prep}.  These corrections
have now been applied to data from all 10 differencing assemblies.

When producing polarization maps, we account for differences in the frequency
pass-band between the two linear polarization channels in a differencing
assembly \citep{page/etal:prep}.  If this difference is not accounted for,
Galactic foreground signals would alias into linear polarization signals.

Due to a combination of 1/f noise and observing strategy, the noise in the \map\
sky maps is correlated from pixel to pixel.  This results in certain low-$l$
modes on the sky being less well measured than others.  This effect can be
completely ignored for temperature analysis since the low-$l$ signal-to-noise
ratio is so high, and the effect is not important at high-$l$
(\S\ref{sec:tt_noise}).  However, it is very important for polarization analysis
because the signal-to-noise ratio is so much lower.  In order to handle this
complexity, the map-making procedure has been overhauled to produce genuine
maximum likelihood solutions that employ optimal filtering of the time-ordered
data and a conjugate-gradient algorithm to solve the linear map-making equations
\citep{jarosik/etal:prep}.  In conjunction with this we have written code to
evaluate the full pixel-to-pixel weight (inverse covariance) matrix at low pixel
resolution. (The HEALPix convention is to denote pixel resolution by the
parameter $N_{\rm side}$, with $N_{\rm pix}=12 N_{\rm side}^2$
\citep{gorski/etal:2004}.  We define a resolution parameter $r$ such that
$N_{\rm side} = 2^r$.  The weight matrices have been evaluated at resolution r4,
$N_{\rm side}=16$, $N_{\rm pix}=3072$.)  The full noise covariance information
is propagated through the power spectrum analysis \citep{page/etal:prep}.

When performing template-based Galactic foreground subtraction, we now use
templates based on \map\ K- and Ka-band data in place of the 408 MHz synchrotron
map \citep{haslam/etal:1981}.  As discussed in \S\ref{sec:gal_template}, this
substitution reduces errors caused by spectral index variations that change the
spatial morphology of the synchrotron emission as a function of frequency.  A
similar model is used for subtracting polarized synchrotron emission from the
polarization maps \citep{page/etal:prep}. 

We have performed an error analysis of the internal linear combination (ILC) map
and have now implemented a bias correction as part of the algorithm.  We believe
the map is now suitable for use in low-$l$ CMB signal characterization, though
we have not performed a full battery of non-Gaussian tests on this map, so we
must still advise users to exercise caution.  Accordingly, we present full-sky
multipole moments for $l=2,3$, derived from the three-year ILC map.

We have improved the final temperature power spectrum ($C_l^{TT}$) by using a
maximum likelihood estimate for low-$l$ and a pseudo-$C_l$ estimate for $l>30$
(see \S\ref{sec:tt}).  The pseudo-$C_l$ estimate is simplified by using only V-
and W-band data, and by reducing the number of pixel weighting schemes to two,
``uniform'' and ``$N_{\rm obs}$'' (\S\ref{sec:tt_high_l}).  With three
individual years of data and six V- and W-band differencing assemblies (DAs) to
choose from, we can now form individual cross-power spectra from 15 DA pairs
within a year and from 36 DA pairs across 3 year pairs, for a total of 153
independent cross-power spectra.  In the first-year spectrum we included Q-band
data, which gave us 8 DAs and 28 independent cross-power spectra.  The arguments
for dropping Q-band from the three-year spectrum are given in
\S\ref{sec:tt_gal}.

We have developed methods for estimating the polarization power spectra
($C_l^{XX}$ for XX = TE, TB, EE, EB, BB) from temperature and polarization
maps.  The main technical hurdle we had to overcome in the process was the
proper handling of low signal-to-noise ratio data with complex noise properties
\citep{page/etal:prep}.  This step, in conjunction with the development of the
new map-making process, was by far the most time consuming aspect of the
three-year analysis.

We have improved the form of the likelihood function used to infer cosmological
parameters from the Monte Carlo Markov Chains \citep{spergel/etal:prep}.  In
addition to using an exact maximum likelihood form for the low-$l$ TT data, we
have developed a method to self-consistently evaluate the joint likelihood of
temperature and polarization data given a theoretical model (described in
Appendix~D of \citet{page/etal:prep}).  We also now account for
Sunyaev-Zeldovich (SZ) fluctuations when estimating parameters. Within the \map\
frequency range, it is difficult to distinguish between a primordial CMB
spectrum and a thermal SZ spectrum, so we adopt the \citet{komatsu/seljak:2002}
model for the SZ power spectrum and marginalize over the amplitude as a nuisance
parameter.

We now use the CAMB code \citep{lewis/challinor/lasenby:2000} to compute angular
power spectra from cosmological parameters.  CAMB is derived from CMBFAST
\citep{seljak/zaldarriaga:1996}, but it runs faster on our Silicon Graphics
(SGI) computers.

\section{OBSERVATIONS AND MAPS}
\label{sec:maps}

The three-year \map\ data encompass the period from 00:00:00 UT, 10 August 2001
(day number 222) to 00:00:00 UT, 9 August 2004 (day number 222).  The observing
efficiency during this time is roughly 99\%; Table~\ref{tab:baddata} lists the
fraction of data that was lost or flagged as suspect.  The Table also gives the
fraction of data that is flagged due to potential contamination by thermal
emission from Mars, Jupiter, Saturn, Uranus, and Neptune.  These data are not
used in map-making, but are useful for in-flight beam mapping
\citep{limon/etal:prep}.

\begin{figure}
\figurenum{1}
\includegraphics[width=3in]{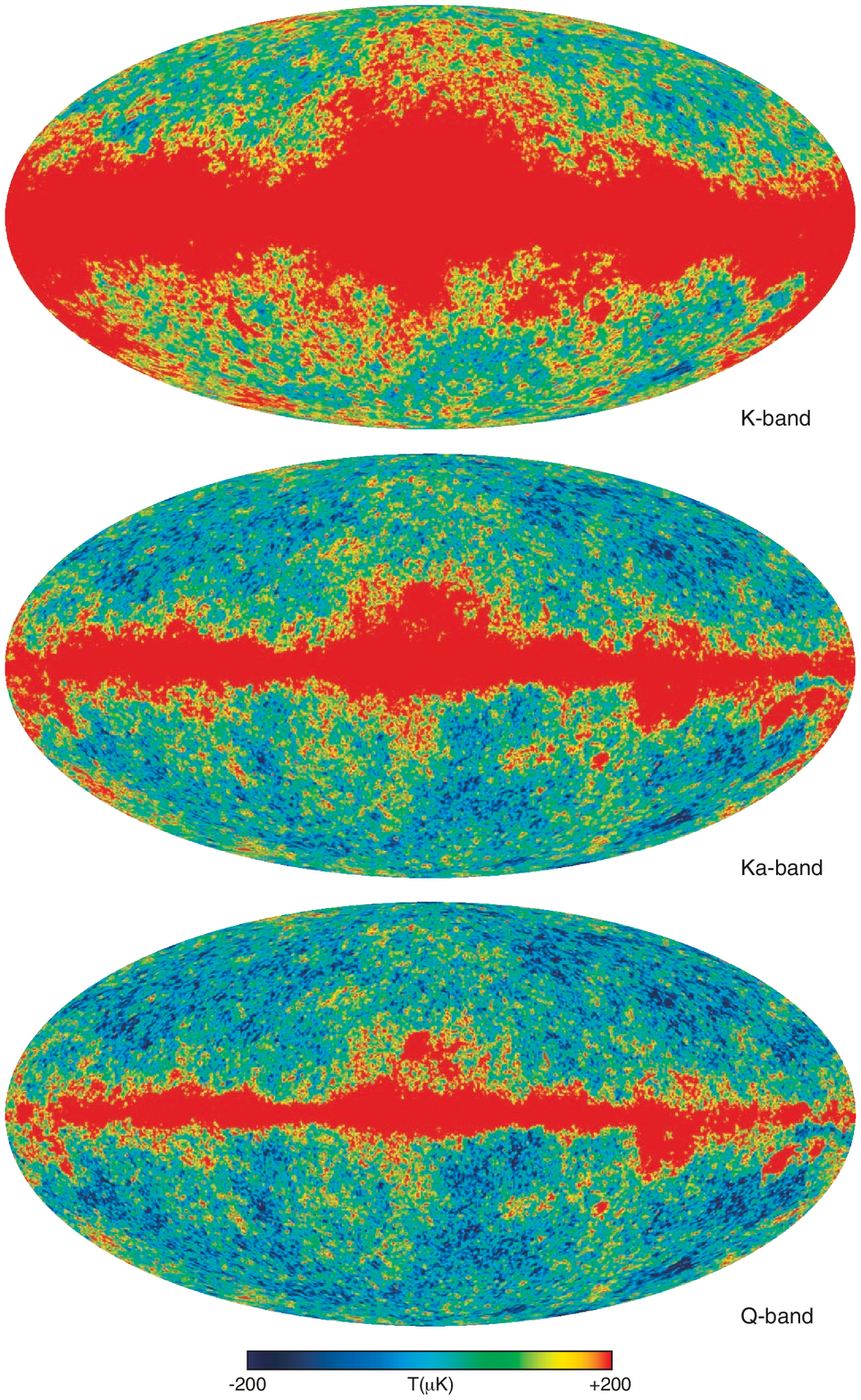}
\includegraphics[width=3in]{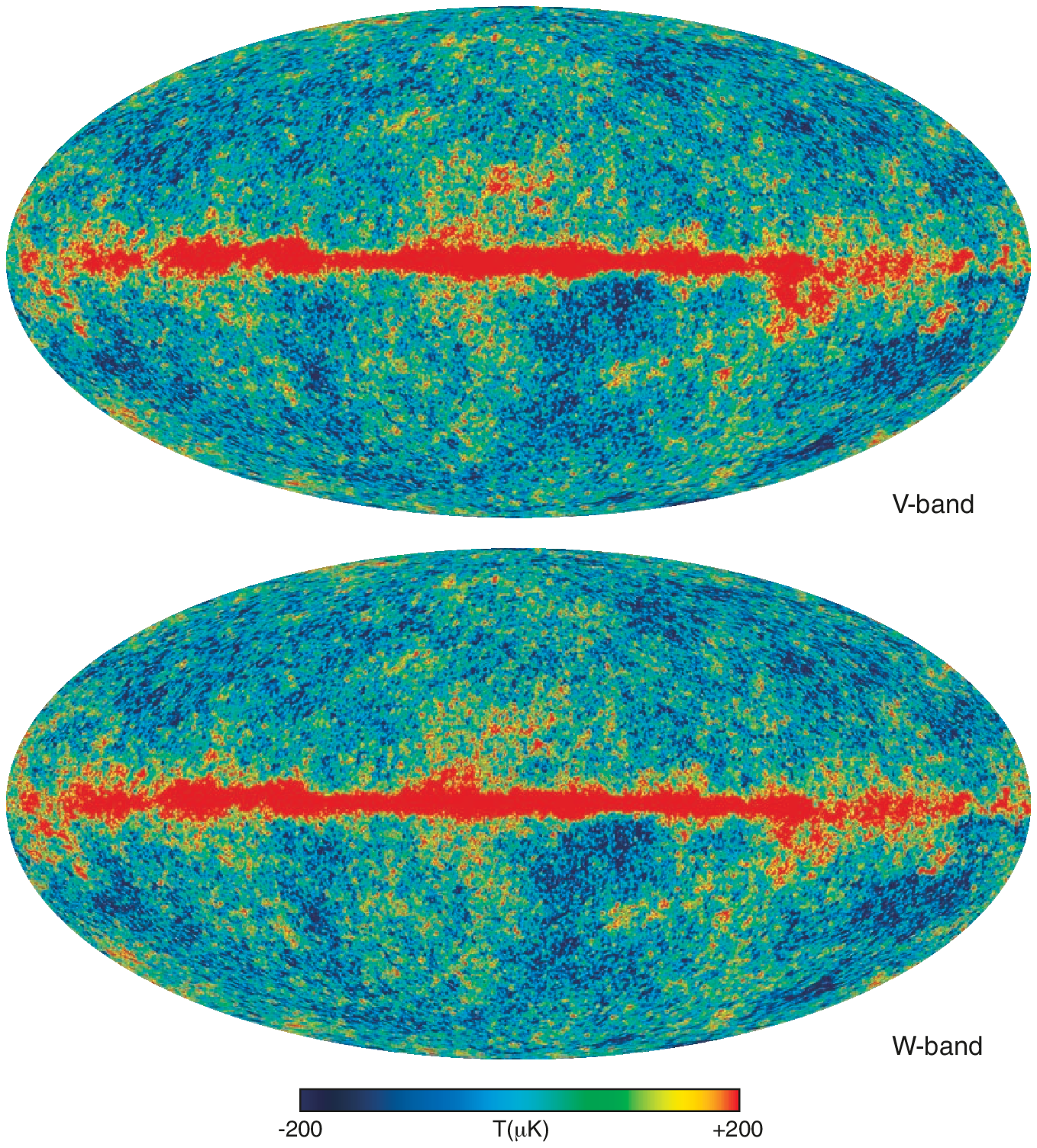}
\caption{Full-sky maps in Galactic coordinates smoothed with a $0\ddeg2$
Gaussian beam, shown in Mollweide projection. {\it top left}: K-band (23 GHz), {\it
middle left}: Ka-band (33 GHz), {\it bottom left}: Q-band (41 GHz),
{\it top right}: V-band (61 GHz),  {\it bottom right}: W-band (94 GHz).}
\label{fig:maps1}
\end{figure}

\begin{figure}
\figurenum{2}
\epsscale{0.7}
\plotone{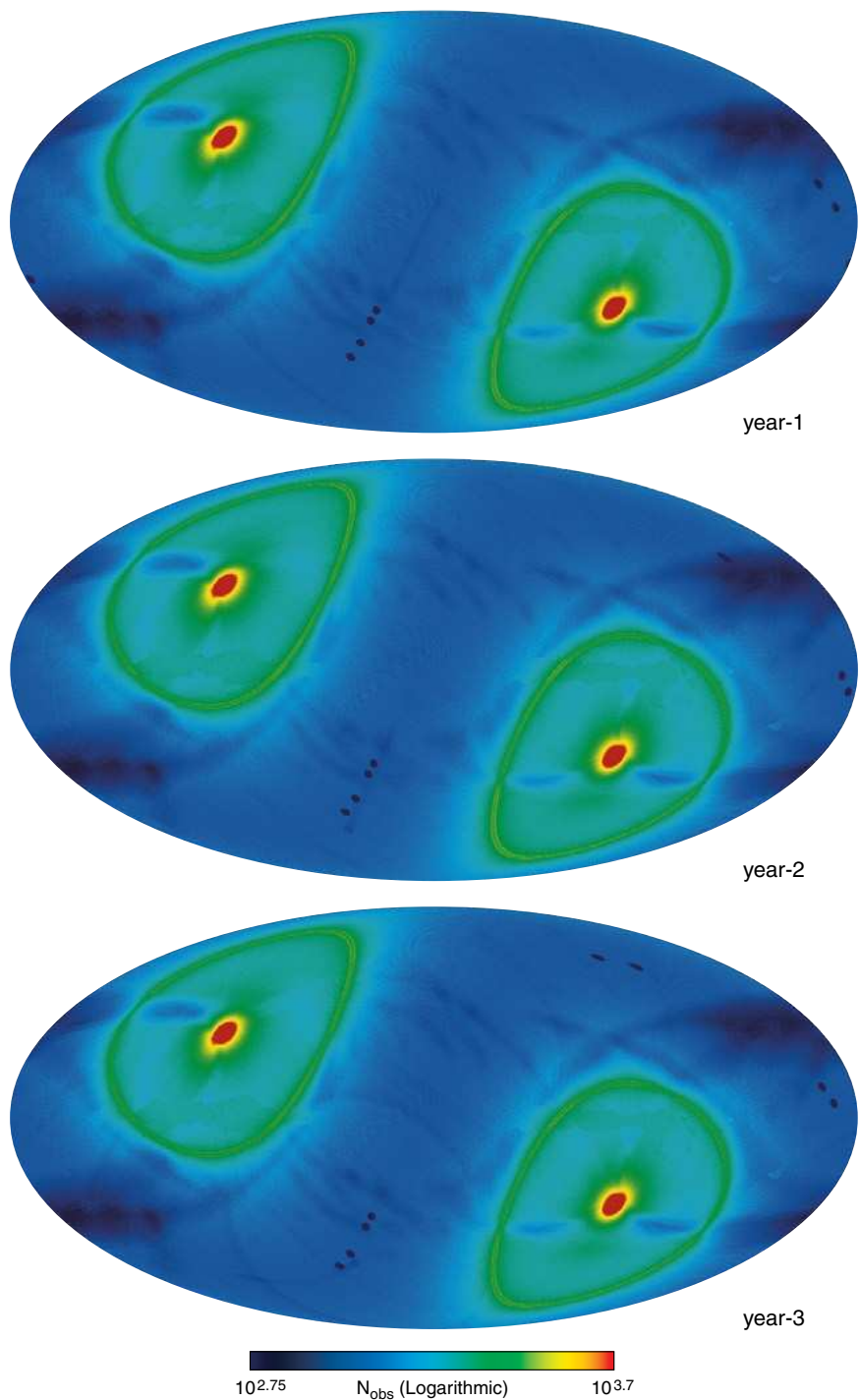}
\caption{The number of independent W-band observations per pixel in Galactic
coordinates, {\it top}: year-1, {\it middle}: year-2, and {\it bottom}: year-3. 
The number of observations is greatest near the ecliptic poles and in rings
approximately 141$^{\circ}$ from each pole (determined by the angular separation
between the two bore-sight directions). The number of observations is least in
the ecliptic plane. The small circular cuts in the ecliptic are where Mars,
Saturn, Jupiter, Uranus, and Neptune are masked so as not to contaminate the CMB
signal.  The coverage is quite consistent from year to year, with the planet
cuts being responsible for the largest fractional variation.}
\label{fig:nobs_123}
\end{figure}

Sky maps are created from the time-ordered data using the procedure described by
\citet{jarosik/etal:prep}.  For several reasons, we produce single-year maps for
each year of the three-year observing period (after performing an end-to-end
analysis of the instrument calibration).  We produce three-year maps by
averaging the annual maps.  Figure~\ref{fig:maps1} shows the three-year maps at
each of the five \map\ observing frequencies: 23, 33, 41, 61, and 94 GHz.  The
number of independent observations per pixel, $N_{\rm obs}$, is displayed in
Figure~\ref{fig:nobs_123}.  The noise per pixel, $p$, is given by $\sigma(p) =
\sigma_0 N_{\rm obs}^{-1/2}(p)$, where $\sigma_0$ is the noise per observation,
given in Table~\ref{tab:radiometers}.  To a very good approximation, the noise
per pixel in the three-year maps is a factor of $\sqrt{3}$ times lower than in
the one-year maps.  The noise properties of the data are discussed in more
detail in  \citet{jarosik/etal:prep}.  

\begin{figure}
\figurenum{3}
\epsscale{1.0}
\plotone{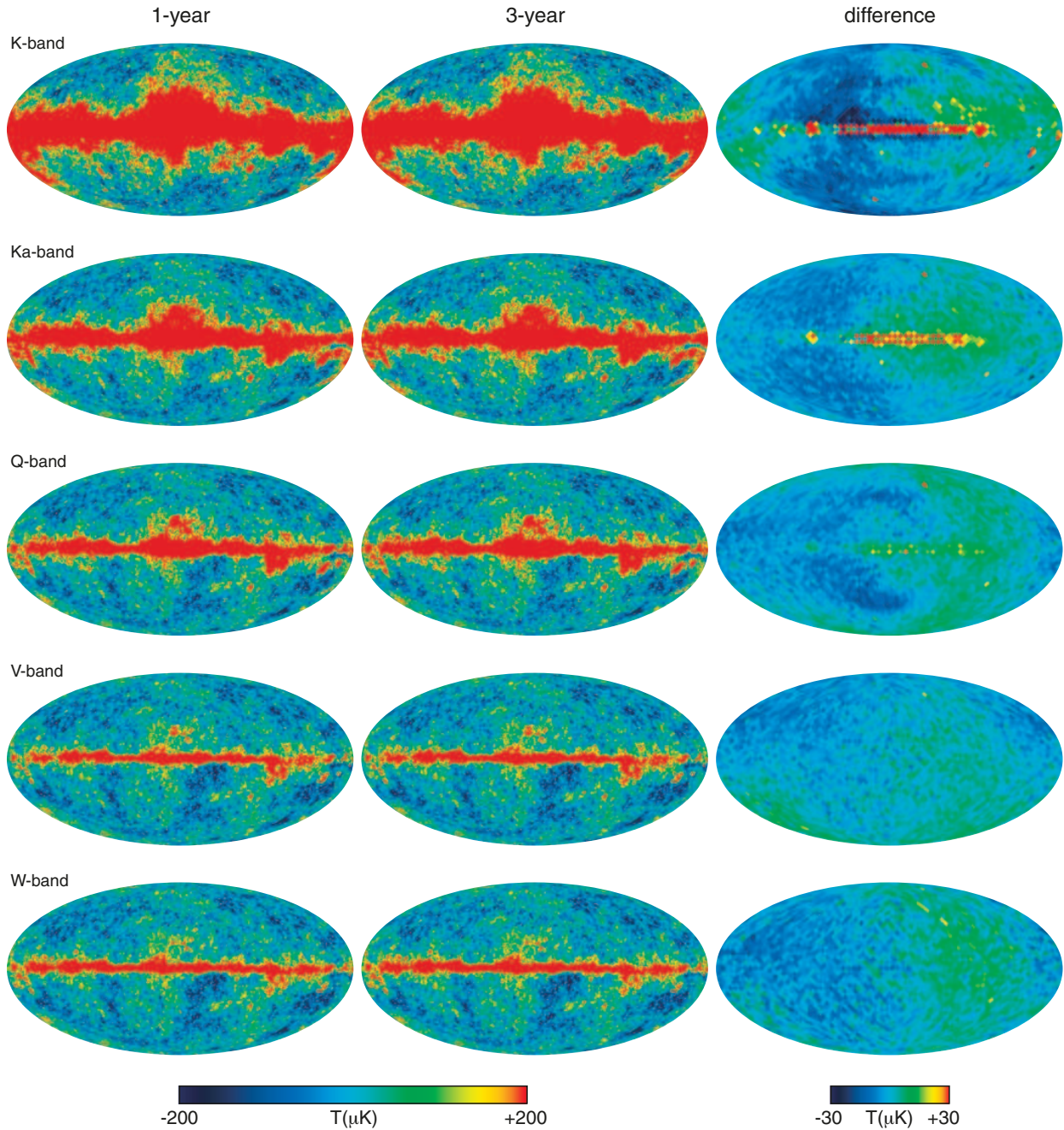}
\caption{Comparison of the three-year maps with the previously released one-year
maps.  The data are smoothed to $1^{\circ}$ resolution and are shown in Galactic
coordinates. The frequency bands K through W are shown top to bottom.  The
first-year maps ({\it left}) and the three-year maps ({\it middle}) are shown
scaled to $\pm$200 $\mu$K.  The difference maps ({\it right}) are degraded to
pixel resolution 4 and scaled to $\pm$20 $\mu$K.  The small difference in
low-$l$ power is mostly due to improvements in the gain model of the instrument
as a function of time \citep{jarosik/etal:prep}.  See \S\ref{sec:maps} and
Table~\ref{tab:3yr_1yr_diff}.}
\label{fig:maps_3yr_1yr}
\end{figure}

The three-year maps are compared to the previously released maps in
Figure~\ref{fig:maps_3yr_1yr}.  Both set of maps have been smoothed to
1$^{\circ}$ resolution to minimize the noise difference between them.  When
viewed side by side they look indistinguishable.  The right column of
Figure~\ref{fig:maps_3yr_1yr} shows the difference of the maps at each frequency
on a scale of $\pm$30 $\mu$K.  Aside from the noise reduction and a few bright
variable quasars, such as 3C279, the main difference between the maps is in the
large-scale (low-$l$) emission.  This is largely due to improvements in our
model of the instrument gain as a function of time, which is made possible by
having a longer time span with which to fit the model
\citep{jarosik/etal:prep}.  In the specific case of K-band, the improved
far-sidelobe pickup correction produced an effective change in the absolute
calibration scale by $\sim$1\%.  This, in turn, is responsible for the
difference seen in the bright Galactic plane signal in K-band
\citep{jarosik/etal:prep}.  We discuss the low-$l$ emission in detail in
\S\ref{sec:tt_low_l} and \S\ref{sec:phase}, but we stress here that the changes
shown in Figure~\ref{fig:maps_3yr_1yr} are small, even compared to the low
quadrupole moment seen in the first-year maps.  Table~\ref{tab:3yr_1yr_diff}
gives the amplitude of the dipole, quadrupole, and octupole moments in these
difference maps.  For comparison, we estimate the CMB power at $l=2,3$ to be
$\Delta T_l^2$ = 236 and 1053 $\mu$K$^2$, respectively (\S\ref{sec:tt_low_l}).

As discussed in \S\ref{sec:phase}, several authors have noted unusual
features in the large-scale signal recorded in the first-year maps.  We have not
attempted to reproduce the analyses presented in those papers, but based on the
small fractional difference in the large-scale signal, we anticipate that most
of the previously reported results will persist when the three-year maps are
analyzed.

\section{GALACTIC FOREGROUND ANALYSIS}
\label{sec_gal_fg}

The CMB signal in the \map\ sky maps is contaminated by microwave emission from 
the Milky Way Galaxy and from extragalactic sources.  In order to use the maps
reliably for cosmological studies, the foreground signals must be understood and
removed from the maps. In this section we present an overview of the mechanisms
that produce significant diffuse microwave emission in the Milky Way and we
assess what can be learned about them using a Maximum Entropy Method (MEM)
analysis of the \map\ data.  We discuss foreground removal in
\S\ref{sec:gal_remove}.

\subsection{Free-Free Emission}
\label{sec:gal_ff}

Free-free emission arises from electron-ion scattering which produces microwaves
with a brightness spectrum $T_A \sim (EM / 1\,{\rm cm}^{-6} {\rm pc})\;
\nu^{-2.14}$ for frequencies $\nu > 10$ GHz, where $EM$ is the emission measure,
$\int n^2_e dl$, and we assume an electron gas temperature $T_e \sim 8000$ K. 
As discussed in \citet{bennett/etal:2003c}, high-resolution maps of H$\alpha$
emission \citep{dennison/simonetti/topasna:1998, haffner/etal:2003,
reynolds/haffner/madsen:2002,  gaustad/etal:2001} can serve as approximate
tracers of free-free emission.  The intensity of H$\alpha$ emission is given by 
\beq
I({\rm R}) = 0.44\; \xi(\tau_d) \; (EM / 1\,{\rm cm}^{-6} {\rm pc}) \; 
(T_e/8000\,{\rm K})^{-0.5} \left[ 1 - 0.34\; \ln (T_e/8000\,{\rm K}) \right],
\eeq
where $I$ is in Rayleighs (1 R $ = 2.42 \times 10^{-7}$ ergs cm$^{-2}$s$^{-1}$
sr$^{-1}$ at the H$\alpha$ wavelength of 0.6563 $\mu$m), the helium contribution
is assumed to be small, and $\xi(\tau_d)$ is an extinction factor that depends
on the dust optical depth, $\tau_d$, at the wavelength of H$\alpha$.  If the
emitting gas is co-extensive with dust, then $\xi(\tau_d) =
[1-\exp(-\tau_d)]/\tau_d$.  H$\alpha$ is in R-band, where  the extinction is
0.75 times visible, $A_R=0.75\,A_V$; thus, $A_R=2.35 E_{B-V}$, and $\tau_d=2.2
E_{B-V}$.  \citet{finkbeiner:2003} assembled a full-sky H$\alpha$ map using data
from several surveys: the Wisconsin H-Alpha Mapper (WHAM), the Virginia Tech
Spectral-Line Survey (VTSS), and the  Southern H-Alpha Sky Survey Atlas
(SHASSA). We use this map, together with the
\citet{schlegel/finkbeiner/davis:1998} (SFD) extinction map, to predict a map of
free-free emission in regions where $\tau_d \lt 1$, under the assumption that
the dust and ionized gas are co-extensive.  As discussed in
\citet{bennett/etal:2003c}, this template has known sources of uncertainty and
error.  We use it as a prior estimate in the MEM analysis (\S\ref{sec:gal_mem}),
and as a free-free estimate in the template-based foreground removal
(\S\ref{sec:gal_template}).

\subsection{Synchrotron Emission}
\label{sec:gal_sync}

Synchrotron emission arises from the acceleration of cosmic ray electrons in  
magnetic fields.  In our Galaxy, discrete supernova remnants contribute only
$\sim$10\% of the total synchrotron emission at 1.5 GHz
\citep{lisenfeld/volk:2000, biermann:1976, ulvestad:1982}, while $\sim$90\% of
the observed emission arises from a diffuse component.
\citet{hummel/dahlem/vanderhulst:1991} present maps of synchrotron emission at
610 MHz and 1.49 GHz from the edge-on spiral galaxy NGC891.  They find that the
synchrotron spectral index varies from $\beta_{\rm s} \approx -2.6$ in most of
the galactic plane to $\beta_{\rm s} \approx -3.1$  in the halo.  Similar
spectral index variations are seen in the Milky Way at $\sim$1 GHz, where the
synchrotron signal is complex.  Variations of the synchrotron spectral index are
both expected and observed.  Moreover, the emission is dominated at low
frequencies by components with steep spectra, whereas at higher frequencies it
is dominated by components with flatter spectra, usually with a different
spatial distribution.  As a result, great care must be taken when using low
frequency maps, like the 408 MHz map of \citet{haslam/etal:1981}, as tracers of
the synchrotron emission at microwave frequencies.

Synchrotron emission can be highly polarized.  Theoretically, the linear
polarization fraction can be as high as $\sim$75\%, though values $\le 30$\% are
more typically observed.  See \citet{page/etal:prep} for a discussion of the new
full-sky observations of polarized synchrotron emission in the three-year \map\
data.

\subsection{Thermal Dust Emission}
\label{sec:gal_dust}

Thermal dust emission has been mapped over the full sky in several infrared
bands by the {\it IRAS} and \cobe\ missions. 
\citet{schlegel/finkbeiner/davis:1998} combined data from both missions to
produce an absolutely calibrated full-sky map of the thermal dust emission.
\citet{finkbeiner/davis/schlegel:1999} (FDS) extended this work to far-infrared
and microwave frequencies using the \cobe-FIRAS and \cobe-DMR data to constrain
the low-frequency dust spectrum.  They fit the data to a particular
two-component model that gives power-law emissivity indices $\alpha_1=1.67$ and
$\alpha_2=2.70$, and temperatures of $T_1=9.4$ K and $T_2=16.2$ K.  The fraction
of power emitted by each component is $f_1=0.0363$ and $f_2=0.9637$, and the
relative ratio of IR thermal emission to optical opacity of the two components
is $q_1/q_2 = 13.0$. The cold component is potentially identified as emission
from amorphous silicate grains while the warm component is plausibly carbon
based.  Independent of the physical interpretation of the model, FDS found that
it fit the data moderately well, with $\chi^2_\nu = 1.85$ for 118 degrees of
freedom.  \citet{bennett/etal:2003} noted that this model, call ``Model 8'' by
FDS, did well predicting the first-year \map\ dust emission.

It is reasonable to assume that the Milky Way is like other spiral galaxies and
that the microwave properties of external galaxies should help to inform our
understanding of the global properties of the Milky Way.  It has long been known
that a remarkably tight correlation exists between the broadband far-infrared
and broadband synchrotron emission in external galaxies.  This relation has been
extensively studied and modeled
\citep{
dickey/salpeter:1984,
dejong/etal:1985,
helou/soifer/rowanrobinson:1985,
sanders/mirabel:1985,
gavazzi/cocito/vettolani:1986,
hummel:1986,
wunderlich/wielebinski/klein:1987,
wunderlich/klein:1988,
beck/golla:1988,
fitt/alexander/cox:1988,
hummel/etal:1988,
mirabel/sanders:1988,
bicay/helou/condon:1989,
devereux/eales:1989,
unger/etal:1989,
voelk:1989,
chi/wolfendale:1990,
wunderlich/klein:1991,
condon:1992,
bressan/silva/granato:2002}.
All theories attempting to explain this tight correlation are tied to the level
of the star formation activity.  During this cycle, stars form, heat, and
destroy dust grains; create magnetic fields and relativistic electrons; and
create the O- and B-stars that ionize the surrounding gas.  However, it is not
clear what these models predict on a ``microscopic'' (cloud by cloud) level
within a galaxy.

\citet{bennett/etal:2003c} showed that the synchrotron and dust emission in our
own Galaxy are spatially correlated at \map\ frequencies.  Many authors have
argued that this correlation is actually due to radio emission from dust grains
themselves, rather than from a tight dust-synchrotron correlation. We review the
evidence for this more fully in the next section.

\subsection{``Anomalous'' Microwave Emission from Dust?}
\label{sec:anom_dust}

With the advent of high-quality diffuse microwave emission maps in the early
1990s, it became possible to study the high-frequency tail of the synchrotron
spectrum and the low-frequency tail of the interstellar dust spectrum. 
\citet{kogut/etal:1996a, kogut/etal:1996b} analyzed foreground emission in the
\cobe-DMR maps and reported a signal that was significantly correlated with 240
$\mu$m dust emission \citep{arendt/etal:1998} but not with 408 MHz synchrotron
emission \citep{haslam/etal:1981}.  The correlated signal was notably brighter
at 31 GHz than at 53 GHz ($\beta \sim -2.2$), hence they concluded it was
consistent with free-free emission that was spatially correlated with dust. The
same conclusion was reached by  \citet{deoliveira-costa/etal:1997}, who found
the Saskatoon 40 GHz data to be correlated with infrared dust, but not with
radio synchrotron emission.

\citet{leitch/etal:1997, leitch:1998}, and \citet{leitch/etal:2000} analyzed
data from the ``RING5m'' experiment.  A likelihood fit to their 14.5 GHz and
31.7 GHz data, assuming CMB anisotropy and a single foreground component,
produced a foreground spectral index of $\beta=-2.58^{+0.53}_{-0.42}$. The data
would have preferred a steeper value had it not been for an assumed prior limit
of $\beta>-3$.  This signal was fully consistent with synchrotron emission. 
However, a puzzle arose in comparing the RING5m data with a the  Westerbork
Northern Sky Survey (WENSS) \citep{rengelink:1997} at 325 MHz: the WENSS data
showed no detectable signal in the vicinity of the RING5m field.  The 325 MHz
limit implies that $\beta>-2.1$ and rules out conventional synchrotron emission
as the dominant  foreground.  (Since the WENSS is an interferometric survey
primarily designed to study discrete sources, the data are insensitive to
zero-point flux from extended emission.  It is not clear how much this affects
the above conclusion.)  As with the DMR and Saskatoon data, the 14.5 GHz
foreground emission was correlated with dust, but it was difficult to attribute
it to spatially correlated free-free emission because there was negligible
H$\alpha$ emission in the vicinity.  To reconcile this, a gas temperature in
excess of a million degrees would be needed to suppress the H$\alpha$.  Flat
spectrum synchrotron was also suggested as a possible source; it had been
previously observed in other sky regions and it would obviate the need for such
a high temperature and pressure.

\citet{draine/lazarian:1998a} dismissed the hot ionized gas explanation on
energetic grounds and instead suggested that the emission (which they described 
as ``anomalous'') be attributed to electric dipole rotational emission from very
small dust grains -- a mechanism first proposed by \citet{erickson:1957} in a
different astrophysical context.  One of the hallmarks of this mechanism is that
it produces a frequency spectrum that peaks in the 10-60 GHz range and falls off
fairly steeply on either side.

\citet{deoliveira-costa/etal:1998a} analyzed the nearly full-sky 19 GHz sky map
\citep{boughn/etal:1992} and found some correlation with the 408 MHz synchrotron
emission, but found a stronger correlation with the \cobe-DIRBE 240 $\mu$m dust
emission. They concluded the 19 GHz data were consistent with either free-free
or spinning dust emission.

\citet{leitch:1998} commented that the preferred model of
\citet{draine/lazarian:1998a} could produce the RING5m foreground component at
31.7 GHz, but that it only accounted for at most 30\% of the 14.5 GHz emission,
even when adopting  unlikely values of the grain dipole moment.   Since
\cite{leitch/etal:2000} were primarily interested in studying the CMB 
anisotropy, they considered using the IRAS 100 $\mu$m map as a foreground
template  to remove the ``anomalous'' emission, regardless of its physical
origin.  They found,  however, that fitting only a CMB component and a
dust-correlated component produced an unacceptably high $\chi^2_\nu=10$ per
degree of freedom.  Thus, while the radio foreground morphology correlates with
dust, the correlation is not perfect.

\citet{finkbeiner/etal:2002} used the Green Bank 140 foot telescope to search
for spinning dust emission in a set of dusty sources selected to be promising
for detection.  Ten infrared-selected dust clouds were observed at 5, 8, and 10
GHz.  Eight of the ten sources yielded negative results, one was marginal, and
one (the only diffuse H II region of the ten, LPH 201.6) was claimed as a
tentative detection based on its spectral index of $\beta > -2$.  Recognizing
that this spectrum does not necessarily imply spinning dust emission, 
\citet{finkbeiner/etal:2002} offer three additional requirements to convincingly
demonstrate the detection of spinning dust, and concluded that none of the three
requirements was met by the existing data.  The absence of a rising spectrum in
most of the sources may be taken as evidence that spinning dust emission is not
typically dominant in this spectral region, at least for this type of
infrared-selected cloud.  The tentative detection in LPH 201.6 has met with
three criticisms: (1) lack of evidence for the premise that its radio emission
is proportional to its far-infrared  dust emission \citep{casassus/etal:2004}, 
(2) the putative spinning dust emission is stronger than theory predicts
\citep{mccullough/chen:2002}, and (3) the positive spectral index may be
accounted for by unresolved  optically thick emission
\citep{mccullough/chen:2002}.  On the latter point, follow-up observations
failed to identify a compact HII region candidate (McCullough, private
communication).

\citet{bennett/etal:2003c} fit the first-year \map\ foreground data to within
$\sim$1\% using a Maximum Entropy Method (MEM) analysis (see also
\S\ref{sec:gal_mem}).  As with the above-cited results, \map\ found that the 22
GHz to 33 GHz foregrounds are dominated by a component with a synchrotron-like
spectrum, but a dust-like spatial morphology.  \citet{bennett/etal:2003c}
suggested that this may be due to spatially varying synchrotron spectral indices
acting over a large frequency range, significantly altering the synchrotron
morphology with frequency.  A spinning dust component with a thermal dust
morphology and the Draine and Lazarian spectrum could not account for more than
$\sim 5$\% of the emission at 33 GHz.  Of course, the \map\ fit did not rule out
spinning dust as a sub-dominant emission source (as it surely must be at some
level), nor did it rule out spinning dust models with other spectra or spatial
morphologies.  

\citet{casassus/etal:2004} report evidence for anomalous microwave emission in
the Helix planetary nebula at 31 GHz, where at least 20\% of the emission is
correlated with 100 $\mu$m dust emission.  They rejected several explanations. 
The observed features are not seen in H$\beta$, ruling out free-free emission as
the source.  Cold grains are also ruled out as the source by the absence of 250
GHz continuum emission.  Very small grains are not expected to survive in
planetary nebulae, and none have been detected in the Helix, but Fe is strongly
depleted in the gas.  Instead, \citet{casassus/etal:2004} favor the notion of
magnetic dipole emission (produced by variations in grain magnetization) from
hot ferromagnetic classical grains \citep{draine/lazarian:1999}.  Although the
derived emissivity per nucleon in the Helix is a factor of $\sim5$ larger than
the highest end of the range predicted by \citet{draine/lazarian:1999}, this
excess could be explained by a high dust temperature, since
\citet{draine/lazarian:1999} assume an ISM temperature of 18 K instead of a
typical planetary nebula dust temperature of $\sim 100$ K. The fraction of 31
GHz Helix emission attributable to free-free is  estimated to be in the range
$36-80$\%.  This low level of free-free emission implies  an electron
temperature of $T_e=4600\pm 1200$ K, which is much lower than the value $T_d\sim
9000$ K based on collisionally excited lines.  This discrepancy may be  due to
strong temperature variations within the nebula.  \citet{casassus/etal:2004}
suggest that the \citet{finkbeiner/etal:2002} measurement of LPH 201.6 may also
be produced by magnetic dipole emission from classical dust grains.

\citet{deoliveiracosta/etal:2004} correlate the Tenerife 10 and 15 GHz data
\citep{gutierrez/etal:2000} with the \map\ non-thermal (``synchrotron'') map
that was produced as part of the first-year Maximum Entropy Method (MEM)
analysis of the \map\ foreground signal.  They detect a low frequency roll-off
in the correlated emission, as shown in Figure~\ref{fig:mem_spec}.  We consider
this result after discussing the three-year MEM analysis in the next section.

\citet{fernadez-cerezo/etal:2006} report new measurements with the COSMOSOMAS
experiment covering 9000 square degrees with $\sim$1$^{\circ}$ resolution at
frequencies of 12.7, 14.7, and 16.3 GHz.  In addition to CMB signal and what is
interpreted to be a population of unresolved radio sources, they find evidence
for diffuse emission that is correlated with the DIRBE 100 $\mu$m and 240 $\mu$m
bands.  As with many of the above-mentioned results, they find that the
correlated signal amplitude rises from 22 GHz (using \map\ data) to 16.3 GHz,
and that it shows signs of flattening below 16.3 GHz ``compatible with
predictions for anomalous microwave emission related to spinning dust.''

The topic of anomalous dust emission remains unsettled, and is likely to remain
so until high-quality diffuse measurements are available over a modest fraction
of sky in the 5-15 GHz frequency range.  We offer some further comments after
presenting the three-year MEM results in the following section.

\subsection{Maximum Entropy Method (MEM) Foreground Analysis}
\label{sec:gal_mem}

\cite{bennett/etal:2003c} described a MEM-based approach to modeling the
multifrequency \map\ sky maps on a pixel-by-pixel basis.  Since the method is
non-linear, it produces maps with complicated noise properties that are
difficult to propagate in cosmological analyses.  As a result, it is not a
promising method for foreground removal.  However, the method is quite useful in
helping to separate Galactic foreground components by emission mechanism, which
in turn informs our understanding of the foregrounds.

We model the temperature map at each frequency, $\nu$, and pixel, $p$, as
\beq
T_m(\nu,p) \equiv T_{\rm cmb}(p)
  + S_{\rm s}(\nu, p)\,T_{\rm s}(p) 
  + S_{\rm ff}(\nu, p)\,T_{\rm ff}(p) 
  + S_{\rm d}(\nu, p)\,T_{\rm d}(p), 
\eeq
where the subscripts cmb, s, ff, and d denote the CMB, synchrotron (including
any anomalous dust component), free-free, and thermal dust components,
respectively.  $T_c(p)$ is the spatial distribution of emission component $c$ in
pixel $p$, and $S_c(\nu, p)$ is the spectrum of emission $c$, which is not
assumed to be uniform across the sky.  We normalize the spectra ($S \equiv 1$)
at K-band for the synchrotron and free-free components, and at W-band for the
dust component.

The model is fit in each pixel by minimizing the functional $H = A + \lambda B$
\citep{press/etal:NRIC:2e}, where $A = \sum_\nu [T(\nu,p) - T_m(\nu,p)]^2 /
\sigma_{\nu}^2$, is the standard $\chi^2$ of the model fit, and $B = \sum_c
T_c(p) \, \ln [T_c(p)/P_c(p)]$ is the MEM functional (see below).  The parameter
$\lambda$ controls the relative weight of $A$ (the data) and $B$ (the prior
information) in the fit.  In the functional $B$, the sum over $c$ is restricted
to Galactic emission components, and $P_c(p)$ is a prior estimate of $T_c(p)$. 
The form of $B$ ensures the positivity of the solution $T_c(p)$ for the Galactic
components, which greatly alleviates degeneracy between the foreground
components.  

Throughout the MEM analysis, we smooth all maps to a uniform  $1^\circ$ angular
resolution.  To improve our ability to constrain and understand the foreground
components, we first subtract a prior estimate of the CMB signal from the data
rather than fit for it.  We use the ILC map described in \S\ref{sec:gal_ilc} for
this purpose and subtract it from all 5 frequency band maps.  Since \map\
employs differential receivers, the zero level of each temperature map is
unspecified.  For the MEM analysis we adopt the following convention.  In the
limit that the Galactic emission is described by a plane-parallel slab, we have
$T(\vert b \vert) = T_p \, \csc \vert b \vert$, where $T_p$ is the temperature
at the Galactic pole.  For each of the 5 frequency band maps, we assign the zero
level such that a fit of the form $T(\vert b \vert) = T_p \, \csc \vert b \vert
+ c$, over the range $-90^{\circ} < b < -15^{\circ}$, yields $c=0$.

We construct a prior estimate for dust emission, $P_{\rm d}(p)$, using Model 8
of \citet{finkbeiner/davis/schlegel:1999}, evaluated at 94 GHz.  The dust
spectrum is modeled as a straight power law, $S_{\rm d}(\nu) =
(\nu/\nu_W)^{+2.0}$.  For free-free emission, we estimate the prior, $P_{\rm
ff}(p)$, using the extinction-corrected H$\alpha$ map \citep{finkbeiner:2003}. 
This is converted to a free-free signal using a conversion factor of 11.4 $\mu$K
R$^{-1}$ (units of antenna temperature at K-band).  We model the spectrum as a
straight power law, $S_{\rm ff}(\nu) = (\nu/\nu_K)^{-2.14}$.  As noted in
\S\ref{sec:gal_ff}, the main source of uncertainty in this free-free estimate is
the level of extinction correction (in addition to any H$\alpha$ photometry
errors).  We reduce $\lambda$ in regions of high dust optical depth to minimize
the effect of errors in the prior.

For the synchrotron emission, we construct a prior estimate, $P_{\rm s}(p)$, by
subtracting an extragalactic brightness of 5.9 K from the Haslam 408 MHz map
\citep{lawson/etal:1987} and scaling the result to K-band assuming
$\beta_s=-2.9$.  Since the synchrotron spectrum varies with position on the sky,
this prior estimate is expected to be imperfect.  We account for this in
choosing $\lambda$, as described  below.  We construct an initial spectral model
for the synchrotron, $S_{\rm s}(\nu, p)$, using the spectral index map
$\beta_{\rm s}(p) \equiv \beta(408\,{\rm MHz},23\,{\rm GHz})$.  Specifically, we
form $S_{\rm s}(\nu, p) = (\nu/\nu_K)^{\beta_{\rm s}-0.25\cdot[\beta_{\rm
s}+3.5]}$ for Ka-band, and $S_{\rm s}(\nu, p) = S_{\rm s}(\nu_{Ka},
p)\cdot(\nu/\nu_{Ka})^{\beta_{\rm s}-0.7\cdot[\beta_{\rm s}+3.5]}$ for Q-, V-,
and W-band ($S_{\rm s} \equiv 1$ for K-band). This allows for a
$\beta$-dependent steepening of the synchrotron spectrum at microwave
frequencies.  In our first-year results, use of this initial spectral model
produced solutions with zero synchrotron signal, $T_{\rm s}(p)$, in a few low
latitude pixels for which the K-Ka spectrum is flatter than free-free emission. 
For the three-year analysis, this problem is handled by setting the initial
model, $S_{\rm s}$, to be flatter than free-free emission in these pixels.

For all three emission components, the priors $P_c$ and the spectra $S_c$ are
fixed during each minimization of $H$.  As described below, we iteratively
improve the synchrotron spectrum model based on the residuals of the fit.

The parameter $\lambda$ controls the degree to which the solution follows the
prior.  In regions where the signal is strong, the data alone should constrain
the model without the need for prior information, so $\lambda$ can be small
(though we maintain $\lambda>0$ to naturally impose positivity on $T_c$ and
reduce degeneracy among the emission components).  As in the first-year
analysis, we base $\lambda(p)$ on the foreground signal strength: $\lambda(p) =
0.6\;[T_{\rm K}(p)/1\;{\rm mK}]^{-1.5}$, where $T_{\rm K}(p)$ is the K-band map
(with the ILC map subtracted) in units of mK, antenna temperature.  This gives
$0.2 \lsim \lambda \lsim 3$.

After each minimization of $H$, we update $S_{\rm s}$ for K- through V-band
according to
\beq
S^{\rm new}_{\rm s}(\nu, p)\,T_{\rm s}(p) 
  = S_{\rm s}(\nu, p)\,T_{\rm s}(p) + g \cdot R(\nu,p)
\label{eq:mem_it}
\eeq
where $R(\nu,p) \equiv T(\nu,p) - T_m(\nu,p)$ is the solution residual, and $g$
is a gain factor that we set to 0.5.  For W-band we update $S_{\rm s}$ by 
power-law extrapolation using the inferred Q,V-band spectral index.  For some
low signal-to-noise ratio pixels, we occasionally find $\beta_{\rm
s}(\nu_Q,\nu_V) > 0$. In a change from our first-year analysis, we now 
extrapolate the synchrotron spectrum from V- to W-band using $\beta_{\rm
s}=-3.3$.  This change affects only 7\% of pixels.

We iterate the minimization of $H$ and the update of $S_{\rm s}$ eleven times. 
At the end of this cycle, the residual, $R(\nu,p)$, is less than 1\% of the
total signal $T(\nu,p)$.  However, there are still several sources of potential
error in the component decomposition, including:

(1) Zero-level uncertainty.  As noted above, we use a plane-parallel slab model
to assign the sky map zero point at each frequency.  We estimate the uncertainty
in this convention by fitting the model separately in the northern and southern
Galactic hemispheres.  These separate fits change the zero levels by as much as
$4$ $\mu$K.  When these differences are propagated into $T(\nu,p)$, the output
maps, $T_c$, change by $\lsim 15$\%, $\lsim 5$\%, and $\lsim 2$\% for the
free-free, synchrotron, and dust components, respectively, at low Galactic
latitudes.

(2) Dust spectrum uncertainty.  Changing $\beta_{\rm d}$ by 0.2 changes the
component maps by $\lsim 10$\%, $\lsim 3$\%, and $\lsim 2$\% for the free-free,
synchrotron, and dust components, respectively, at low Galactic latitudes.

(3) Dipole subtraction uncertainty.  A 0.5\% dipole error would result in a
systematic 0.5\% gain error in all bands.  The dominant effect would be to
rescale each component map by 0.5\%. 

(4) CMB signal subtraction uncertainty.  Errors in the ILC estimate of the CMB
signal will produce errors in the corrected Galactic data, $T(\nu,p) - T_{\rm
ILC}(p)$, used in the MEM analysis.  We quantify ILC errors in
\S\ref{sec:gal_ilc}, but note here that they are small compared to the total
Galactic signal at low latitudes.  However, they plausibly dominate the total
uncertainty at high Galactic latitudes, where the \map\ data are (fortunately)
dominated by CMB signal.  Fractional uncertainties of $\sim$100\% are not
unlikely in the foreground model at high latitudes.

\begin{figure}
\figurenum{4}
\epsscale{0.9}
\plotone{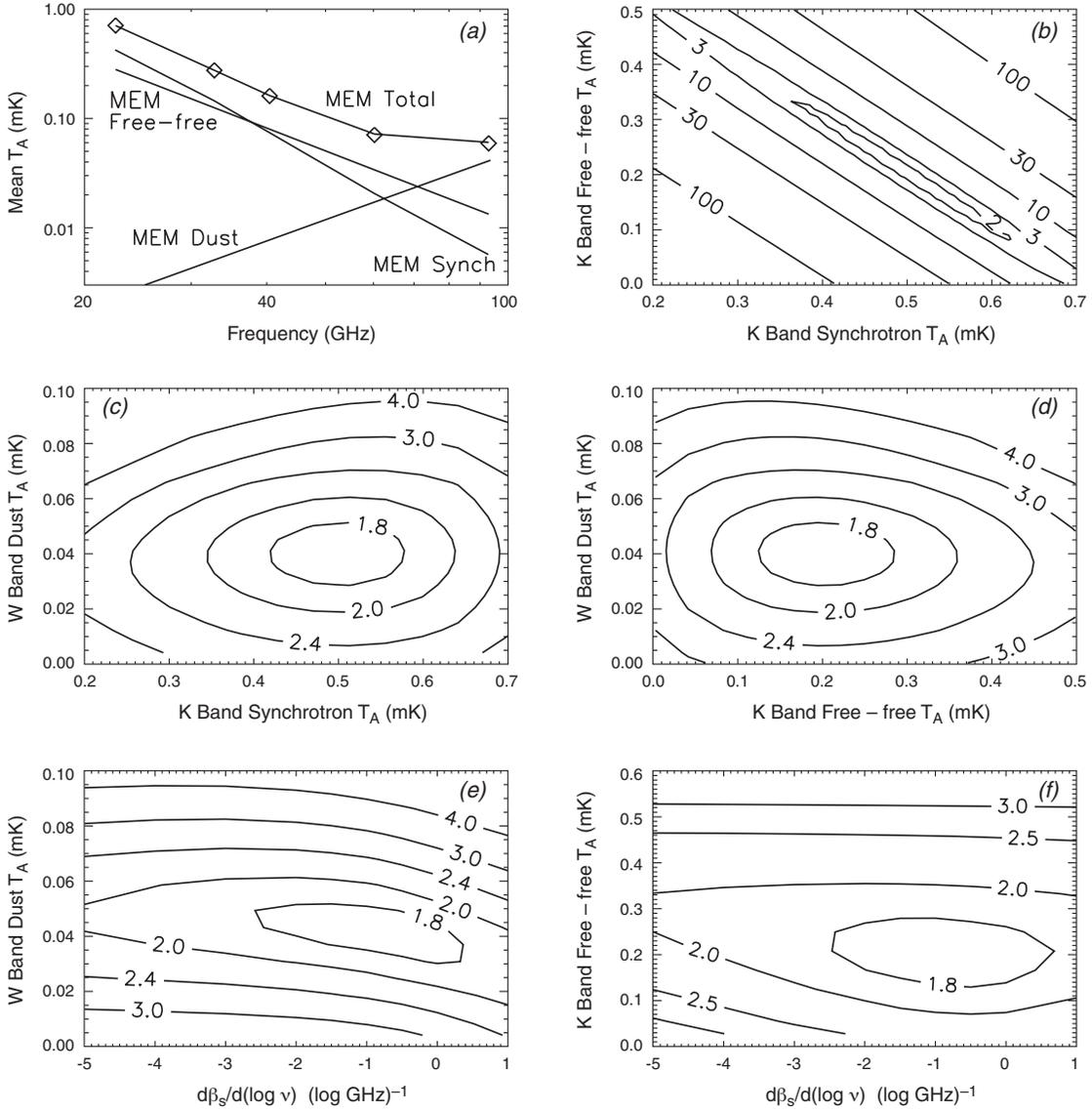}
\caption{Results from the MEM foreground degeneracy analysis. (a) The spectrum
of total foreground emission (diamonds) compared to the sum of the MEM
components (lines connecting the diamonds), averaged over the full sky. The
three component spectra of the model are also shown, as indicated. (b-f) The
contour plots illustrate the degeneracy between selected components.  The panels
show the change in the MEM functional (in units of $\Delta \chi^2$) as a
function of two global model parameters. For example, panel (b) shows the change
in the MEM functional that results from adding a constant value to the
synchrotron and/or free-free solutions, while panel (f) shows the dependence of
the functional on a global steepening of the synchrotron spectrum, modeled as
$\beta_{\rm s} = \beta_0 + d\beta_{\rm s}/d\log\nu \cdot (\log\nu -
\log\nu_K)$.  The only strong degeneracy is between the synchrotron and
free-free amplitudes in (b).  This effect is mitigated by the prior
distributions assumed for each component (see text).  The dust spectral index,
$\beta_d$, was found to be essentially unconstrained, so contour plots for that
quantity are not shown.}
\label{fig:mem_degen}
\end{figure}

(5) While the total model residuals are small, there is still potentially
significant uncertainty in the individual foreground components.  The program
outlined above produces three component maps, $T_{\rm s}$, $T_{\rm ff}$, and
$T_{\rm d}$, and the synchrotron spectrum model $S_{\rm s}(\nu,p)$. To
illustrate the degeneracy between these outputs, we consider a simplified
single-pixel model of the form
\beq
T_m(\nu) \equiv \EV{S_{\rm s}(\nu, p)}\,T_{\rm s} 
  + S_{\rm ff}(\nu)\,T_{\rm ff} + S_{\rm d}(\nu)\,T_{\rm d}, 
\eeq
where the angle brackets indicate a full-sky average, and we explicitly evaluate
the average MEM functional,
\beq
H = \sum_\nu \frac{[\EV{T(\nu,p)}-T_m(\nu)]^2}{\sigma_{\nu}^2} 
  + \EV{\lambda(p)}\,\sum_c T_c \, \ln [T_c/\EV{P_c(p)}],
\eeq
for selected pairs of parameters, while marginalizing over the rest.  Contour
plots of $H$ are shown in Figure~\ref{fig:mem_degen}.  For the panels that
explore the shape of the synchrotron spectrum, $S_{\rm s}$, we parameterize it
as a power law with a steepening parameter, $\beta_{\rm s}(\nu) = \beta_0 +
d\beta_{\rm s}/d\log\nu \cdot [\log\nu - \log\nu_K]$ and evaluate $H$ as a
function of $d\beta_{\rm s}/d\log\nu$, while marginalizing over $\beta_0$.  For
the rest, we iteratively update $S_{\rm s}$ as per equation~(\ref{eq:mem_it}). 
For the most part, the output parameters are only weakly correlated; the most
notable degeneracy is between the free-free amplitude and the synchrotron
amplitude.  \map\ data tightly constrain the {\it sum} of the two, but their
difference is determined by the relative amplitude of the prior estimates for
these two components, and by the initial synchrotron spectrum model.  The
synchrotron spectrum is found with modest significance to be steepening with
increasing frequency.  However, the dust index, $\beta_{\rm d}$ (not shown), is
poorly constrained; thus our conclusion in the first-year analysis that
$\EV{\beta_{\rm d}} = 2.2$ was not well founded upon further investigation.

\begin{figure}
\figurenum{5}
\epsscale{1.0}
\plotone{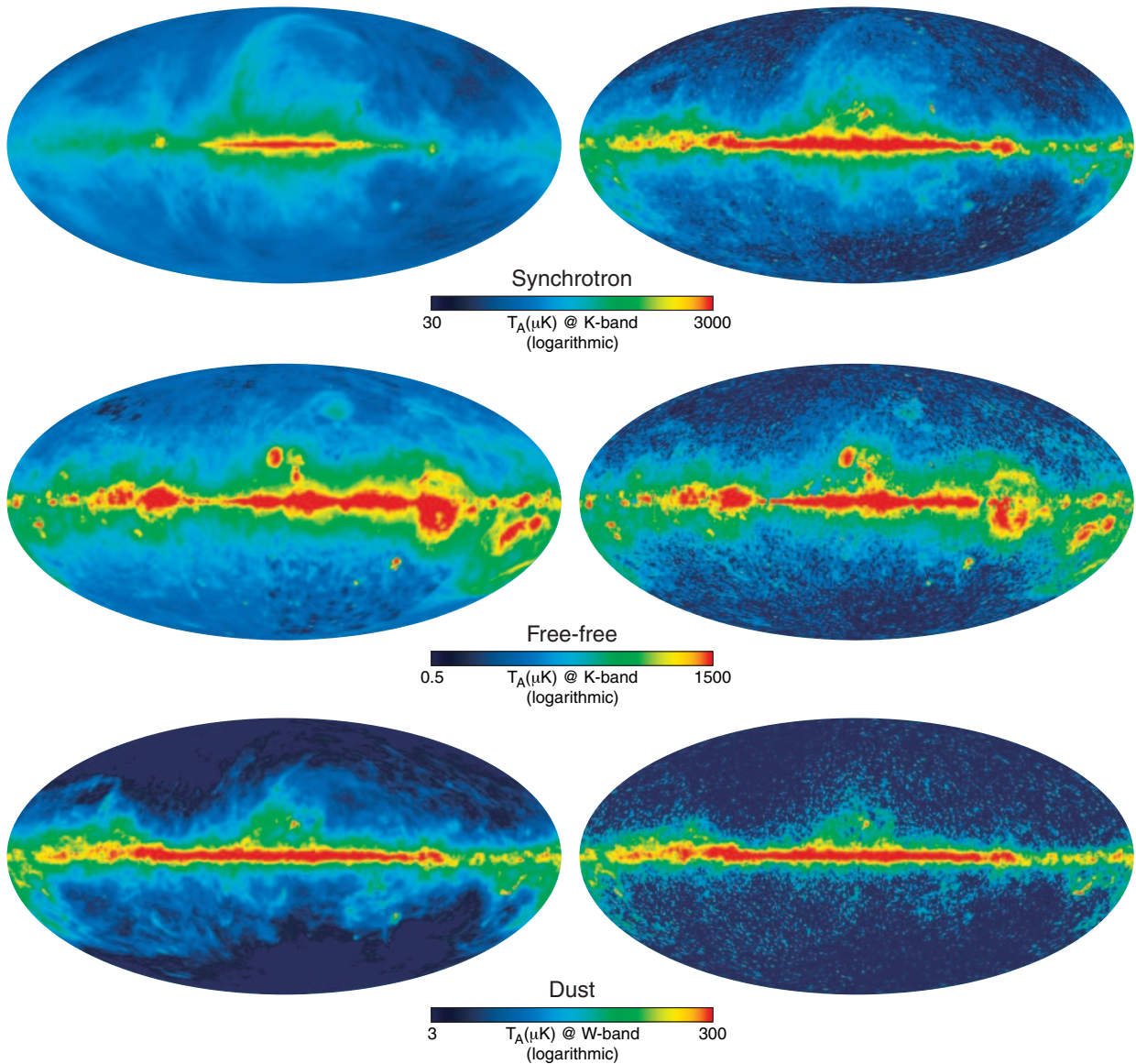}
\caption{Galactic signal component maps from the Maximum Entropy Method (MEM)
analysis (\S\ref{sec:gal_mem}).  {\it top-bottom}: synchrotron, free-free, and
dust emission with logarithmic temperature scales, as indicated.  {\it left}:
Input prior maps for each component.  {\it right}: Output maps based on
three-year \map\ data for each component.  See text for discussion.}
\label{fig:mem_maps}
\end{figure}

\begin{figure}
\figurenum{6}
\epsscale{1.0}
\plotone{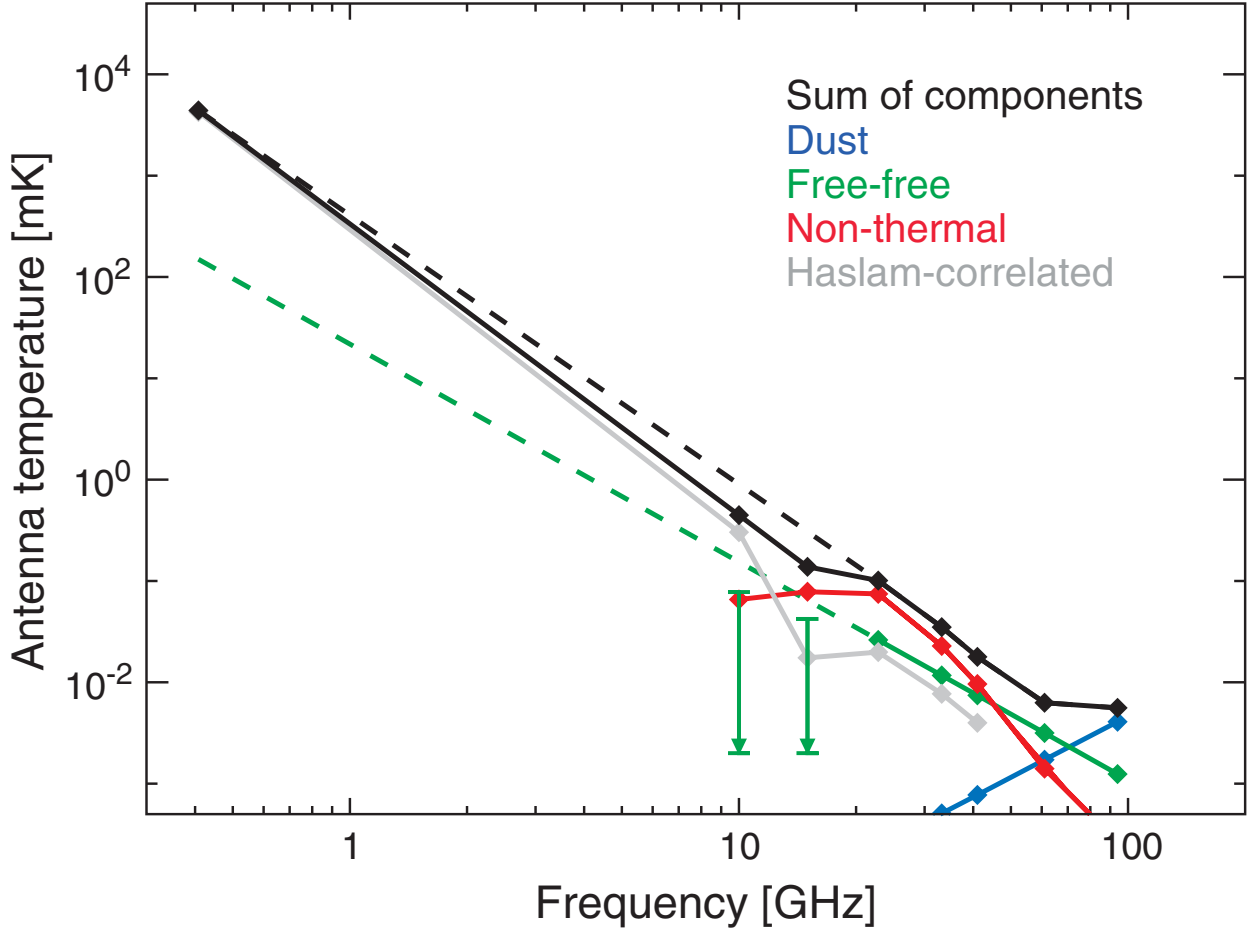}
\caption{The mean Galaxy spectrum from 408 MHz to 94 GHz, based on data from
Haslam, Tenerife and \map.  Each point is the mean signal in the Galactic
latitude range $20^{\circ} < \vert b \vert < 50^{\circ}$ with the mean at the
Galactic poles subtracted.  The \map\ values (23-94 GHz) were derived from the
MEM dust model (blue), the MEM free-free model (green), and the MEM non-thermal
component (red), and the sum of the three (black).  The Haslam (408 MHz) value
is computed directly from the map (black).  The dashed black curve is the
interpolation of the total signal between 408 MHz and 23 GHz.  The dashed green
curve is the extrapolation of the \map\ MEM free-free result to 408 MHz assuming
$\beta_{\rm ff} = -2.14$.  The Tenerife values were obtained by taking the
template correlation coefficients reported by \citet{deoliveira-costa/etal:2002}
and \citet{deoliveiracosta/etal:2004} (fit to $\vert b \vert > 20^{\circ}$
data), scaling the templates to obtain model emission maps, then evaluating the
mean signal in the same way as with the \map\ and Haslam data.  The Tenerife
correlation with H$\alpha$ provided only upper limits for the free-free signal. 
See \S\ref{sec:gal_mem} for more detail.}
\label{fig:mem_spec}
\end{figure}

Figure~\ref{fig:mem_maps} shows the three input prior maps, $P_c(p)$, and the
corresponding output component maps, $T_c(p)$, obtained from the three-year
data.  These maps are available on LAMBDA as part of the three-year data
release.  The maps are displayed using a logarithmic color stretch to highlight
a range of intensity levels.  The morphology and amplitude of the thermal dust
emission are well predicted by the prior (FDS) dust map (see also
\S\ref{sec:gal_template}).  The free-free emission is generally over-predicted 
by the prior discussed above, especially in regions of high extinction.  But
even in regions where the extinction is low, we find the mean free-free to
H$\alpha$ ratio at K-band to be closer to $\sim$8 $\mu$K R$^{-1}$ than the value
of 11.4 $\mu$K R$^{-1}$ assumed in generating the prior.  Moreover, we find
considerable variation in this ratio (a factor of $\sim$2) depending on
location.  

The most notable discrepancy between prior and output maps is seen in the
synchrotron emission.  Specifically, the K-band signal has a much more extended
Galactic longitude distribution than does the 408 MHz emission, and it is
remarkably well correlated with the thermal dust emission.  Is this K-band
non-thermal component due to anomalous dust emission or to mostly flat-spectrum
synchrotron emission that dominates at microwave frequencies and is well
correlated with dusty active star-forming regions?  We cannot answer this
question with \map\ data alone because the frequency range of \map\ does not
extend low enough to see the predicted rollover in the low-frequency anomalous
dust spectrum.  However, we note the following points.

In Figure~\ref{fig:mem_spec} we show the mean spectra of the three Galactic
emission components observed by \map, in addition to the sum of the three.  For
comparison, we also show the signal observed at 408 MHz and we infer the signals
at 10 and 15 GHz based on correlation analyses of the Tenerife CMB data by
\citet{deoliveira-costa/etal:2004} and \citet{deoliveira-costa/etal:1999}.  The
curves show the mean signal in the range $20^{\circ} < \vert b \vert <
50^{\circ}$, as computed from the output MEM component maps: blue is dust, green
is free-free, red is the non-thermal signal, and black is the sum of the three. 
The total intensity of the 408 MHz emission is remarkably well matched to a
simple power-law extrapolation of the total \map\ signal measured from K-band to
V-band.  The spectral index of the dashed black curve is $-$2.65 between 408 MHz
and K-band, and $-$2.69 between 408 MHz and Q-band.  If we interpret the
non-thermal emission as synchrotron, the implied spectral index between 408 MHz
and the K-band non-thermal component is $-$2.73.

As noted in \S\ref{sec:anom_dust}, \citet{deoliveiracosta/etal:2004} correlate
the Tenerife 10 and 15 GHz data with the first-year \map\ non-thermal MEM map. 
We infer the non-thermal signal at 10 and 15 GHz (shown in red) by scaling the
three-year non-thermal map using the reported correlation coefficients.  We find
a frequency rollover consistent with \citet{deoliveiracosta/etal:2004}.  Using
the same scaling method, we also infer the 10 and 15 GHz signals derived from
correlations with H$\alpha$ emission (green arrows) and the Haslam 408 MHz
(synchrotron) map (grey points).  The derived free-free emission is lower than
the extrapolation of the free-free emission inferred from \map\ -- which is not
physically tenable.  Similarly, the Haslam-correlated emission at 15 GHz is
substantially lower than it is at either 10 or 23 GHz. Thus, the sum of all
correlated components in the 10 and 15 GHz data requires a substantial dip in
the spectrum of the total emission (solid black points), which is not hinted at
by the Haslam or \map\ data (dashed black points).  This may be a sign of
substantial anomalous emission, but one must be cautious interpreting the
spectra of correlation coefficients.

\citet{page/etal:prep} have used the three-year \map\ polarization data to
construct a novel decomposition of the intensity foreground signal at each \map\
frequency band.  In brief, they predict a synchrotron polarization fraction from
a model of the Galaxy's magnetic field strength and electron density.  This
fraction is de-rated by an empirical factor to account for missing structure in
the model, then a ``high-polarization-fraction'' component of the intensity
signal is formed as $T_{\rm high}(\nu,p) \equiv P(\nu,p)/f(p)$, where $P(\nu,p)$
is the polarization intensity at frequency $\nu$, in mK, and $f(p)$ is the model
polarization fraction.  After subtracting an estimate of the CMB and free-free
signal from the intensity maps, the remaining non-thermal signal is attributed
to a ``low-polarization fraction'' component, $T_{\rm low}(\nu,p) \equiv
T'(\nu,p) - T_{\rm high}(\nu,p)$, where $T'$ is the temperature map corrected
for CMB and free-free emission.  The high-fraction component has a morphology
similar to the 408 MHz emission (Figure~9 in \citet{page/etal:prep}), while
low-fraction component has a very dust-like morphology (Figure~13 in
\citet{page/etal:prep}).  While the accuracy of this decomposition depends on
the model fraction, $f(p)$, the basic picture should inform future studies of
anomalous emission.

\section{GALACTIC FOREGROUND REMOVAL}
\label{sec:gal_remove}

The primary goal of foreground removal is to provide a clean map of the CMB for
cosmological analysis; an improved understanding of foreground astrophysics is a
secondary goal.  Removal techniques typically rely on the fact that the
foreground signals have quite different spatial and spectral distributions than
the CMB.  In this section we describe two-and-a-half approaches to foreground
removal that use complementary information.  The first is an update of the
Internal Linear Combination (ILC) method we employed in the first-year analysis
\citep{bennett/etal:2003c}.  The second is an updated approach to fitting
Galactic emission templates to each \map\ frequency band map.  The remaining
strategy is to mask regions of the sky that are too contaminated to be reliably
cleaned.  Extragalactic sources are treated in \S\ref{sec:exfg}.  For our
primary CMB results, we analyze the masked, template-subtracted maps, but for
some low-$l$ applications we also analyze the ILC map as a consistency check.

\subsection{Temperature Masks}
\label{sec:gal_mask}

Many regions of the sky are so strongly contaminated by foreground signals that
reliable cleaning cannot be assured.  These regions are masked for cosmological
analysis, though the extent of the masking required depends on the type of
analysis being done.  \citet{bennett/etal:2003c} defined a set of pixel masks
based on the first-year K-band temperature map.  Since these masks were based on
high signal-to-noise ratio Galactic signal contours in the K-band data, we have
not modified the diffuse emission masks for the three-year analysis.

In addition to diffuse Galactic emission, point sources also contaminate the
\map\ data.  A point source mask was constructed for the first-year analysis
that included all of the sources from  \citet{stickel/meisenheimer/kuehr:1994};
sources with 22 GHz fluxes $\ge 0.5$ Jy from \citet{hirabayashi/etal:2000};
flat  spectrum objects from \citet{terasranta/etal:2001}; and sources from the
blazar survey of \citet{perlman/etal:1998} and  \citet{landt/etal:2001}.  The
mask contained nearly 700 objects, including all 208 of the sources directly
detected by \map\ in the first-year data.  Each source was masked to a radius of
$0.6^\circ$.  For the three-year analysis, we have supplemented the source mask
with objects from the three-year \map\ source catalog discussed in
\S\ref{sec:src}.  Of the newly detected sources, 81 were not included in the
previous mask and have been added to the three-year mask.  Weaker, undetected
sources still contribute to the high-$l$ angular power spectrum.  As discussed
in \S\ref{sec:tt_gal}, we fit and subtract a residual source contribution to the
multi-frequency power spectrum data.

\begin{figure}
\figurenum{7}
\epsscale{1.0}
\plotone{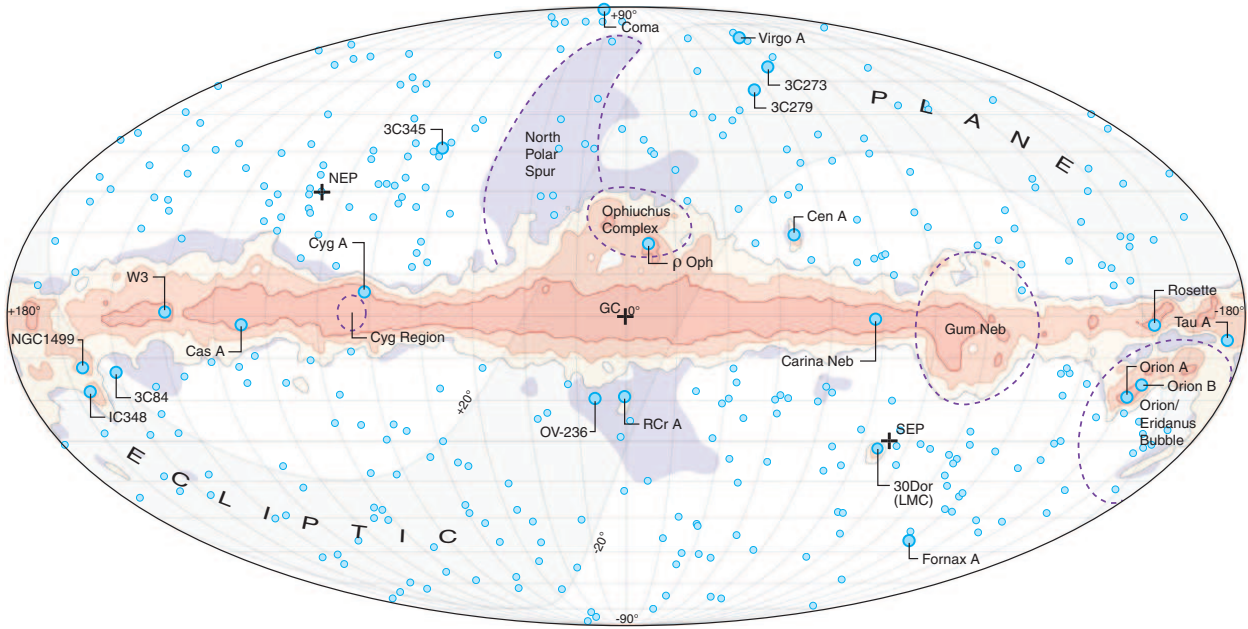}
\caption{An overview of the microwave sky.  The yellow, salmon, and red shaded
regions indicate the Kp0, Kp2, and Kp8 diffuse emission masks where the Galactic
foreground signal is especially strong.  See \citet{bennett/etal:2003c} for a
discussion of how these masks were constructed. The Kp0 and Kp2 masks are useful
for cosmological analysis, and the Kp8 mask closely follows the ``processing''
mask described in \citet{jarosik/etal:prep}, which is used for reducing
systematic errors in the sky maps.  The violet shading shows the ``P06''
polarization analysis mask described in  \citet{page/etal:prep}.  The small blue
dots indicate the position of point sources detected by \map.  Some well-known
sources and regions are specifically labeled.}
\label{fig:gal_overview}
\end{figure}

Figure~\ref{fig:gal_overview} gives an overview of the microwave sky and
indicates the extent of the various foreground masks.  The yellow, salmon, and
red shaded bands indicate the diffuse masks defined in
\citet{bennett/etal:2003c}.  The violet shading shows the ``P06'' polarization
analysis mask described in \citet{page/etal:prep}.  The small blue dots indicate
point sources detected by \map\ (to alleviate crowding, the full source mask
described above is not shown).  In addition, some well-known sources and regions
are specifically called out.

\subsection{The Internal Linear Combination (ILC) Method}
\label{sec:gal_ilc}

Linear combinations of the multi-frequency \map\ sky maps can be formed using 
coefficients that approximately cancel Galactic signals while preserving the CMB
signal.  This approach exploits the fact that the frequency spectrum of
foreground emission is different from that of the CMB.  The method is
``internal'' in that it relies only on \map\ data, so the calibration and
systematic errors of other experiments do not enter.  There are a number of ways
the coefficients can be determined, some of which require only minimal
assumptions about the nature of the foreground signals.  In the first year \map\
papers we introduced a method in which the coefficients were determined by
minimizing the variance of the resulting map subject to the constraint that the
coefficients sum to unity, in order to preserve the CMB signal.  We called the
resulting map the ``ILC'' map.  In this section we elaborate on the strengths
and limitations of the ILC method and quantify the uncertainties in the ILC
map.  

\citet{eriksen/etal:2004d} have also analyzed the method as an approach to
foreground removal.  They devised an approach to variance minimization that
employed a Lagrange multiplier to linearize the problem and dubbed the resulting
map the ``LILC'' map, where the first L denotes Lagrange.  They found their LILC
map differed somewhat from the ILC map in certain regions of the sky.  We have
since verified that the two minimization methods produce identical results for a
given set of inputs and that the differences were due to an ambiguity in the way
the regions were defined in the original ILC description.  Because the
linearized algorithm is considerably faster than our original nonlinear
minimization, we have adopted it in the present work.

\subsubsection{Uniform Foreground Spectra}
\label{sec:gal_ilc1}

In order to better understand how errors arise in the ILC map, we first consider
a simple scenario in which instrument noise is negligible and the spectrum of
the foreground emission is uniform across the sky, or within a defined region of
the sky.  In this case, a frequency map, $T_i(p) \equiv T(\nu_i,p)$, may be
written as a superposition of a CMB term, $T_{\rm c}(p)$, and a foreground term,
$S_i T_{\rm f}(p)$, where $S_i \equiv S(\nu_i)$ describes the composite
frequency spectrum of the foreground emission, and $T_{\rm f}(p)$ describes the
spatial distribution, so that $T_i(p) = T_{\rm c}(p) + S_i T_{\rm f}(p)$.  A
linear combination map has the form
\beq
T_{\rm ILC}(p) = \sum_i \zeta_i  T_i(p)
               = \sum_i \zeta_i \left[T_{\rm c}(p) + S_i T_{\rm f}(p)\right] 
               = T_{\rm c}(p) + \Gamma \, T_{\rm f}(p),
\eeq
where we have imposed the constraint $\sum_i \zeta_i = 1$, and have defined
$\Gamma \equiv \sum_i \zeta_i S_i$.

Suppose we choose to determine the coefficients $\zeta_i$  by minimizing the 
variance of $T_{\rm ILC}$.  Then, 
\beqa
\sigma_{\rm ILC}^2 
 & = & \left< T_{\rm ILC}^2(p) \right> - \left< T_{\rm ILC}(p) \right>^2 \\
 & = & \left< T_{\rm c}^2 \right> - \left< T_{\rm c} \right>^2 
       + 2\Gamma \left[\left< T_{\rm c} T_{\rm f} \right> 
               - \left< T_{\rm c} \right>\left< T_{\rm f} \right>\right]
       + \Gamma^2 \left[\left< T_{\rm f}^2 \right> - \left< T_{\rm f} \right>^2\right] \\
 & = & \sigma_{\rm c}^2 + 2\Gamma \sigma_{\rm cf} + \Gamma^2 \sigma_{\rm f}^2 
\label{eq:var}
\eeqa
where the angle brackets indicate an average over pixels, and we have defined
the variance and covariance in terms of these averages.  Note that this
expression would still hold if we added an arbitrary constant to each frequency
map, $T_i \rightarrow T_i + T_{0,i}$. The ILC variance will be minimized when
\beq
0 = \frac{\partial \sigma_{\rm ILC}^2}{\partial \zeta_i}
  = 2 \frac{\partial\Gamma}{\partial\zeta_i} \sigma_{\rm cf}
  + 2 \Gamma \frac{\partial\Gamma}{\partial\zeta_i} \sigma_{\rm f}^2.
\label{eq:ilc_min_0}
\eeq
Thus the coefficients $\zeta_i$ that minimize $\sigma_{\rm ILC}^2$ give $\Gamma
= -\sigma_{\rm cf}/\sigma_{\rm f}^2$, and in the absence of noise, the
corresponding ILC solution is
\beq
T_{\rm ILC}(p) = T_{\rm c}(p) - \sigma_{\rm cf}/\sigma_{\rm f}^2 \, T_{\rm f}(p),
\label{eq:ilc_bias_0}
\eeq
with
\beq
\sigma_{\rm ILC}^2 
 = \sigma_{\rm c}^2 - \sigma_{\rm cf}^2 / \sigma_{\rm f}^2.
\eeq
In this ideal case, the frequency maps combine in such a way as to maximize the
cancellation between CMB signal and foreground signal, producing a biased CMB
map with $\sigma_{\rm ILC}^2 \le \sigma_{\rm c}^2$.  We have tested this result
with ideal simulations in which we generate 5 frequency maps, $T_i$, which
include a Galaxy signal with a constant spectrum, $S_i$, and random realizations
of CMB signal and instrument noise.  We then generate ILC maps from each
realization and compare the residual map, $T_{\rm ILC} - T_{\rm c}$, to the bias
prediction, $-\sigma_{\rm cf}/\sigma_{\rm f}^2 \, T_{\rm f}$.  The results
confirm that the above description is correct, and that instrument noise is not
a significant concern in this situation.  The level of the bias is typically
$\sim$10 $\mu$K in the Galactic plane.

\subsubsection{Non-uniform Foreground Spectra}
\label{sec:gal_ilc2}

To minimize the anti-correlation bias we should choose regions that minimize the
covariance between the CMB and the foreground, $\left< T_{\rm c}T_{\rm f}
\right>$.  However, in the previous analysis we assumed that the spectra of the
foreground signals were constant over the sky.  In reality these will vary as
the ratio of synchrotron, free-free, and dust emission varies across the sky
(and as the intrinsic synchrotron and dust spectra vary).  In this case, the
bias analysis becomes more complex.  Specifically, the foreground component at
each frequency may be written as $S_i(p)T_{\rm f}(p)$, and the ILC map takes the
form
\beq
T_{\rm ILC}(p) = T_{\rm c}(p) + \Gamma(p) T_{\rm f}(p),
\eeq
where $\Gamma(p) \equiv \sum_i \zeta_i S_i(p)$.  The ILC variance then
generalizes to
\beq
\sigma_{\rm ILC}^2 = \left< T_{\rm c}^2 \right> - \left< T_{\rm c} \right>^2 
       + 2 \left[\left< T_{\rm c} \Gamma T_{\rm f} \right> 
               - \left< T_{\rm c} \right>\left< \Gamma T_{\rm f} \right>\right]
       + \left[\left< \Gamma^2 T_{\rm f}^2 \right> - \left< \Gamma T_{\rm f} \right>^2\right].
\eeq
Using the same reasoning that led to equation~(\ref{eq:ilc_min_0}), we obtain
the following result for the minimum variance solution
\beq
\left< \Gamma T_{\rm f} \cdot S_i T_{\rm f} \right>
= - \left< T_{\rm c} \cdot S_i T_{\rm f} \right>.
\label{eq:ilc_min_1}
\eeq
This has the same interpretation as equation~(\ref{eq:ilc_min_0}), in the sense
that it relates the foreground variance to the CMB-foreground covariance. We can
solve this equation for $\Gamma(p)$ by noting that $\Gamma(p) \equiv \sum_i
\zeta_i S_i(p)$, so that
\beq
\sum_j \left< S_i T_{\rm f} \cdot S_j T_{\rm f} \right> \, \zeta_j
= - \left< T_{\rm c} \cdot S_i T_{\rm f} \right>.
\eeq
Now define $F_{ij} \equiv \left< S_i T_{\rm f} \cdot S_j T_{\rm f} \right>$ and
$C_i \equiv \left< T_{\rm c} \cdot S_j T_{\rm f} \right>$, whereby
\beq
\Gamma = \sum_i \zeta_i S_i = -\sum_{ij} S_i \cdot (F^{-1})_{ij} \cdot C_j,
\label{eq:ilc_bias_1}
\eeq
which is the multi-frequency analog of equation~(\ref{eq:ilc_bias_0}).  Once
again though, the bias in the ILC solution is proportional to (minus) the
CMB-foreground covariance.

We have tested this expression with simulations like the ones described above,
except this time we employ a three-component Galaxy model with variable spectra,
$S(\nu_i,p)$, based on the first-year MEM model.  As we discuss in more detail
below, the simulations verify that the output ILC map is biased, $T_{\rm ILC}(p)
- T_{\rm c}(p) = \Gamma(p)T_{\rm f}(p)$, with $\Gamma$ as given in
equation~(\ref{eq:ilc_bias_1}).  Unfortunately, we do not know $T_{\rm f}$ and
$T_{\rm c}$ {\it a priori}, and it has proven difficult to relate this bias
expression to the frequency band maps, $T_i$, in a way the can be used to
minimize the bias.  As a result, we have primarily resorted to correcting the
bias with Monte Carlo simulations, as we describe below.

\begin{figure}
\figurenum{8}
\epsscale{0.75}
\plotone{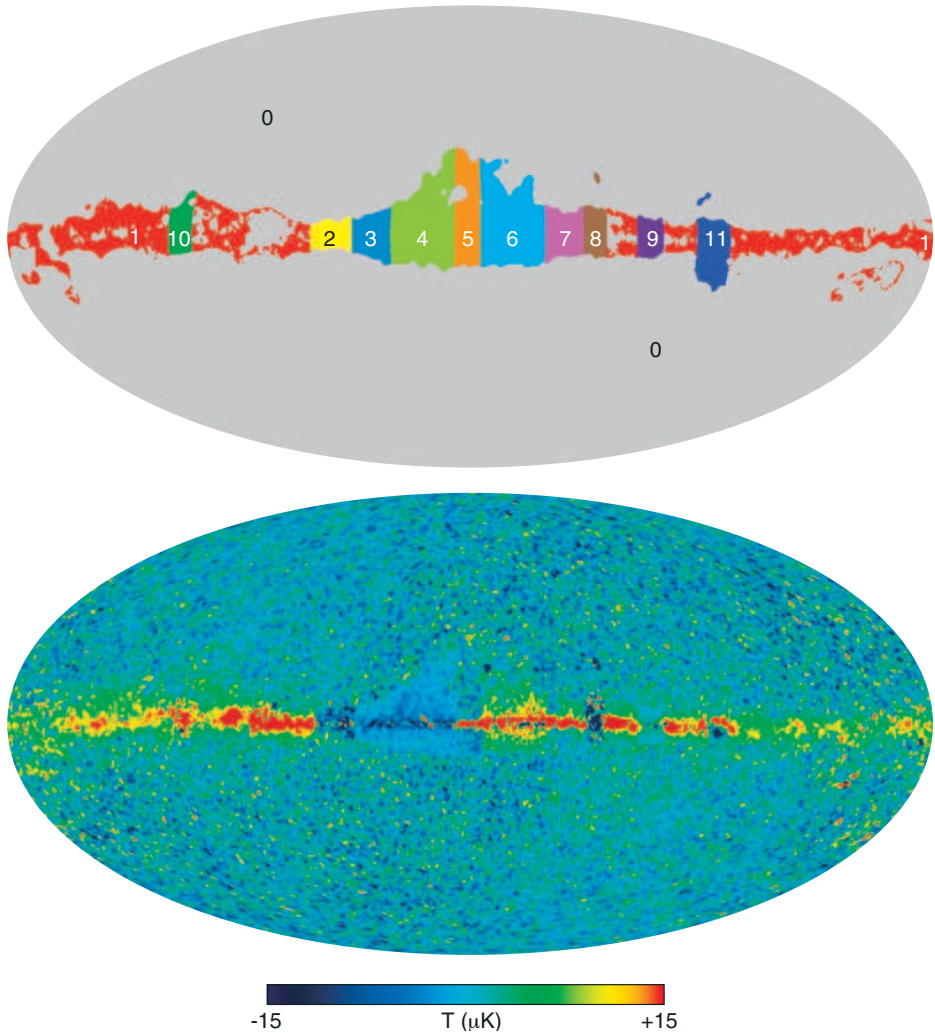}
\caption{{\it top}: Full-sky map color-coded to show the 12 regions that were
used to generate the three-year ILC map (see \S\ref{sec:gal_ilc}). {\it bottom}: The
mean ILC residual map from 100 Monte Carlo simulations of CMB signal, Galactic
foreground signal and instrument noise. The CMB signal was drawn from a
$\Lambda$CDM power spectrum that was modified to reproduce the power measured in
the first-year spectrum at $l=2,3$. The first-year MEM foreground model
was used to generate the (fixed) Galactic maps.  In the simulations, systematic
errors arose in the recovered ILC maps due to a combination of effects: (i) a
tendency for the minimum variance method to exploit (anti) alignments between
the CMB and foreground signal, and (ii) variations in the spectra of the
foreground signal, due mostly to changes in the admixture of synchrotron,
free-free, and dust emission (see \S\ref{sec:gal_ilc}).  This bias map was used to
correct the three-year ILC map (Figure~\ref{fig:ilc_3yr_1yr}).}
\label{fig:ilc_bias}
\end{figure}

Given the results above, and our previous experience with the ILC method, it is
clear that one should subdivide the sky into regions selected by foreground
spectra, in order to reduce bias prior to correcting it.  We have carried out
such a program for the three-year analysis and have found it very difficult to
improve on the region selection made in the first-year analysis
\citep{bennett/etal:2003c}.  Nonetheless, we have adopted a few changes: (1) we
eliminate the Taurus A region, as it is too small to ensure a reliable
CMB-foreground separation \citep{eriksen/etal:2004d}; (2) we add a new region to
minimize dust residuals in the Galactic plane.  This region is based on a $T_V -
T_W$ color selection and encompasses the outer Galactic plane within the Kp2
cut.  The new ILC region map is shown in Figure~\ref{fig:ilc_bias}.  The region
designated 1, shown in red, replaces the old Taurus A region; the remaining
regions are unchanged.  This map is available on LAMBDA as part of the
three-year data release.

For each region $n$, we determine a set of band weights, $\zeta_{n,i}$, by
minimizing the variance of the linear combination map $T_n(p) = \sum_{i=1}^5
\zeta_{n,i} T_i(p)$ in that region, subject to the constraint $\sum_i
\zeta_{n,i} = 1$.  There are two exceptions to note.  The coefficients for
region 0 were derived from a subset of the data in that region, specifically
pixels inside the Kp2 cut with $\vert l \vert > 60^{\circ}$.  The coefficients
for region 1 were derived from a slightly larger region of data, specifically
pixels inside the Kp2 cut  with $\vert l \vert > 50^{\circ}$, that pass the
$T_V-T_W$ color cut.  To ensure uniformity, the band maps have been smoothed to
a common resolution of $1^{\circ}$ FWHM.  The coefficients $\zeta_{n,i}$ are
given in Table~\ref{tab:ilc_wt}.  

We form a full-sky map by combining the $N$ region maps, $T_n$; but to minimize
edge effects, we blend the region maps as follows.  We create a set of $N$
full-sky weight maps, $w_n$ such that $w_n(p)=1$ for $p\in R_n$ and $w_n(p)=0$
otherwise.  We smooth these maps (which contain only ones and zeros) with a
$1\ddeg5$ smoothing kernel, to get smoothed weight maps, $\tilde{w}_n(p)$.  The
final full-sky map is then given by
\beq
T(p) = \frac{\sum_n \tilde{w}_n(p)\,T_n(p)}
            {\sum_n \tilde{w}_n(p)},
\eeq
where the sum is over the twelve sky regions.

\begin{figure}
\figurenum{9}
\epsscale{0.75}
\plotone{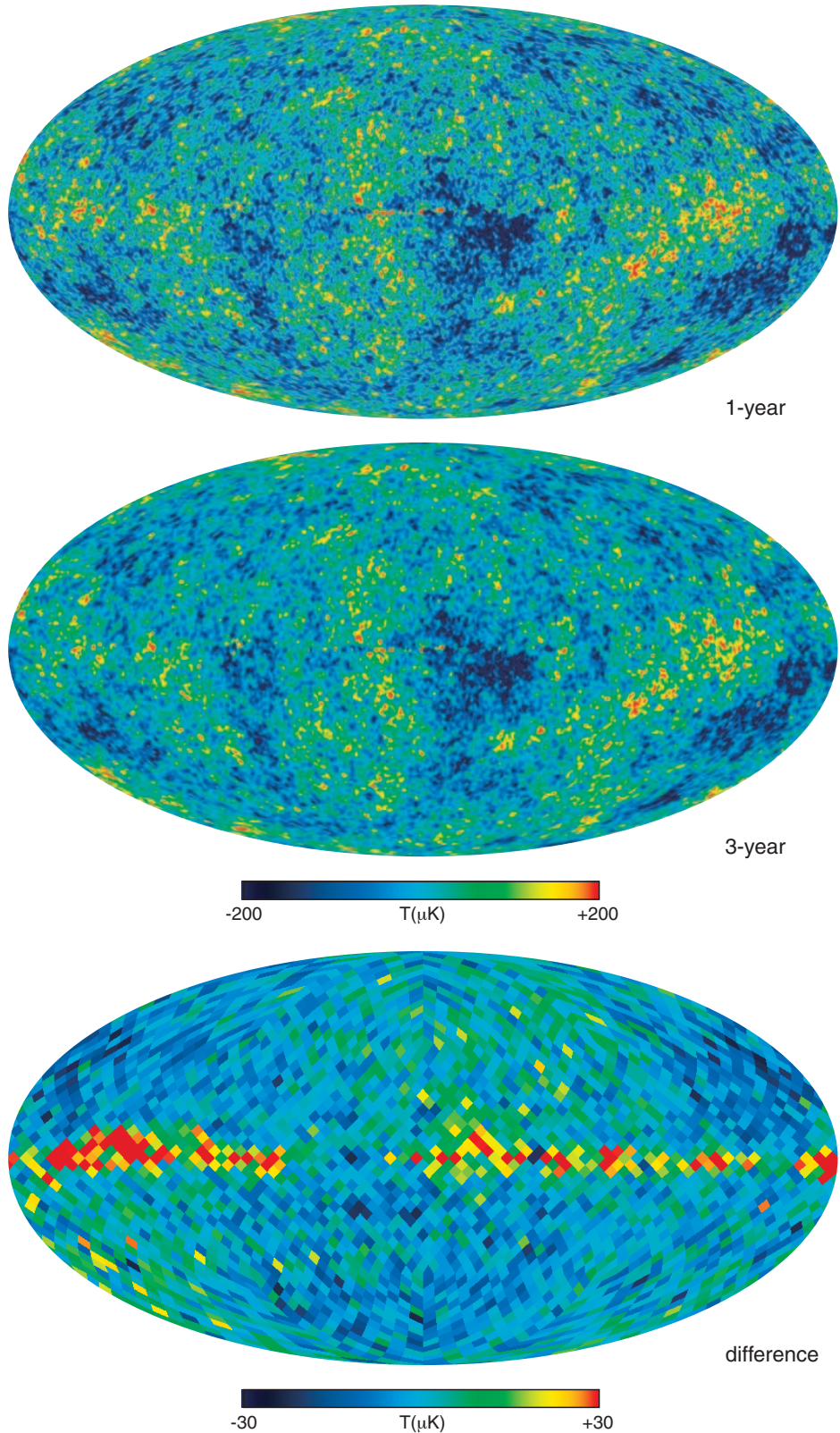}
\caption{{\it top}: The first-year ILC map reproduced from
\citet{bennett/etal:2003c}.  {\it middle}: The three-year ILC map produced
following the steps outlined in \S\ref{sec:gal_ilc}.  {\it bottom}: The difference
between the two (1-yr $-$ 3-yr).  The primary reason for the  difference is the
new bias correction (Figure~\ref{fig:ilc_bias}).  The low-$l$ change noted in
\S\ref{sec:maps} and shown in Figure~\ref{fig:maps_3yr_1yr} is also apparent.}
\label{fig:ilc_3yr_1yr}
\end{figure}

In order to obtain the final bias correction, we generate multi-region Monte
Carlo simulations, using the variable spectrum, MEM-based Galaxy model as
input.  We evaluate the error, $T_{\rm ILC} - T_{\rm c}$, for each realization
and compute a bias from the mean error averaged over 100 realizations.  The
result, shown in Figure~\ref{fig:ilc_bias}, is roughly 20-30 $\mu$K in the
Galactic plane, but substantially less off the plane.  This map was used to
correct the three-year ILC map, which is shown in the middle panel of Figure
~\ref{fig:ilc_3yr_1yr}.  This Figure also shows the first-year ILC map (top
panel) for comparison.  The difference between the two (bottom panel) is
primarily due to the new bias correction, but a small quadrupole difference, due
to the changes noted in Figure~\ref{fig:maps_3yr_1yr}, is also visible.

Based on the Monte Carlo simulations carried out for this ILC study, we estimate
that residual Galactic removal errors in the three-year ILC map are less than 5
$\mu$K on angular scales greater than $\sim10^{\circ}$.  But we caution that on
smaller scales, there is significant structure in the bias correction map that 
is still uncertain.  On larger scales, we believe the three-year ILC map
provides a reliable estimate of the CMB signal, with negligible instrument
noise, over the full sky.  We analyze the low-$l$ multipole moments of this map
in \S\ref{sec:tt_ilc}.

\subsection{Foreground Template Subtraction}
\label{sec:gal_template}

The ILC method discussed above produces a CMB map with complicated noise
properties, while the MEM method discussed in \S\ref{sec:gal_mem} is primarily
used to identify and separate  foreground components from each other.  For most
cosmological analyses one must retain the well-defined noise properties of the
\map\ frequency band maps.  To achieve this we form low-noise model templates of
each foreground emission component and fit them to the \map\ sky maps at each
frequency.  After subtracting the best-fit model we mask regions that cannot be
reliably cleaned because of limitations in the template models.  In this section
we describe the model templates we use for synchrotron, free-free, and dust
emission and we estimate the residual foreground uncertainties that remain after
these templates have been fit and subtracted.  The \map\ band maps are
calibrated in thermodynamic temperature units; where appropriate, we convert
Galactic signals to units of antenna temperature using the factors $g_{\nu}$
given in Table~\ref{tab:radiometers}.

In our first-year model we used the Haslam 408 MHz map as a template for 
synchrotron emission. We now use the \map\ K- and Ka-band data to provide a
synchrotron template, as described below.  This is preferable because: (1) the
intrinsic systematic measurement errors are smaller in the \map\ data than in
the Haslam data, and (2) the non-uniform synchrotron spectrum produces
morphological changes in the brightness as a function of frequency
\citep{bennett/etal:2003c}, so that the low frequency Haslam map is less
reliable at tracing microwave synchrotron emission than the \map\ data.

There are two potential pitfalls associated with using the K- and Ka-band data
for cleaning: (1) the data are somewhat noisy, and since the template
subtraction will be common to all cleaned channels, there can be a noise bias
introduced in the inferred angular power spectrum. (But note that we use
separate templates for each year of data, so the correlation only acts across
frequency bands within a single year.)  (2) Since the K- and Ka- band data are
contaminated with point sources, this signal could interfere with the primary
goal of cleaning the diffuse emission.  Using the fitting coefficients obtained
below and the known noise properties of the K- and Ka-band data, we estimate the
noise bias in the final power spectrum to be $<$5 $\mu$K$^2$ near the 1st
acoustic peak ($<$0.1\% of the CMB signal), and even smaller at lower and higher
multipoles.  Further, assuming the point source model given in
equation~(\ref{eq:source_model}), and the fact that the template has been
smoothed to an effective resolution of $1\ddeg0$ FWHM, we estimate that sources
contribute $<$1 $\mu$K$^2$ to the power spectrum at $l=400$ in the $T_{\rm K} -
T_{\rm Ka}$ template, and thus may be safely ignored.  In the end, these
pitfalls are not a source of concern for the three-year analysis.

The difference map $T_{\rm K}-T_{\rm Ka}$, in thermodynamic units, cancels CMB
signal while it contains a specific linear combination of synchrotron and
free-free emission (and a minimal level of thermal dust emission).  We use this
map as the first template in the model.  For the second template we use the
full-sky H$\alpha$ map compiled by \citet{finkbeiner:2003} with a correction for
dust extinction \citep{bennett/etal:2003c}.  This template independently traces
free-free emission, allowing the model to produce an arbitrary ratio of
synchrotron to free-free emission at a given frequency (the limitations of
H$\alpha$ as a proxy for free-free are discussed below).  For dust emission, we
adopt ``Model 8'' from the  \citet{finkbeiner/davis/schlegel:1999} analysis of
{\it IRAS} and \cobe\ data, evaluated at 94 GHz (see \S\ref{sec:gal_dust}).  The
full model has the form
\beq
M(\nu,p) = b_1(\nu) (T_{\rm K} - T_{\rm Ka}) + b_2(\nu) I_{H\alpha} + b_3(\nu) M_{\rm d},
\label{eq:gal_template}
\eeq
where $b_i(\nu)$ are the fit coefficients for each template at frequency $\nu$,
and $M_{\rm d}$ is the dust map.  As discussed below, this model is
simultaneously fit to the Q-, V-, and W-band maps, and we constrain the
coefficients $b_2$ and $b_3$ to follow specified frequency spectra to minimize
component degeneracy.

To clarify the physical interpretation of $b_1$ and $b_2$ we first note that
$T_{\rm K} - T_{\rm Ka}$ may be rewritten in terms of synchrotron and free-free
emission as
\beq
T_{\rm K} - T_{\rm Ka} =  R_{\rm s}\,T_{\rm s} + R_{\rm ff}\,T_{\rm ff},
\eeq
where $T_{\rm s}$ and $T_{\rm ff}$ are the synchrotron and free-free maps in
antenna temperature at K-band, $R_c \equiv g_{\rm K} S_c(\nu_{\rm K}, p) -
g_{\rm Ka} S_c(\nu_{\rm Ka}, p)$ is the surviving fraction of emission component
$c$ (synchrotron or free-free) in $T_{\rm K} - T_{\rm Ka}$, and $S_c$ is the
spectrum of component $c$, in antenna temperature, relative to K-band.  To a
very good approximation, the spectrum of free-free emission is $S_{\rm ff} =
(\nu/\nu_{\rm K})^{-2.14}$ (\S\ref{sec:gal_ff}), so that $R_{\rm ff} = 0.552$. 
For synchrotron emission, variations in the spectrum as a function of position
will produce variations in $R_{\rm s}$.  For spectral indices in the range
$\beta_{\rm s} = -2.9 \pm 0.2$ we have $R_{\rm s} = 0.667 \pm 0.026$.  In the
remainder of this section we assume $R_{\rm s} \equiv 0.67$ and neglect the
$\sim$2.5\% error introduced by spectral index variations between K and Ka
bands.  Note that this value of $R_{\rm s}$ is only used to estimate the level
of synchrotron emission in the template $T_{\rm K} - T_{\rm Ka}$; we do {\em
not} constrain the fit coefficients $b_1$ to follow a specified frequency
spectrum.

By adding the H$\alpha$ map to the model, we allow the synchrotron to free-free
ratio to vary as a function of frequency, but we must be cognizant of potential
errors introduced by the use of H$\alpha$ as a proxy for the free-free
emission.  Nominally we have $T_{\rm ff} = h_{\rm ff} I_{\rm H\alpha}$, where
$I_{\rm H\alpha}$ is the H$\alpha$ intensity in Rayleighs and $h_{\rm ff}$ is
the free-free to H$\alpha$ ratio.  At K-band $h_{\rm ff}$ is predicted to be
$\sim$11.4 $\mu$K R$^{-1}$ \citep{bennett/etal:2003c} but the actual ratio is
both uncertain and dependent locally on extinction and reflection.  These
effects make the H$\alpha$ proxy unacceptable in the Galactic plane, and force
one to mask these regions for CMB analysis.  Outside the masked regions the
variations in $h_{\rm ff}$ are primarily due to residual extinction and to
variations in the temperature of the emitting gas.  Here higher fractional
errors can be tolerated because the total free-free signal is fainter.

Due to the uncertainties in the free-free to H$\alpha$ ratio, and the fact that
$T_{\rm K} - T_{\rm Ka}$ contains a mixture of synchrotron and free-free
emission, care must be taken to interpret the model correctly.  Let the combined
synchrotron and free-free emission in the data at frequency $\nu$ be
\beq
T(\nu,p) = g(\nu) \left[ S_{\rm s}(\nu)T_{\rm s}(p) + S_{\rm ff}(\nu)T_{\rm ff}(p) \right],
\label{eq:sff_nu}
\eeq
where the terms are as defined above.  The synchrotron and free-free terms in
the model may be written as
\beqa
b_1(\nu)\,(T_{\rm K} - T_{\rm Ka}) + b_2(\nu)\,I_{H\alpha} 
& = & b_1(\nu) [R_{\rm s}\,T_{\rm s} + R_{\rm ff}\,T_{\rm ff}] + [b_2(\nu)/h_{\rm ff}]T_{\rm ff} \\
& = & [b_1(\nu) R_{\rm s}]\,T_{\rm s} + [b_1(\nu) R_{\rm ff} + b_2(\nu)/h_{\rm ff}]\,T_{\rm ff}.
\label{eq:b1b2_interp}
\eeqa
Comparing the synchrotron terms in equations~(\ref{eq:b1b2_interp}) and
(\ref{eq:sff_nu}) we can infer the mean synchrotron spectral index returned by
the fit
\beq
\beta_{\rm s}(\nu_K,\nu) = \frac{\log[S_{\rm s}(\nu)]}{\log(\nu/\nu_K)}
 = \frac{\log[b_1(\nu)R_{\rm s}/g(\nu)]}{\log(\nu/\nu_K)}.
\label{eq:beta_synch}
\eeq
Comparing the free-free terms, and assuming $S_{\rm ff}$ is known, we can solve
for the free-free to H$\alpha$ ratio at K-band
\beq
h_{\rm ff} = \frac{b_2(\nu)}{g(\nu)S_{\rm ff}(\nu) - b_1(\nu)R_{\rm ff}}.
\eeq

We fit the template model, equation~(\ref{eq:gal_template}), simultaneously to
each of the eight Q- through W-band differencing assembly (DA) maps (the
three-year maps smoothed to 1$^{\circ}$ resolution) by minimizing $\chi^2$
\beq
\chi^2 = \sum_{i,p}
  \frac{\left[ T(\nu_i,p) - b_1(\nu) (T_{\rm K} - T_{\rm Ka}) 
                          - b_2(\nu) I_{H\alpha} 
                          - b_3(\nu) M_{\rm d} \right]^2}
       {\sigma^2_i}
\eeq
where $T(\nu_i,p)$ is the \map\ sky map from DA $i$ (in thermodynamic units),
$\sigma^2_i$ is the mean noise variance per pixel for DA $i$, and the second sum
is over pixels outside the Kp2 sky cut.  To regularize the model, we impose the
following constraints on the fit coefficients: (1) all coefficients must be
positive-definite, (2) the dust coefficients must follow a spectrum $b_3(\nu_i)
\equiv b_3 \cdot g(\nu_i)(\nu_i/\nu_{\rm W1})^{+2.0}$, and (3) the free-free
coefficients for each DA must follow a free-free spectrum, which leads to the
following form
\beq
b_2(\nu_i) \equiv b_2 \cdot \left[ g(\nu_i)(\nu_i/\nu_{\rm K})^{-2.14} - b_1(\nu_i)R_{\rm ff} \right].
\eeq
The synchrotron coefficients are fit separately for each differencing assembly. 
Given the 10 coefficients from the three-year fit, we subtract the model from
each single-year DA map to produce a set of cleaned maps.  In doing so, we form
separate single-year maps of $T_{\rm K} - T_{\rm Ka}$ to maintain rigorously
independent noise between separate years of data.

The fit coefficients $b_i$ are given in Table~\ref{tab:gal_template} along with
derived values for $\beta_{\rm s}$ and $h_{\rm ff}$.  To facilitate model
subtraction, we tabulate values for $b_2$ and $b_3$ for each DA using the
above constraints.  Note that the FDS dust model, which predates \map\ by a few
years, predicts the 94 GHz dust signal remarkably well.  The synchrotron
emission shows a steady steepening with increasing frequency, as seen in the
first-year data \citep{bennett/etal:2003c}.  Also, the free-free to H$\alpha$
ratio is seen to be $\sim$6.5 $\mu$K R$^{-1}$, which is roughly half of the 11.4
$\mu$K R$^{-1}$ prediction.  Taken together, this fit finds a remarkably low
total Galactic foreground amplitude at V-band. 

\begin{figure}
\figurenum{10}
\epsscale{1.0}
\plotone{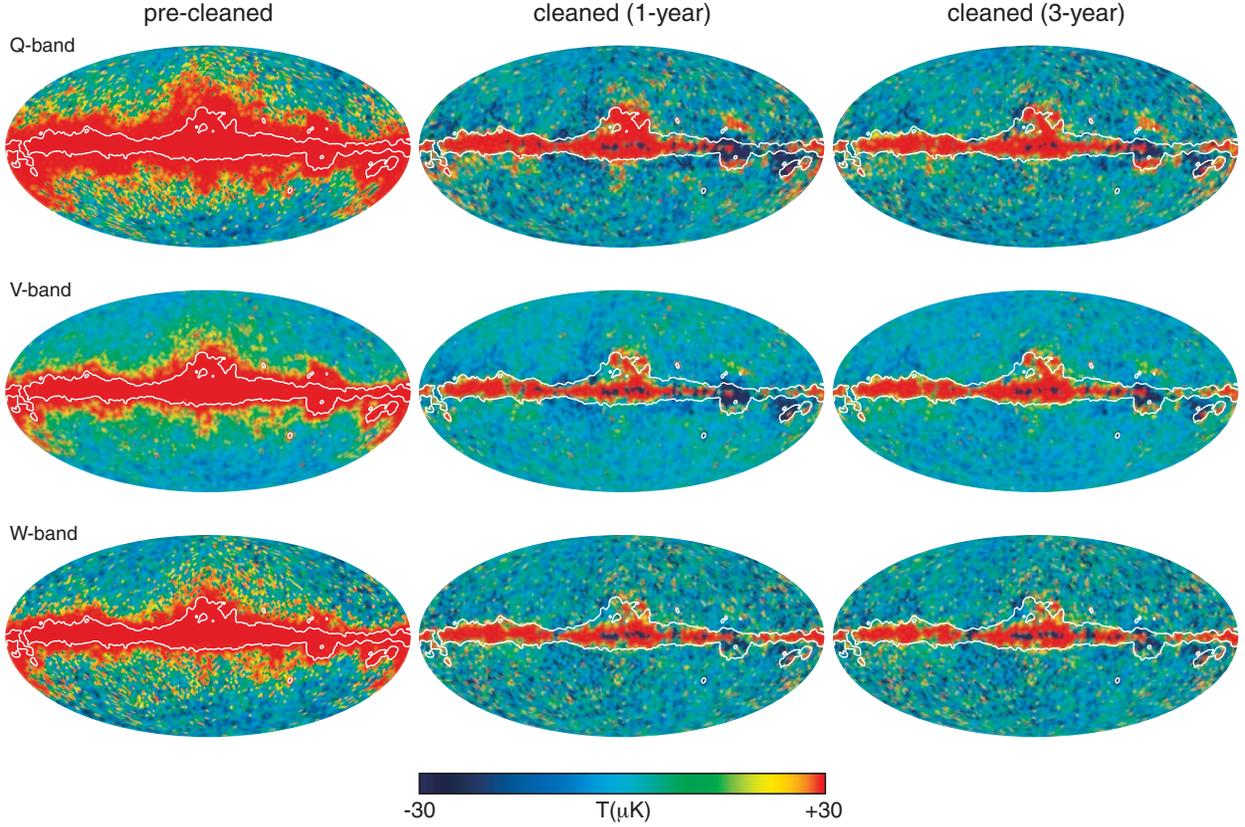}
\caption{Galactic foreground removal with spatial templates.  All maps in this
figure are three-year maps that have had the ILC estimate of the CMB signal
subtracted off to highlight the foreground emission. The maps have been degraded
to pixel resolution 5, are displayed in Galactic coordinates, and are scaled to
$\pm$30 $\mu$K.  The white contour indicates the perimeter of the Kp2 sky cut,
outside of which the template fits were evaluated.  The frequency bands Q
through W are shown top to bottom.  ({\it left}) Sky maps prior to the
subtraction of the best-fit foreground model (\S\ref{sec:gal_template}).  ({\it
middle}) The same sky maps with the first-year template-based model subtracted. 
Note the high-latitude residuals in the vicinity of the North Polar Spur and
around the inner Galaxy due to the use of the Haslam 408 MHz map as a
synchrotron template.  ({\it right}) The same sky maps with the three-year
template-based model subtracted.  This model substitutes K- and Ka-band data for
the Haslam data which produces lower residuals outside the Kp2 sky cut.  There
are still isolated spots with residual emission of order 30 $\mu$K in the
vicinity of the Gum Nebula and the Ophiuchus Complex (see
Figure~\ref{fig:gal_overview}).  Note also that substantial errors ($\ge$30
$\mu$K) remain inside the Kp2 cut due to limitations in the template model.}
\label{fig:gal_template}
\end{figure}

Figure~\ref{fig:gal_template} shows the three-year band maps before and after
subtracting the above model.  For comparison, the Figure also shows the same
three-year maps after subtracting the first-year template-based model
\citep{bennett/etal:2003c}.  In all panels an estimate of the CMB signal (the
ILC map) has been subtracted to better show residual foreground errors.  The
main visible difference between the first-year and three-year residual maps is
the synchrotron subtraction error in the first-year model due to the use of the
Haslam 408 MHz map.  This is especially visible in the region of the North Polar
Spur and around the inner Galaxy.  Note also the significant model errors
visible inside the Kp2 sky cut.  This is presumably caused by a combination of
synchrotron spectral index variations and errors in the extinction correction
applied to the H$\alpha$ template. Indeed, errors of up to 30 $\mu$K are also
clearly visible in isolated regions outside the cut, especially in the vicinity
of the Gum Nebula and the Ophiuchus complex.  In \S\ref{sec:tt_gal} we compare
the foreground signal to the CMB and assess the degree to which these residual
errors contaminate the CMB power spectrum.

\section{EXTRAGALACTIC FOREGROUNDS}
\label{sec:exfg}

\subsection{Point Sources}
\label{sec:src}

Extragalactic point sources contaminate the \map\ anisotropy data and a few
hundred of them are strong enough that they should be masked and discarded prior
to undertaking any CMB analysis.  In this section we describe a new direct
search for sources in the three-year \map\ band maps.  Based on this search, we
update the source mask that was used in the first-year analysis.  In
\S\ref{sec:tt_gal} we describe our approach to fitting and subtracting residual
sources in the data.  \citet{page/etal:prep} discuss the treatment of polarized
sources.

For the first-year analysis, we constructed a catalog of sources surveyed at
4.85 GHz using the northern hemisphere GB6 catalog \citep{gregory/etal:1996} and
the southern hemisphere PMN catalog  \citep{griffith/etal:1994,
griffith/etal:1995, wright.a/etal:1994, wright.a/etal:1996}.  The GB6 catalog
covers the declination range $0^\circ \lt \,\delta\, \lt +75^\circ$ to a flux
limit of 18 mJy, while the PMN catalog covers $-87^\circ \lt \,\delta\, \lt
+10^\circ$ to a flux limit between 20 and 72 mJy.  Combined, these catalogs
contain 119,619 sources, with 93,799 in the region $\vert b \vert \gt
10^\circ$.  We have examined the three-year \map\ sky maps for evidence of these
sources as follows: we bin the catalog by source brightness and, for each bin,
we cull the corresponding sky map pixels that contain those sources.  The data
show a clear correlation between source strength and mean sky map temperature
that disappears if the sky map pixels are randomized.  The multi-frequency \map\
data suggest that the detected sources are primarily flat-spectrum, with
$\alpha\sim 0$.

In the first-year analysis, we produced a catalog of bright point sources in the
\map\ sky maps, independent of their presence in external surveys.  This process
has been repeated with the three-year maps as follows. We filter the weighted
maps, $N_{\rm obs}^{1/2} T$ ($N_{\rm obs}$ is the number of observations per
pixel) in harmonic space by $b_l/(b^2_lC^{\rm cmb}_l + C^{\rm noise}_l)$,
\citep{tegmark/deoliveira-costa:1998, refregier/spergel/herbig:2000}, where
$b_l$ is the transfer function of the \map\ beam response
\citep{page/etal:2003b, jarosik/etal:prep}, $C^{\rm cmb}_l$ is the CMB angular
power spectrum, and $C^{\rm noise}_l$ is the noise power. Peaks that are $\gt
5\sigma$ in the filtered maps are fit in the unfiltered maps to a Gaussian
profile plus a planar baseline.  The Gaussian amplitude is  converted to a
source flux density using the conversion factors given in Table~5 of
\citet{page/etal:2003b}.  When a source is identified with $\gt 5\sigma$
confidence in any band, the flux densities for other bands are given if they are
$\gt 2\sigma$ and the fit source width is within a factor of 2 of the true beam
width.  We cross-correlate detected sources with the GB6, PMN, and 
\citet{kuehr/etal:1981} catalogs to identify 5 GHz counterparts.  If a 5 GHz
source is within $11\arcmin$ of the \map\ source position (the \map\ source
position uncertainty is $4\arcmin$) we tag the \map\ source. When more than one
source lies within the cutoff radius the brightest one is assumed to be the
\map\ counterpart.  

\begin{figure}
\figurenum{11}
\epsscale{1.0}
\plotone{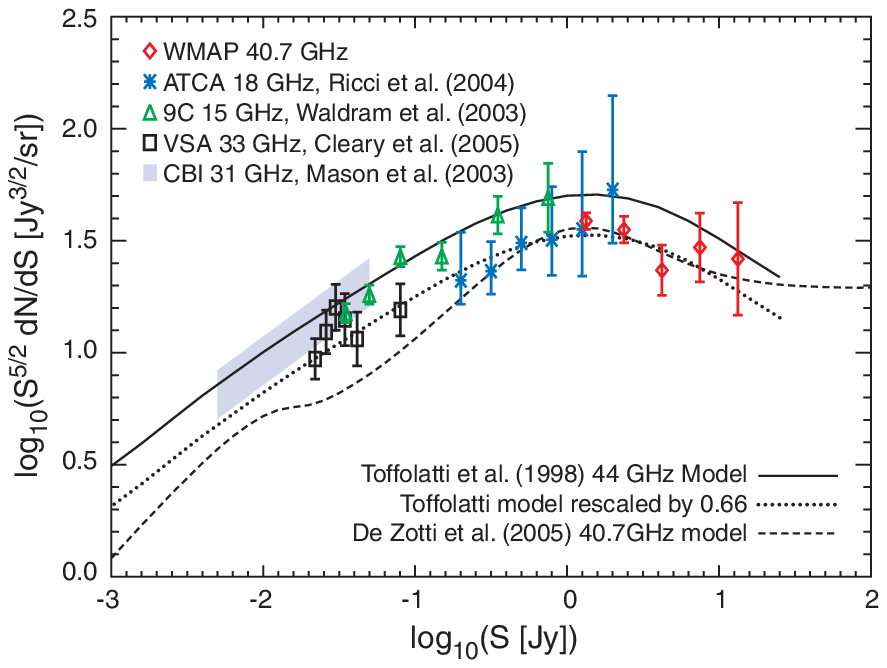}
\caption{Measurement of the source number count distribution $dN/dS$.  {\it red
diamonds}: from the \map\ three-year point source catalog at 40.7 GHz (Q-band).
{\it stars}: from the Australian Telescope Compact Array (ATCA) 18 GHz pilot
survey \citep{ricci/etal:2004}; {\it triangles}: from the 9C survey at 15 GHz
\citep{waldram/etal:2003}; and {\it squares}: from the Very Small Array (VSA) at
33 GHz \citep{cleary/etal:2005}.  The parallelogram is from the Cosmic  
Background Imager (CBI) experiment at 31 GHz \citep{mason/etal:2003}.  Several
models for $dN/dS$ are shown for comparison: the solid curve is the 44 GHz
$dN/dS$ model from \citet{toffolatti/etal:1998}, the dotted curve is the
Toffolatti model rescaled by 0.66 (found to be a good fit to the first-year
data), and the dashed curve is an updated 40.7 GHz model from
\citet{dezotti/etal:2005}.}
\label{fig:dnds}
\end{figure}

The catalog of 323 sources obtained from the three-year maps is listed in
Table~\ref{tbl:sources}.  In the first-year catalog, source ID numbers were
assigned on the basis of position (sorted by galactic longitude).  Now, rather
than assigning new numbers to the newly detected sources, we recommend that WMAP
sources be referred to by their coordinates, e.g., WMAP J0006-0622.  For
reference, we give the first-year source ID in column 3 of
Table~\ref{tbl:sources}.  The 5 GHz IDs are given in the last column.  The
source count distribution, $dN/dS$, obtained from the Q-band data is shown in
Figure~\ref{fig:dnds}.

The first-year catalog contained 208 sources.  Given the increased sensitivity
in the three-year maps, the number of new sources detected is consistent with
expectations based on differential source count models.  By the same token, 6
sources from the first-year catalog were dropped from the three-year list 
(numbers 15, 31, 61, 96, 156, and 168).  Simulations of the first-year catalog
suggested that it contained $5\pm4$ false detections, so the number of dropped
sources is consistent with expectations.  Five of the 208 sources in the
first-year catalog could not be identified with 5 GHz counterparts; now 6 out of
323 can not be.  Of the 6 sources that dropped off the first-year catalog,
source numbers 15 and 61 did not have 5 GHz identifications.  The remaining 4
may have dropped off because of variable (declining) flux density.

\citet{trushkin:2003} has compiled multifrequency radio spectra and high
resolution radio maps of the sources in the first-year \map\ catalog.  Reliable
identifications are claimed for 205 of the 208 sources.  Of the 203 sources with
optical identifications, \citet{trushkin:2003} finds 141 quasars, 29 galaxies,
19 active galactic nuclei, 19 BL Lac-type objects and one planetary nebula,
IC418.  Forty percent of the sources are identified as having flat and inverted
radio spectra, 13\% have GHz-peaked spectra, 8\% are classical power-law
sources, and 7\% have a classical low frequency power-law combined with a flat
or inverted spectrum component (like 3C84). \citet{trushkin:2003} suggests that
\map\ source number 116 is likely to be spurious and, for source 61 no radio
component was found.

\subsection{Sunyaev-Zeldovich (SZ) Effect}
\label{sec:sz}

Hot gas in galaxy clusters produces secondary anisotropy in the CMB via the
Sunyaev-Zeldovich effect: a systematic frequency shift of CMB photons produced
by Compton scattering off hot electrons in the cluster gas.  At frequencies less
than 217 GHz, this produces a temperature decrement in the CMB along sight lines
that pass through clusters.  The effect is relatively small for \map\, due to
its moderate angular resolution; nevertheless, nearby clusters are large enough
to produce a measurable effect in the sky maps.  In our first year analysis we
reported detections of an SZ signal from the Coma cluster and from an aggregate
sample of X-ray selected Abell clusters, the XBAC catalog.  Since then, there
have been numerous additional reports of SZ signal detection in the first-year
\map\ data  \citep{fosalba/gaztanaga:2004, fosalba/gaztanaga/castander:prep, 
myers/etal:2004, afshordi/loh/strauss:2004, hernandez/etal:prep, 
hernandez/martin:2004, afshordi/lin/sanderson:2005, hansen/etal:2005a, 
Atrio-Barandela/Muecket:2006}.  In this section, we update our results for the
Coma cluster and the XBAC catalog; in subsequent cosmological studies, we mask
these clusters from the data.

The brightest SZ source is the Coma cluster.  For SZ analysis we model its
temperature profile with an isothermal $\beta$-model of the form
\beq
\Delta T_{\rm SZ}(\theta) = \Delta T_{\rm SZ}(0)
	\left[ 1+(\theta/\theta_c)^2 \right]^{(1-3\beta)/2},
\label{eq:sz:comaprofile}
\eeq
where $\theta = r/D_A(z)$ is the angular distance from the core, located at
$(l,b) = (56\ddeg75,88\ddeg05)$; \citet{briel/henry/bohringer:1992} give
$\theta_c = 10.5'\pm0.6 \arcmin$ and $\beta=0.75\pm0.03$. Using these values, we
convolve the profile with the \map\ beam response at each frequency
\citep{page/etal:2003b,jarosik/etal:prep} and fit for $\Delta T_{\rm SZ}(0)$ by
minimizing
\beq
\chi^2 = \sum_{ij}\left[T_i - T_0 - \Delta T_{\rm SZ}(\theta_i)\right]
	({\bf M}^{-1})_{ij}
        \left[T_j - T_0 - \Delta T_{\rm SZ}(\theta_j)\right],
\eeq
where $T_i$ is the temperature in pixel $i$, $T_0$ is a temperature baseline,
$\theta_i$ is the angular distance of pixel $i$ from the cluster core, and 
${\bf M}_{ij}$ is the pixel-pixel covariance matrix of the sky map.  We model
${\bf M}_{ij}$ as $C(\hat{n}_i\cdot\hat{n}_j) + \sigma^2_i \delta_{ij}$,  where
$C(\theta)$ is the 2-point correlation function of the CMB and $\sigma^2_i$ is
the variance due to instrument noise in pixel $i$.

For W-band we find a Coma decrement of $\Delta T_{\rm SZ}(0) = -0.46 \pm 0.16$
mK, with a reduced $\chi^2$ of 1.05 for 196 degrees of freedom. In V-band the
decrement is $-0.31 \pm 0.16$ mK. \cite{herbig/etal:1995} measured a
Rayleigh-Jeans decrement of $-0.51 \pm 0.09$ mK at 32~GHz.  Given the SZ
frequency spectrum, their result predicts $-0.47 \pm 0.08$ mK at V-band and
$-0.42 \pm 0.07$ mK at W-band, consistent with what we observe.

The XBAC catalog of X-ray clusters produced by \citet{ebeling/etal:1996} is an
essentially complete flux-limited sample of 242 Abell clusters selected from the
ROSAT All-Sky Survey data.  We cross-correlate this catalog with the \map\ 94
GHz map.  Treating the XBAC clusters as point sources, we estimate their flux
density at 94 GHz using the expression from
\citet{refregier/spergel/herbig:2000}
\beq
S_{94} = 11.44 \left({{300\,{\rm Mpc}}\over{D(z)}}\right)^2
        \left({{f_{\rm gas}}\over{0.11}}\right)
        \left({{kT_e}\over{1\,{\rm keV}}}\right)^{5/2}
        \quad[{\rm mJy}],
\eeq
where $D(z)$ is the angular diameter distance to the cluster, $f_{\rm gas}\equiv
M_{\rm gas}/M$ is the gas mass fraction, and $T_e$ is the electron temperature.
The overall normalization of this relation is uncertain due to our ignorance of
the correct gas fraction and cluster virialization state. We fix $f_{\rm gas} =
0.11$. Extended clusters (more than one-third the extent of Coma) are omitted;
the remaining fluxes are all $<1$ Jy. A template map is constructed by
convolving the clusters with the \map\ W-band beam response
\citep{page/etal:2003b} and fitting the result to the \map\ 94 GHz map. We find
a template normalization of $-0.32\pm 0.14$ for the three-year data (compared to
$-0.36\pm 0.14$ in the first year data).  Since the fluxes used to construct the
template are positive, the negative scaling is consistent with a 2.5$\sigma$ SZ
decrement.

CMB photons that travel to us through the plane of the Milky Way undergo an SZ 
distortion of $y \approx k n_e T_e \sigma_T L / m_e c^2$, where $\sigma_T$ is
the Thomson  scattering cross-section and $L$ is the electron-pressure-weighted
path length through our Galaxy.  Taking $n_e T_e = 10^3$ K cm$^{-3}$ and $L=50$
kpc we find $y \approx 2 \times 10^{-8}$; thus our Galaxy does not significantly
affect CMB photons.  The SZ effect can safely be ignored as a diffuse
contaminating foreground signal.

\section{THE ANGULAR POWER SPECTRUM}
\label{sec:tt}

Full-sky maps provide the most compact record of CMB anisotropy without loss of
information.  They permit a wide variety of statistics to be computed from the
data, the most fundamental of which is the CMB angular power spectrum. Indeed,
if the  temperature fluctuations are Gaussian distributed, with random phase,
the angular power spectrum provides a {\em complete} description of the
statistical properties of the CMB.  While there have been a number of papers
based on the first-year data that claim evidence of non-Gaussianity and/or
non-random phases in the fluctuations (see below) it is clear that the
fluctuations are not {\em strongly} deviant from Gaussian, random phase. Thus,
the measured power spectrum does provide the primary point of contact between
data and cosmological parameters. This section presents the angular power
spectrum obtained from the first three years of \map\ observations.

A sky map $T(\n)$ defined over the full sky may be decomposed in spherical
harmonics as
\beq
T(\n) = \sum_{l=0}^{\infty}\sum_{m=-l}^{l} a_{lm} Y_{lm}(\n)
\eeq
with
\beq
a_{l m}= \int d{\n} \, T(\n) Y^*_{lm}(\n),
\label{eq:int_alm_full}
\eeq
where $\n$ is a unit direction vector. If the CMB anisotropy is Gaussian
distributed with random phases, then each $a_{lm}$ is an independent Gaussian
deviate with $\EV{a_{lm}} = 0$, and
\beq
\EV{a_{lm} a^*_{l'm'}} = \delta_{ll'}\,\delta_{mm'}\,C_l,
\eeq
where $C_l$ is the angular power spectrum and $\delta$ is the Kronecker symbol.
$C_l$ is the mean variance per $l$ that would be observed by a hypothetical
ensemble of observers distributed throughout the universe. The actual power
spectrum observed in {\em our} sky, by a supposedly typical member of this
ensemble, would be
\beq
C^{\rm sky}_l 
= \frac{1}{2l+1}\sum_{m=-l}^{l}\left|{{a}_{lm}}\right|^2.
\label{eq:sum_alm_full}
\eeq
If we had noiseless CMB data over the full sky, equation~(\ref{eq:int_alm_full})
could be evaluated exactly and equation~(\ref{eq:sum_alm_full}) would give an
unbiased estimate of the true power spectrum, in the sense that $\EV{C^{\rm
sky}_l} = C_l$ when averaged over the ensemble.   However, we do not have ideal
data (see below), and even if we did, since we only measure $2l+1$ modes per $l$
(per sky), the above estimate of the variance has an intrinsic uncertainty (or
``cosmic variance'') of
\beq
\frac{\sigma_l}{C_l} = \sqrt{\frac{2}{2l+1}}.
\eeq
Beyond cosmic variance, there are two effects that preclude using
equations~(\ref{eq:int_alm_full}) and (\ref{eq:sum_alm_full}) to estimate the
power spectrum; (1) real CMB data contains noise and other sources of error that
cause the quadratic expression in equation~(\ref{eq:sum_alm_full}) to be biased;
and (2) data near the Galactic plane must be masked because Galactic emission
cannot be reliably cleaned there.  Masking precludes the proper evaluation of
the integral in equation~(\ref{eq:int_alm_full}), so other methods must be found
to estimate $a_{lm}$ and $C_l$.

In the first-year analysis, we addressed problem (1) by adopting a
``cross-power'' estimator in which we replace $\vert a_{lm} \vert^2$ by
$(a_{lm}^i a_{lm}^{j*})$, where $i$ and $j$ denote data channels with
uncorrelated noise \citep{hinshaw/etal:2003}, see also \citet{polenta/etal:2004,
patanchon:2003, tristram/etal:2004a}. This form removes the noise bias from the
estimate of $\vert a_{lm} \vert^2$.  We addressed problem (2) by using a
``pseudo-$C_l$'' estimator \citep{peebles/hauser:1974, hivon/etal:2002} that
statistically corrects for the aliasing introduced by non-uniform pixel weights,
of which the sky cut is an extreme case. 

An alternative approach is to invoke a maximum likelihood estimator that employs
Bayes' theorem to relate the likelihood of the power spectrum to the likelihood
of the data,
\beq
L(C_l\,\vert\,{\bf d}) \propto L({\bf d}\,\vert\,C_l) P(C_l).
\eeq
Here $L(C_l\,\vert\,{\bf d})$ is the likelihood of the model (the underlying
power spectrum, $C_l$) given the data ${\bf d}$ (usually a sky map, or its
transform, the $a_{lm}$ coefficients), $L({\bf d}\,\vert\,C_l)$ is the
likelihood of the data given the model, and $P(C_l)$ is the prior probability of
the power spectrum. It is standard to assume that the data follow a multivariate
Gaussian distribution (as predicted by most inflationary models, for example) in
which case the likelihood takes the form
\beq
L(C_l\,\vert\,{\bf d}) \propto \frac{\exp(-\frac{1}{2}\,{\bf d}^T{\bf C}^{-1}{\bf d})}
                                  {\sqrt{\det{\bf C}}} P(C_l),
\label{eq:gauss_like}
\eeq
where ${\bf C}$ is the covariance matrix of the data, including contributions
from both signal and noise, ${\bf C}(C_l) = {\bf S}(C_l) + {\bf N}$.  Since the
covariance matrix includes both terms, the confidence regions on $C_l$ deduced
from $L(C_l\,\vert\,{\bf d})$ automatically incorporate the uncertainty from
both cosmic variance and instrument noise.  In addition, if the data are
restricted to lie outside a Galaxy cut, the confidence regions on $C_l$ will
incorporate the uncertainty due to aliasing. Finally, it is also standard to
assume a uniform prior distribution for the model, $P(C_l) = 1$, though given
sufficiently robust data, this choice will be unimportant.  There are many
technical challenges one must overcome when evaluating
equation~(\ref{eq:gauss_like}) and most of these have been well covered in the
literature \citep{peebles:1973, peebles/hauser:1974, hamilton:1997a,
hamilton:1997b, tegmark:1997, bond/jaffe/knox:1998, borrill:1999,
oh/spergel/hinshaw:1999}.  See \citet{page/etal:prep} for an extension of this
methodology to a simultaneous analysis of the temperature and polarization data.

Since the \map\ first-year data release, several authors have re-derived the
angular power spectrum from the sky maps using a variety of other estimators. 
Some of these analyses cover the full $l$ range to which \map\ is sensitive,
while others restrict attention to the low $l$ regime where certain technical
issues matter most, and where hints of unusual behavior have been suggested.  Of
the analyses covering the full $l$ range, \citet{fosalba/szapudi:2004} employ a
pixel-space estimator called SpICE that produces a spectrum that agrees
remarkably well with the released \map\ spectrum.  These authors do find small
discrepancies at the very highest $l$ values (where the \map\ data is
noise-dominated) but these differences do not significantly affect parameter
determination. \citet{odwyer/etal:2004} and \citet{eriksen/etal:2004e} use a
novel method based on Gibbs Sampling to undertake a full Bayesian analysis of
the power spectrum.  This method also produces results that are in good
agreement with the \map\ spectrum.  In addition, the method is able to generate
a full conditional likelihood for the spectrum that allows one to rigorously
evaluate cosmological models without resorting to likelihood approximations.

\citet{efstathiou:2004b} has surveyed a variety of methods for estimating the
power spectrum of large data sets such as \map\, and concluded that the best
approach is a hybrid one which employs a maximum likelihood estimate at low $l$
and a pseudo-$C_l$ based estimate at high $l$.  We agree with this approach and
adopt it for the three-year analysis, making the transition at $l=30$.  This
gives results that agree well with the first-year spectrum for $l>30$, and
reasonably well for low-$l$, as detailed below.  In the original version of this
paper, we chose to make the transition at $l=12$, but a detailed study of the
three-year spectrum by \citet{eriksen/etal:2006} suggested that the transition
needed to be at $l \sim 30$ to avoid biasing cosmological parameter estimates,
particularly the scalar spectral index, $n_s$.  \citet{eriksen/etal:2006} and
\citet{spergel/etal:prep} (Appendix~A) discuss the effect of this choice on the
spectral index.

In the following sections we discuss the instrumental properties that are
important to know for accurate power spectrum estimation, and assess the ability
of the new Galactic foreground models to clean the power spectrum data.  We then
analyze the low and high-$l$ power spectrum in detail and compare the result
with the previous \map\ spectrum and with other contemporary CMB data.

\subsection{Instrumental properties}
\label{sec:tt_instrument}

The temperature measured on the sky is modified by the properties of the 
instrument. The most important properties that affect the angular power spectrum
are finite resolution and instrument noise. Let $C^{iji'j'}_l$ denote the auto
or cross-power spectrum evaluated from two sky maps, $ij$ and  $i'j'$, where
$ij$ is a composite DA/year index ($i$ denotes the DA, $j$ the year).  Further,
define the shorthand ${\bf i} \equiv  (ij,i'j')$ to denote a pair of composite
indices, e.g., (Q1/yr-1,V2/yr-1), (Q1/yr-1,Q1/yr-2), etc. This spectrum will
have the form
\beq
C^{\bf i}_l = w^{\bf i}_l\,C^{\rm sky}_l + N^{\bf i}_l,
\label{eq:cl_obs}
\eeq
where $w^{\bf i}_l \equiv b^i_l\,b^{i'}_l\,p^2_l$ is the window function that
describes the combined smoothing effects of the beam and the finite sky map
pixel size.  Here $b^i_l$ is the beam transfer function for DA $i$, defined by
\citet{page/etal:2003b} and updated by \citet{jarosik/etal:prep} [note that we
reserve the term ``beam window function'' for $(b^i_l)^2$], and $p_l$ is the
pixel transfer function supplied with the HEALPix package 
\citep{gorski/etal:2004}.  $N^{\bf i}_l$ is the noise spectrum realized
in this particular measurement. On average, the observed spectrum estimates the
underlying power spectrum, $C_l$, 
\beq
\EV{C^{\bf i}_l} = w^{\bf i}_l\,C_l + \EV{N^i_l}\, \delta_{ii'}\, ,
\label{eq:cl_obs_bias}
\eeq
where $\EV{N^i_l}$ is the noise bias for differencing assembly $i$, and the
Kronecker symbol indicates that the noise is uncorrelated between differencing
assemblies and between years of data. To estimate the underlying power spectrum
on the sky, $C_l$, the  effects of the noise bias and beam convolution must be
removed.  The ability to determine these terms accurately is a critical element
of any CMB experiment design.

In \S\ref{sec:tt_windows} we summarize the results of \citet{jarosik/etal:prep}
on the \map\ window functions and their uncertainties. We propagate these
uncertainties through to the final Fisher matrix for the  angular power
spectrum. In \S\ref{sec:tt_noise} we present a model of the \map\ noise 
properties appropriate to power spectrum evaluation. For cross power spectra ($i
\ne i'$ above), the noise bias term drops out of  the signal estimate if the
noise between the two DAs (or between years of data from a single DA) is
uncorrelated.  However, the noise bias term still enters into the error
estimate, even for cross spectra.  Therefore, the noise model is used primarily
in error propagation.

\subsubsection{Window Functions}
\label{sec:tt_windows}

The instrument beam response was  mapped in flight using observations of the
planet Jupiter \citep{jarosik/etal:prep, page/etal:2003b}.  The beam widths,
measured in flight, range from $0\ddeg82$ (FWHM) at K band down to $0\ddeg20$ in
some of the W band channels.  The signal-to-noise ratio is such that the
response, relative to the peak of the beam, is measured to approximately $-35$
dB in W band.  As part of the three-year analysis, we have produced a physical
model of the A-side optics based on simultaneous fits to all 10 A-side beam
pattern measurements \citep{jarosik/etal:prep}.  We use this model to augment
the beam response data at very low signal-to-noise ratio ($-30$ to $-38$ dB,
depending on frequency band) which in turn allows us to better determine the
total solid angle and window function of each A-side beam.  For the B-side
response we scale the A-side model by fitting it to the B-side Jupiter
measurements in the high signal-to-noise regime.  We then form similar hybrid
response maps by augmenting the B-side data with the scaled model in the low
signal-to-noise regime.  (We plan to update the A-side model and produce an
independent B-side model in the near future.)  These hybrid beam response maps
are available on LAMBDA as part of the \map\ three-year data release. The radial
beam profiles obtained from these maps have been fit to a model consisting of a
sum of Hermite  polynomials that accurately characterize the main Gaussian lobe
and small  deviations from it.  The model profiles are then Legendre transformed
to obtain the beam transfer functions $b^i_l$ for each DA $i$
\citep{jarosik/etal:prep}.  These transfer functions are also provided with the
three-year data release.  

The constraints imposed by the physical optics model have allowed us to extend
the beam analysis out to much larger radii than was possible with the first-year
analysis.  As a result, we have determined that the beam solid angles were being
systematically underestimated by $\sim$1\%. Since we normalize the transfer
functions to 1 at $l=1$ this meant that the normalized functions were being
systematically over-estimated by a comparable amount for $l \gsim 50$.
Consequently, as we discuss in \S\ref{sec:tt_full}, the final three-year power
spectrum is $\sim$1-2\% higher than the first-year spectrum for $l \gsim 50$. We
believe this new determination of the beam response is more accurate than that
given in the first-year analysis, but until we can complete the B-side model
analysis, we have held the window function uncertainties at the first-year level
\citep{page/etal:2003b}.  The first-year uncertainties are large enough to
encompass the differences between the previous and current estimates of $b_l$. 
The propagation of random beam errors through to the power spectrum Fisher
matrix is discussed in Appendix~\ref{app:beam_src_err}.

An additional source of systematic error in our treatment of the beam response
arises from non-circularity of the main beam
\citep{wu/etal:2001,souradeep/ratra:2001,mitra/sengupta/souradeep:2004}.  The
effects of non-circularity are mitigated by \map's scan strategy, which causes
most sky pixels to be observed over a wide range of azimuth angles.  However,
residual asymmetry in the effective beam response on the sky will, in general,
bias estimates of the power spectrum at high $l$. In
Appendix~\ref{app:beam_asymm} we compute the bias induced in a cross-power
spectrum due to residual beam asymmetry. In principle the formalism is exact,
but in order to make the calculation tractable, we make two approximations: (1)
that \map's scan coverage is independent of ecliptic longitude, and (2) that the
azimuthal structure in the \map\ beams is adequately described by modes up to
azimuthal quantum number $m=16$.  We define the bias as the ratio of the true
measured power spectrum to that which would be inferred assuming a perfectly
symmetric beam,
\beq
\alpha_l \equiv \frac{1}{b^i_l b^{i'}_l C^{\rm fid}_l} 
                \sum_{l'} G^{\bf i}_{ll'} C^{\rm fid}_{l'},
\eeq
where $C^{\rm fid}_l$ is a fiducial power spectrum, $b^i_l$ is the symmetrized
beam transfer function noted above, and $G^{\bf i}_{ll'}$ is the coupling
matrix, defined in Appendix~\ref{app:beam_asymm}, that accounts for the full
beam structure. 

Plots of $\alpha_l$ for selected channel combinations are shown in
Appendix~\ref{app:beam_asymm}. These results assume three-year sky coverage and
beam multipole moments fit to the hybrid beam maps with $\lmax=1500$ and
$\mmax=16$. As shown in the Figure, the Q-band cross-power spectra (e.g.,
Q1$\times$Q2) require the greatest correction (6.3\% at $l=600$). This arises
because the Q-band beams are farthest from the optic axis of the telescope and
are thus the most elliptical of the three high frequency bands; for example, the
major and minor axes of the Q2A beam are $0\ddeg60$ and $0\ddeg42$ (FWHM)
respectively. The corrections to the V and W-band spectra are all $\lsim$1\% for
$l < 1000$. Because of the relatively large bias in Q-band, and because of the
potential to confuse it with the frequency-dependent point source correction,
Q-band data was excluded from the final three-year power spectrum. The remaining
V and W-band cross-power spectra were {\em not} corrected because the bias is
already substantially smaller than the random instrument noise in the pertinent
$l$ range.

\subsubsection{Instrument Noise Properties}
\label{sec:tt_noise}

The noise bias term in equation~(\ref{eq:cl_obs_bias}) is the noise per 
$a_{lm}$ coefficient on the sky. If auto-power spectra are used in the final
power spectrum estimate, the noise bias term must be known very accurately
because it exponentially dominates the convolved power spectrum at high $l$. If
only cross-power spectra are used, the noise bias is only required for
estimating errors. Our final combined spectrum is based only on cross-power
spectra for $l>30$; while for $l<30$, the noise bias is negligible compared to
the sky signal, so that it need not be known to high precision for the maximum
likelihood estimation (see \S\ref{sec:tt_low_l}).

In the limit that the time-domain instrument noise is white, the noise bias will
be a constant, independent of $l$.  If the noise has a 1/f component, the bias
term will rise at low $l$.  While the \map\ radiometer noise is nearly white by
design, with 9 out of 10 differencing assemblies having 1/f knee frequencies of
$<$10 mHz \citep{jarosik/etal:2003}, deviations of $\EV{N^i_l}$ from a constant
must be accounted for, especially for the low-$l$ polarization analysis
\citep{page/etal:prep}.  In the first-year analysis, we used end-to-end Monte
Carlo simulations to generate realizations of the noise bias, then fit the mean
of the realizations to a parameterized model.  In the present analysis we use
two sets of independent single-year sky maps from each DA to form the following
estimate of the noise bias
\beq
\EV{N^i_l} = \frac{1}{2} \sum_{j,j'=1}^{2} \left[ C_l^{(ij,ij)} - C_l^{(ij,ij')} \right].
\label{eq:noise_fit}
\eeq
In words, we take the difference between the auto-power spectra (years $j=j'$)
and the cross-power spectra (years $j \ne j'$) for a given DA to estimate the
noise bias.  We then fit the estimate to a model of the form
\beq
\EV{N^i_l} = N_0 + N_1 \left( \frac{450}{l} \right),
\label{eq:noise_bias_model}
\eeq
where $N_0$ and $N_1$ are fit coefficients (we enforce $N_1 \ge 0$), and the fit
is performed over the range $33 \le l \le 1024$.  This model is used to compute
the weights that enter into the final combined spectrum
(\S\ref{sec:tt_high_l}) and to estimate its noise properties.

\subsubsection{Systematic Errors}
\label{sec:tt_systematics}

\citet{jarosik/etal:prep} present limits on systematic errors in the three-year
sky maps.  They consider the effects of absolute and relative calibration
errors, artifacts induced by environmental disturbances (thermal and
electrical), errors from the map-making process, pointing errors, and other
miscellaneous effects.  The combined uncertainty due to relative calibration
errors, environmental effects, and map-making errors are limited to $<$29
$\mu$K$^2$ ($2\sigma$) in the quadrupole moment $C_2$ in any of the 8
high-frequency DAs.  We estimate the absolute calibration uncertainty in the
three-year \map\ data to be 0.5\%.

\begin{figure}
\figurenum{12}
\epsscale{1.0}
\plotone{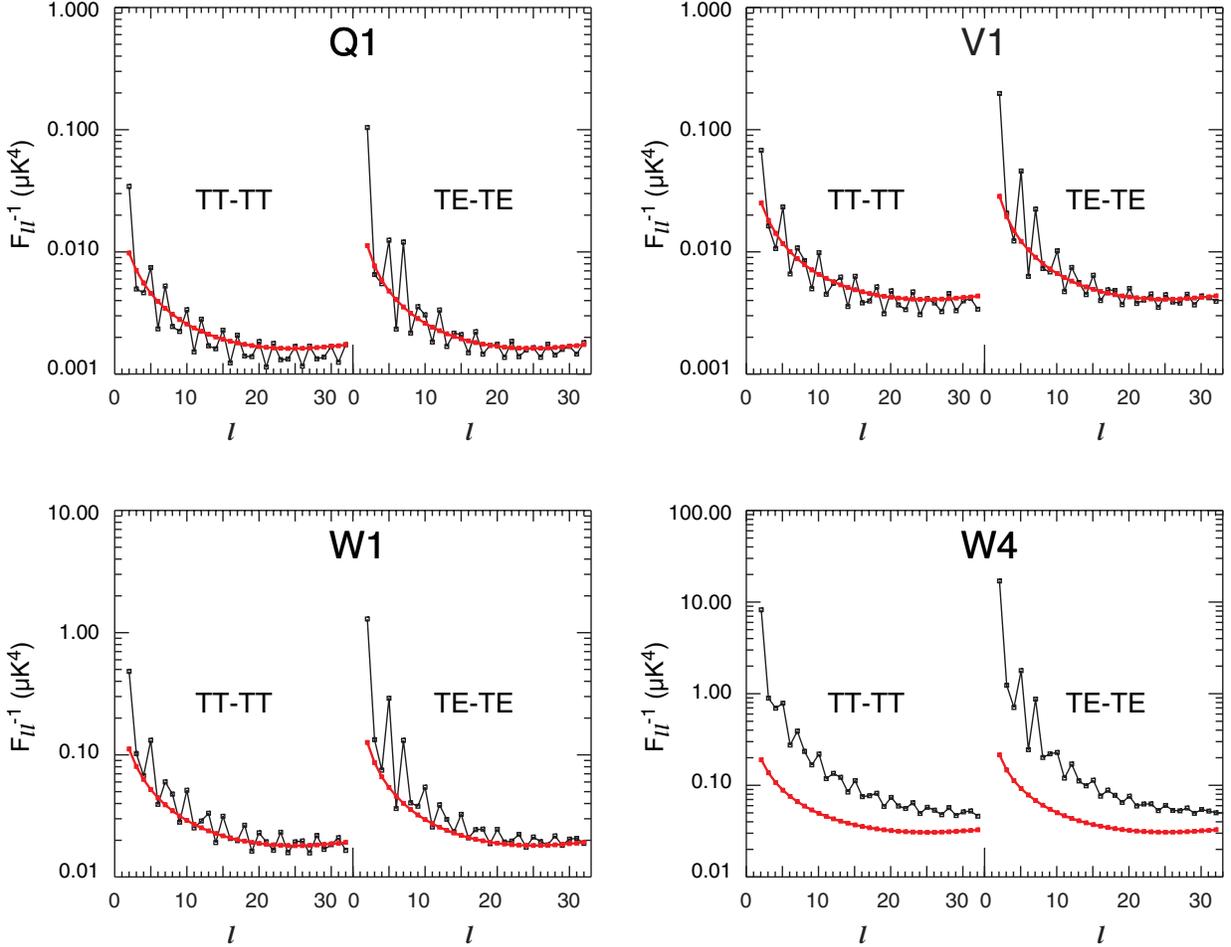}
\caption{The predicted $C_l$ uncertainty (inverse Fisher matrix) at low $l$, in
$\mu$K$^4$.  The red curves show the result when pixel-pixel noise correlations
are ignored.  The smooth rise at low $l$ reflects our approximate representation
of 1/f noise in the noise bias model, equation~(\ref{eq:noise_bias_model}).  The
black curves account for the full structure of the pixel-pixel inverse
covariance matrix, including 1/f noise and map-making covariance.  For the TT
spectrum (left pair of curves in each panel), the noise is negligible compared
to the signal, so this structure can be safely ignored.  However, the TE
analysis must account for it \citep{page/etal:prep}.  See Figure~16 of
\citet{page/etal:prep} for an analogous plot of the EE and BB uncertainties.}
\label{fig:tt_te_noise}
\end{figure}

The noise in the \map\ sky maps is weakly correlated from pixel-to-pixel due to
1/f noise in the radiometers, and to the map-making process that infers pixel
temperatures from filtered differential data.  Neglecting these correlations is
a form of systematic error that must be quantified.  As part of the three-year
analysis, we have developed code to compute the pixel-pixel inverse covariance
matrix at pixel resolution r4.  The resulting information is propagated through
the computation of the Fisher matrix to estimate power spectrum uncertainties
\citep{page/etal:prep}.  Sample $C_l$ uncertainty results for single-year
cross-power spectra are shown in Figure~\ref{fig:tt_te_noise}.   The red curves
show the result when pixel-pixel noise correlations are ignored.  The smooth
rise at low $l$ reflects our approximate representation of 1/f noise in the
noise bias model, equation~(\ref{eq:noise_bias_model}).  The black curves
account for the full structure of the pixel-pixel inverse covariance matrix,
including 1/f noise and map-making covariance.  For the TT spectrum (left pair
of curves in each panel), the noise is negligible compared to the signal
($\sim$1000 $\mu$K$^2$), so this structure can be safely ignored.

Random pointing errors are accounted for in the beam mapping procedure; the beam
transfer functions presented in \citet{jarosik/etal:prep} incorporate random
pointing errors automatically. A systematic pointing error of $\sim$1\amin at
the spin period is suspected in the quaternion solution that defines the
spacecraft pointing. This is much smaller than the smallest beam width
($\sim$12\amin at W band), and we estimate that it would produce $<$1\% error in
the angular power spectrum at $l=1000$; thus we do not attempt to correct for
this effect. \citet{barnes/etal:2003} place limits on spurious contributions due
to stray light pickup through the far sidelobes of the instrument.  As shown in
Figure~4 of that paper, they place limits of $<$10  $\mu$K$^2$ on spurious
contributions to $l(l+1)C_l/2\pi$, at Q through W band, due to far sidelobe
pickup.

Our template-based approach to Galactic foreground subtraction is detailed in
\S\ref{sec:gal_template}; the effect of this frequency-by-frequency cleaning on
the power spectra is shown in \S\ref{sec:tt_gal}.  Diffuse foreground emission
is a modest source of contamination at large angular scales ($\gsim$2\dg).
Systematic errors on these angular scales are negligible compared to the
(modest) level of foreground emission. On intermediate angular scale
($\lsim$2\dg), the 1-2\% uncertainty in the beam transfer functions is the
largest source of uncertainty, while for multipole moments greater than
$\sim$400, random white noise from the instrument is the largest source of 
uncertainty.

\subsection{Galactic and Extragalactic Foregrounds}
\label{sec:tt_gal}

In this subsection we determine the level of foreground contamination in the
angular power spectrum.  On large angular scales ($l \lsim 50$) the primary
source of contamination is diffuse emission from the Milky Way, while on small
scales ($l \gsim 400$) the primary culprit is extragalactic point source
emission.  To simplify the discussion, we present foreground results using the
pseudo-$C_l$ estimate of the power spectra over the entire $l$-range under
study, despite the fact that we use a maximum likelihood estimate for the final
low-$l$ power spectrum.  Thus, while the low-$l$ estimates presented in the
section may not be optimal, they are consistently applied as a function of
frequency, so that the relative foreground contamination is reliably determined.

\begin{figure}
\figurenum{13}
\epsscale{1.0}
\plotone{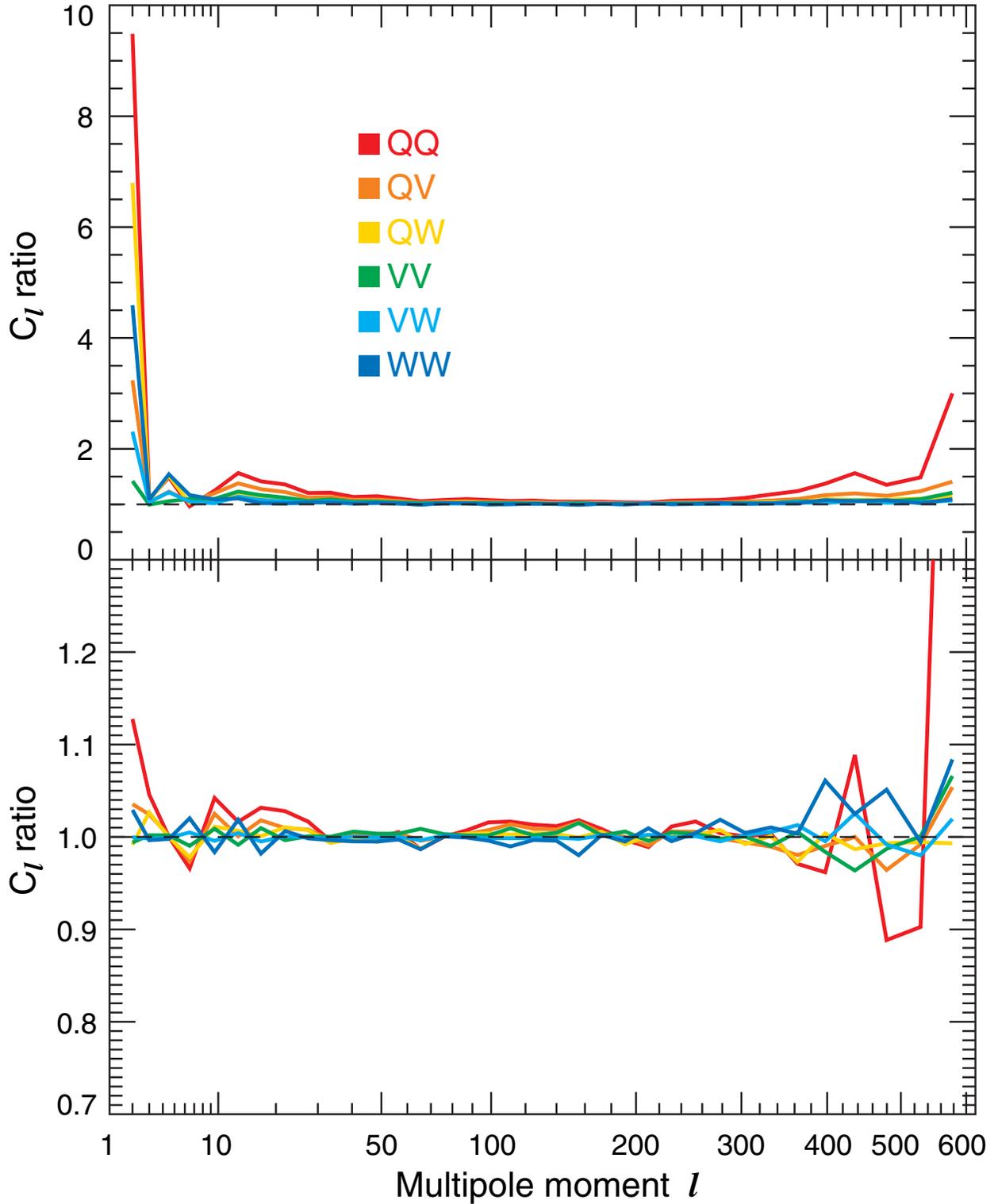}
\caption{({\it top}) Q-, V-, \& W-band cross-power spectra before foreground
subtraction, evaluated outside the Kp2 sky cut, relative to the final combined
spectrum.  ({\it bottom}) Same cross-power spectra after subtracting a
template-based diffuse foreground model (\S\ref{sec:gal_template}) and the
best-fit residual point source contamination (\S\ref{sec:tt_gal}).}
\label{fig:tt_raw_clean}
\end{figure}

Figure~\ref{fig:tt_raw_clean} shows the cross-power spectra obtained from the
high-frequency band maps (Q-W) prior to any foreground subtraction.  All DA/yr
combinations within a band pair were averaged to form these spectra, and the
results are color coded by effective frequency, $\sqrt{\nu^i\nu^{i'}}$, where 
$\nu^i$ is the frequency of differencing assembly $i$.  As indicated in the plot
legend, the low frequency data are shown in red, and so forth.  To clarify the
result, we plot the ratio of each frequency band spectrum to the final combined
spectrum (see \S\ref{sec:tt_high_l}).  The red trace in the top panel shows very
clearly that the low frequency data are contaminated by diffuse Galactic
emission at low $l$ and by point sources at higher $l$.  The contamination
appears to be least at the highest frequencies, as expected in this frequency
range.  This is consistent with the dominance of synchrotron and free-free
emission over thermal dust emission.

Not surprisingly, the most contaminated multipoles are $l=2$ \& 4, which most
closely trace the Galactic plane morphology.  Specifically, the total
quadrupolar emission at Q-band is nearly 10 times brighter (in power units) than
the CMB signal, while at W-band it is nearly 5 times brighter.  The foreground
emission reaches a minimum near V-band where the total $l=2$ emission is less
than a factor of 2 brighter than the CMB, and even less for $l>2$.  Note also
the modest foreground features in the range $l \sim 10-30$. 

The contribution from extragalactic radio sources can be seen in
Figure~\ref{fig:tt_raw_clean} as excess emission in Q-band at high $l$.  As
discussed in \S\ref{sec:src}, we performed a direct search for sources in the
\map\ data and found 323.  Based on this search, we augmented the mask we use
for CMB analysis; so the contribution shown in the Figure is mostly due to
sources just below our detection threshold.  In the limit that these sources
are not clustered, their contribution to the cross-power spectra has the form
\beq
C_l^{{\bf i},{\rm src}} = A \, g_i g_{i'} \left(\frac{\nu_i}{\nu_{\rm Q}}\right)^\beta 
    \left(\frac{\nu_{i'}}{\nu_{\rm Q}}\right)^\beta w_{l}^{\bf i}, 
\label{eq:source_model} 
\eeq
where $A$ is an overall amplitude, measured antenna temperature, the factors
$g_i$ convert the result to thermodynamic temperature, $\nu_{\rm Q} \equiv 40.7$
GHz (note that this differs from our first-year definition of 45 GHz), and we
assume a frequency spectrum $\beta = -2.0$.  The amplitude $A$ is determined by
fitting to the Q-, V-, and W-band cross-power spectra.  Following the procedure
outlined in Appendix~B of \citet{hinshaw/etal:2003}, we find $A = 0.014 \pm
0.003$ $\mu$K$^2$ sr, in antenna temperature.  (As discussed below, this value
has been revised down from $0.017 \pm 0.002$ $\mu$K$^2$ sr since the original
version of this paper appeared.)  In the first-year analysis we found $A =
0.022$ $\mu$K$^2$ sr (0.015 $\mu$K$^2$ sr, referred to 45 GHz).  The three-year
source amplitude is expected to be lower than the first-year value due to the
enlargement of the three-year source mask and the consequent lowering of the
effective flux cut applied to the source population.  \citet{spergel/etal:prep}
evaluate the bispectrum of the \map\ data and are able to fit a non-Gaussian
source component to a particular configuration of the bispectrum data.  They
find the source model, equation~(\ref{eq:source_model}), fits that data as
well.  We thus adopt the model given above with $A$ = 0.014 $\mu$K$^2$ sr and
use the procedure given in Appendix~\ref{app:beam_src_err} to marginalize the
over the uncertainty in $A$.  At Q band, the correction to $l(l+1)C_l/2\pi$ is
608 and 2431 $\mu$K$^2$ at $l$=500 and 1000, respectively.  At W band, the
correction is only 32 and 128 $\mu$K$^2$ at the same $l$ values.  For
comparison, the CMB power in this $l$ range is $\sim$ 2000 $\mu$K$^2$. 

Since the release of the three-year data, \citet{huffenberger/etal:2006} have
reanalyzed the multi-frequency spectra for residual sources using a similar
methodology. They find a somewhat smaller amplitude of $0.011 \pm 0.001$
$\mu$K$^2$ sr.  Differences in the inferred source amplitude at this level
affect the final power spectrum by about 1\%, which is also about the level at
which second-order effects due to beam asymmetry can alter the spectrum.  We
expect our understanding of these effects to improve significantly with
additional years of data.  In the meantime, our revised estimate of $A = 0.014
\pm 0.003$ $\mu$K$^2$ sr encompasses both our original estimate and the new
\citet{huffenberger/etal:2006} estimate.  The effect of these estimates on
cosmological parameters is discussed in  \citet{huffenberger/etal:2006} and in
Appendix~A of \citet{spergel/etal:prep}.

The bottom panel of Figure~\ref{fig:tt_raw_clean} shows the band-averaged
cross-power spectra after subtraction of the template-based Galactic foreground
model and the above source model.  The Q-band spectra exhibit clear deviations
of order 10\% from the V and W-band spectra at low $l$, while the higher
frequency combinations all agree with each other to better than 5\%.  Also,
while subtracting the source model in equation~(\ref{eq:source_model}) brings
the high-$l$ Q-band spectrum into good agreement with the higher frequency
results up to $l \sim 400$, beyond this point the $0\ddeg48$ angular resolution
of Q-band limits the sensitivity of this band.  This is also the $l$ range where
the effects of the Q-band beam asymmetry start to bias our pseudo-$C_l$-based
power spectrum estimates.  Since the V and W-band data alone provide a cosmic
variance limited measurement of the power spectrum up to $l=400$ (see
\S\ref{sec:tt_high_l}) we have decided to omit Q-band data entirely from our
final power spectrum estimate.  But note that Q band serves a valuable role in
fixing the amplitude of the residual point source contribution, and in helping
us to assess the quality of the Galactic foreground subtraction.

\subsection{Low-$l$ Moments From The ILC Map ($l=1,2,3$)}
\label{sec:tt_ilc}

Based on our analysis of the ILC method presented in \S\ref{sec:gal_ilc} we
conclude that the newly debiased three-year ILC map is suitable for analysis
over the full sky up to $l \la 10$, though we have not performed a full battery
of non-Gaussian tests on this map, so we still advise users to exercise
caution.  \citet{tegmark/deoliveira-costa/hamilton:2003} arrived at a similar
conclusion based on their foreground analysis of the first-year data.  In this
section we use the ILC map to evaluate the low-$l$ $a_{lm}$ coefficients by
direct full-sky integration of equation~(\ref{eq:int_alm_full}).  In
\S\ref{sec:tt_low_l} we estimate the low-$l$ power spectrum using a maximum
likelihood estimate based on the ILC map using only data outside the Kp2 sky
cut.  We compare the two results, as a cross-check, in the latter section.  

Sky maps of the modes from $l=2-8$, derived from the ILC map, are shown in
Figure~\ref{fig:ilc_2_8}.  There  has been considerable comment on the
non-random appearance of these modes.  We discuss this topic in more detail in
\S\ref{sec:phase}, but note here that some non-random appearances may be
deceiving and that high-fidelity simulations and a critical assessment of
posterior bias are very important in assessing significance.

\subsubsection{Dipole (l=1)}
\label{sec:tt_l1}

Owing to its large amplitude and its role as a \map\ calibration source, the
$l=1$ dipole signal requires special handling in the \map\ data processing.  The
calibration process is described in detail in \citet{hinshaw/etal:2003} and in
\citet{jarosik/etal:prep}.  After establishing the calibration, but prior to
map-making, we subtract a dipole term from the time-ordered data to minimize
signal aliasing that would arise from binning a large differential signal into
two finite size pixels.  In the first-year analysis we removed the {\it
COBE}-determined dipole from the data, then fit the final maps for a residual
dipole.  From this, a best-fit \map\ dipole was reported, which was limited by
the 0.5\% calibration  uncertainty.  For the three-year data analysis we
subtract the \map\ first-year dipole from the time-ordered data and fit for a
residual dipole in the final ILC map.  The results are reported in
Table~\ref{tab:tt_123}.  The errors reported for the $a_{lm}$ coefficients in
the body of the Table are obtained from Monte Carlo simulations of the ILC
procedure that attempt to include the effect of Galactic removal uncertainty. In
the Table end notes we report the dipole direction and magnitude.  The direction
uncertainty is dominated by Galactic removal errors, while the magnitude
uncertainty is dominated by the 0.5\% calibration uncertainty.

\subsubsection{Quadrupole (l=2)}
\label{sec:tt_l2}

The quadrupole moment computed from the full-sky decomposition of the ILC map is
given in Table~\ref{tab:tt_123}. (The released ILC map, and the quadrupole
moment reported here have {\em not} been corrected for the 1.2 $\mu$K
``kinematic'' quadrupole, the second-order Doppler term.)  As with the dipole
components, the errors reported for the $a_{lm}$ coefficients are obtained from
Monte Carlo simulations that attempt to include the effect of Galactic removal
uncertainty.  The magnitude of this quadrupole moment, computed using
equation~(\ref{eq:sum_alm_full}), is $\Delta T_2^2 = 249$ $\mu$K$^2$ [$\Delta
T_l^2 \equiv l(l+1)C_l/2\pi$].  Here, we do not attempt to correct for the bias
associated with this estimate, and we postpone a more complete error analysis of
$\Delta T_2^2$ to \S\ref{sec:tt_low_l}.  However, we note that this estimate is
quite consistent with the maximum likelihood estimate presented in
\S\ref{sec:tt_low_l}.

A corresponding analysis of the first-year ILC map gives $\Delta T_2^2 = 196$
$\mu$K$^2$.  The increase in the three-year amplitude relative to this value
comes mostly from the ILC bias correction and, to a lesser degree, from the
improved three-year gain model.  When our new ILC algorithm is applied to the
first-year sky maps, we get $\Delta T_2^2 = 237$ $\mu$K$^2$.  While this $41$
$\mu$K$^2$ bias correction is small in absolute terms, it produces a relatively
large {\em fractional} correction to the small quadrupole.  Further discussion
of the low quadrupole amplitude is deferred to \S\ref{sec:tt_low_l}.

\subsubsection{Octupole (l=3)}
\label{sec:tt_l3}

The octupole moment computed from the full-sky decomposition of the ILC map is
given in Table~\ref{tab:tt_123}.  The reported errors attempt to include the
uncertainty due to Galactic foreground removal errors.  The magnitude of this
octupole moment, computed using equation~(\ref{eq:sum_alm_full}), is $\Delta
T_3^2 = 1051$ $\mu$K$^2$.  As above, we do not attempt to correct for the bias
associated with this estimate, but again we note that it is quite consistent
with the maximum likelihood estimate presented in \S\ref{sec:tt_low_l}.

It has been noted by several authors that the orientation of the quadrupole and
octopole are closely aligned, e.g., \citet{deoliveira-costa/etal:2004b}, and
that the distribution of power amongst the $a_{lm}$ coefficients is possibly
non-random.  We discuss these results in more detail in \S\ref{sec:phase},
but we note here that the basic structure of the low $l$ modes is largely
unchanged from the first-year data. Thus we expect that most, if not all, of the
``odd'' features claimed to exist in the first-year maps will survive.

\subsection{Low-$l$ Spectrum From Maximum Likelihood Estimate ($l=2-30$)}
\label{sec:tt_low_l}

If the temperature fluctuations are Gaussian, random phase, and the {\it a
priori} probability of a given set of cosmological parameters is uniform, the
power spectrum may be optimally estimated by maximizing the multi-variate
Gaussian likelihood function, equation~(\ref{eq:gauss_like}).  This approach to
spectrum estimation gives the optimal (minimum variance) estimate for a given
data set, and a means for assessing confidence intervals in a rigorous way. 
Unfortunately, the approach is computationally expensive for large data sets,
and for $l \gsim 30$, the results do not significantly improve on the quadratic
pseudo-$C_l$ based estimate used in \S\ref{sec:tt_high_l}
\citep{efstathiou:2004b,eriksen/etal:2006}.  In this section we present the
results of a pixel-space approach to spectrum estimation for the multipole range
$l=2-30$ and compare the results with the pseudo-$C_l$ based estimate.  In
Appendix~\ref{app:tt_ml} we discuss several aspects of maximum likelihood
estimation in the limit that instrument noise may be ignored (a very good
approximation for the low-$l$ \map\ data).  In this case, several simple results
can be derived analytically and used as a guide to a more complete analysis.

We begin with equation~(\ref{eq:gauss_like}) and work in pixel space, where the
data, ${\bf d}={\bf t}$ is a sky map.  As in \citet{slosar/seljak/makarov:2004},
we account for residual Galactic uncertainty by marginalizing the likelihood
over one or more Galactic template fit coefficient(s).  That is, we compute the
joint likelihood
\beq
L(C_l,\alpha\,\vert\,{\bf t}) \propto 
  \frac{\exp[-\frac{1}{2}\,({\bf t} - \alpha\cdot{\bf t}_{\rm f})^T
  {\bf C}^{-1}({\bf t}- \alpha\cdot{\bf t}_{\rm f})]}
  {\sqrt{\det{\bf C}}},
\label{eq:gauss_like_fg}
\eeq
where ${\bf t}_{\rm f}$ is one or more foreground emission templates, and
$\alpha$ is the corresponding set of fit coefficients, and we marginalize over
$\alpha$
\beq
L(C_l\,\vert\,{\bf t}) = \int d\alpha \, L(C_l,\alpha\,\vert\,{\bf t}).
\eeq
This can be evaluated analytically, as in Appendix~B of
\citet{hinshaw/etal:2003}.

In practice, we start with the r9 ILC map, we further smooth it with a Gaussian
kernel of width 9\ddeg183 (FWHM); degrade the map to pixel resolution r4; mask
it with a degraded Kp2 mask (accepting all r4 pixels that have more than 50\% of
its r9 pixels outside the original Kp2 cut); then add 1 $\mu$K rms of white
noise to each pixel to aid numerical regularization of the likelihood
evaluation.  (See Appendix~A of \citet{eriksen/etal:2006} for a complete
discussion of the numerical aspects of likelihood evaluation.)  We evaluate the
marginalized likelihood function with the following expression for the
covariance matrix
\beq
{\bf C}({\bf n}_i,{\bf n}_j) = \frac{(2l+1)}{4\pi} 
 \sum_l C_l w_l^2 P_l({\bf n}_i \cdot {\bf n}_j) + {\bf N}.
\eeq
Here $w_l^2$ is the effective window function of the smoothed map evaluated at
low pixel resolution, $P_l(\cos\theta)$ is the Legendre polynomial of order $l$,
and ${\bf N}$ is a diagonal matrix that describes the 1 $\mu$K rms white noise
that was added to the map.  This expression is evaluated to the Nyquist sampling
limit of an r4 map, namely $l_{\rm max}=47$. The foreground template we
marginalize over is the map $T_{V1} - T_{\rm ILC}$, where $T_{V1}$ is the V1 sky
map and $T_{\rm ILC}$ is the ILC map.  As discussed below, we have tried using
numerous other combinations of data and sky cut and obtain consistent estimates
of $C_l$.  \citet{eriksen/etal:2006} have also studied this question and obtain
similar results using a variety of data combinations and likelihood estimation
methodologies.

The full likelihood curves, $L(C_l)$ for $l=2-10$, are shown in
Figure~\ref{fig:lowl_tt_like} and listed in Table~\ref{tab:tt_low_l_like}. 
(Note that the likelihood code delivered with the three-year data release
employs the above method up to $l=30$, then uses the quadratic form employed in
the first-year analysis \citep{verde/etal:2003} for $l>30$.)  The predicted
$C_l$ values from the best-fit $\Lambda$CDM model (fit to \map\ data only) are
shown as vertical red lines, while the values derived from the pseudo-$C_l$
estimate (\S\ref{sec:tt_high_l}) are shown as vertical blue lines.  Maximum
likelihood estimates from the ILC map with and without a sky cut are also shown,
as indicated in the Figure.  There are several comments to be made in connection
with these results:

(1) All of the maximum likelihood estimates (the black curve and the vertical
dot-dashed lines) are all consistent with each other, in the sense that they all
cluster in a range that is much smaller than the overall 68\% confidence
interval, which is largely set by cosmic variance.  We conclude from this that
foreground removal errors and the effects of masking are not limiting factors in
cosmological model inference from the low-$l$ power spectrum.  However, they may
still play an important role in determining the significance of low-$l$ features
beyond the power spectrum.

(2) The $C_l$ values based on the pseudo-$C_l$ estimates (shown in blue) are
generally consistent with the maximum likelihood estimates.  However, the
results for $l=2,3,7$ are all nearly a factor of 2 lower than the ML values, and
lie where the likelihood function is roughly half of its peak value.  This
discrepancy may be related to the different assumptions the two methods make
regarding the distribution of power in the $a_{lm}$ when faced with cut-sky
data.  Both methods have been demonstrated to be unbiased, as long as the noise
properties of the data are correctly specified, but the maximum likelihood
estimate has a smaller variance.  Additionally, we note that the two methods are
identical in the limit of uniformly-weighted, full-sky data; and the maximum
likelihood estimate based on cut-sky data is consistent with the full-sky
estimate.  In light of this, we adopt the maximum likelihood estimate of $C_l$
for $l=2-30$.

(3) The best-fit $\Lambda$CDM model lies well within the 95\% confidence
interval of $L(C_l)$ for all $l \le 10$, including the quadrupole.  Indeed,
while the observed quadrupole amplitude is still low  compared to the best-fit
$\Lambda$CDM prediction (236 $\mu$K$^2$ vs. 1252 $\mu$K$^2$, for $\Delta T_2^2$)
the probability that the ensemble-average value of $\Delta T_2^2$ is as large or
larger than than the model value is 16\%.  This relatively high probability is
due to the long tail in the posterior distribution, which reflects the
$\chi^2_{\nu}$ distribution ($\nu = 2l+1$) that Gaussian models predict the
$C_l$ will follow.  Clearly the residual uncertainty associated with the exact
location of the maximum likelihood peak in Figure~\ref{fig:lowl_tt_like} will
not fundamentally change this situation.

There have been several other focused studies of the low-$l$ power spectrum in
the first-year maps, especially $l=2$ \& 3 owing to their somewhat peculiar
behavior and their special nature as the largest observable structure in the
universe \citep{efstathiou:2003b, gaztanaga/etal:2003, efstathiou:2003d,
bielewicz/gorski/banday:2004, efstathiou:2004, slosar/seljak/makarov:2004}. 
These authors agree that the quadrupole amplitude is indeed low, but not low
enough to rule out $\Lambda$CDM.

\subsection{High-$l$ Spectrum From Combined Pseudo-$C_l$ Estimate ($l=30-1000$)}
\label{sec:tt_high_l}

The final spectrum for $l>30$ is obtained by forming a weighted average of the
individual cross-power spectra, equation~(\ref{eq:cl_obs}), computed as per
Appendix~A in \citet{hinshaw/etal:2003}.  For the three-year analysis, we
evaluate the constituent pseudo-$a_{lm}$ data using uniform pixel weights for
$l<500$ and $N_{\rm obs}$ weights for $l>500$.  For the latter we use the
high-resolution r10 maps to reduce pixelization smearing.  Given the
pseudo-$a_{lm}$ data, we evaluate cross-power spectra for all 153 independent
combinations of V- and W-band data (\S\ref{sec:tt_instrument}).

The weights used to average the spectra are obtained as follows.  First, the
diagonal elements of a Fisher matrix, $F^{ii}_l$, are computed for each DA $i$,
following the method outlined in Appendix~D of \citet{hinshaw/etal:2003}. Since
the sky coverage is so similar from year to year (see Figure~\ref{fig:nobs_123})
we compute only one set of $F^{ii}_l$ for all three years using the year-1
$N_{\rm obs}$ data.  To account for 1/f noise, the low-$l$ elements of each
Fisher matrix are decreased according to the noise bias model described in
\S\ref{sec:tt_noise}.  $F^{ii}_l$ gives the relative noise per DA (as measured
by the noise bias) and the relative beam response at each $l$.  We weight the
cross-power spectra using the product $F^{ij}_l = (F^{ii}_lF^{jj}_l)^{1/2}$.

The noise bias model, equation~(\ref{eq:noise_bias_model}), is propagated
through the averaging to produce an effective noise bias
\beq
N^{\rm eff}_l = \sqrt{\frac{1}{\sum_{ij}{(N^i_l)^{-1} (N^j_l)^{-1}}}},
\eeq
where $N^i_l$ is the noise bias model for DA $i$, and the sum is over all DA
pairs used in the final spectrum.  In the end, the noise error in the combined
spectrum is expressed as
\beq
\Delta C_l = \sqrt{\frac{a}{2l+1}}\,\frac{N^{\rm eff}_l}{f^{\rm sky}_l}.
\eeq
Here, $f^{\rm sky}_l$ is the effective sky fraction observed, which we calibrate
using Monte Carlo simulations \citep{verde/etal:2003}.  The factor $a$ is 1 for
TE, TB, and EB spectra, and 2 for TT, EE, and BB spectra.  Note that the final
error estimates are independent of the original Fisher elements, $F^{ii}_l$,
which are only used as relative weights.  Note also that $N^{\rm eff}_l$ is
separately calibrated for both regimes of pixel weighting, uniform and $N_{\rm
obs}$.

The full covariance matrix for the final spectrum includes contributions from
cosmic variance, instrument noise, mode coupling, and beam and point source
uncertainties.  We have revised our handling of beam and source errors in the
three-year analysis, as discussed in Appendix~\ref{app:beam_src_err}.  However,
the remaining contributions are treated as we did in the first-year analysis
\citep{hinshaw/etal:2003, verde/etal:2003}.   A good estimate of the uncertainty
per $C_l$ is given by
\beq
\Delta C_l = \frac{1}{f^{\rm sky}_l} \, \sqrt{\frac{2}{2l+1}} \,
             (C^{\rm fid}_l + N^{\rm eff}_l),
\eeq
where $C^{\rm fid}_l$ is a fiducial model spectrum.  This estimate includes the
effects of cosmic variance, instrument noise, and mode coupling, but not beam
and point source uncertainties.  However, these effects are accounted for in the
likelihood code delivered with the three-year release.

\subsection{The Full Power Spectrum}
\label{sec:tt_full}

To construct the final power spectrum we combine the maximum likelihood results
from Table~\ref{tab:tt_low_l_like} for $l \le 30$ with the pseudo-$C_l$ based
cross-power spectra, discussed above, for $l>30$.  The results are shown in
Figure~\ref{fig:tt_final} where the \map\ data are shown in black with
noise-only errors, the best-fit $\Lambda$CDM model, fit to the three-year data,
is shown in red, and the 1$\sigma$ error band due to cosmic variance shown in
lighter red.  The data have been averaged in $l$ bands of increasing width, and
the cosmic variance band has been binned accordingly.  To see the effect that
binning has on the model prediction, we average the model curve in the same $l$
bins as the data and show the results as dark red diamonds.  For the most part,
the binned model is indistinguishable from the unbinned model except in the
vicinity of the second acoustic peak and trough.

Based on the noise estimates presented in \S\ref{sec:tt_noise}, we determine that the
three-year spectrum is cosmic variance limited to $l=400$.  The signal-to-noise
ratio per $l$-mode exceeds unity up to $l=850$, and for bins of width $\Delta l
/ l = 3$\%, the signal-to-noise ratio exceeds unity up to $l=1000$.  In the
noise-dominated, high-$l$ portion of the spectrum, the three-year data are more
than three times quieter than the first-year data due to (1) the additional
years of data, and (2) the use of finer pixels in the V and W band sky maps,
which reduces pixel smearing at high $l$.  The $\chi^2_{\nu}$ of the full power
spectrum relative to the best-fit $\Lambda$CDM model is 1.068 for 988 D.O.F.
($13<l<1000$) \citep{spergel/etal:prep}.  The distribution of $\chi^2$ vs. $l$
is shown in Figure~\ref{fig:tt_chi2}, and is discussed further below.

The first two acoustic peaks are now measured with high precision in the
three-year spectrum.  The 2nd trough and the subsequent rise to a 3rd peak are
also well established.  To quantify these results, we repeat the
model-independent peak and trough fits that were applied to the first-year data
by \citet{page/etal:2003}.  The results of this analysis are listed in
Table~\ref{tab:tt_peak}.  We note here that the first two acoustic peaks are
seen at $l=220.8 \pm 0.7$ and $l=530.9 \pm 3.8$, respectively, while in the
first-year spectrum, they were located at $l=220.1 \pm 0.8$ and $l=546 \pm 10$. 
Table~\ref{tab:tt_peak} also shows that the second trough is now well measured
and that the rise to the third peak is unambiguous, but the position and
amplitude of the 3rd peak are not yet well constrained by \map\ data alone.

Figure~\ref{fig:tt_final_ext} shows the three-year \map\ spectrum compared to a
set of recent balloon and ground-based measurements that were selected to most
complement the \map\ data in terms of frequency coverage and $l$ range.  The
non-\map\ data points are plotted with errors that include both measurement
uncertainty and cosmic variance, while the \map\ data in this $l$ range are
largely noise dominated, so the effective error is comparable.  When the \map\
data are combined with these higher resolution CMB measurements, the existence
of a third acoustic peak is well established, as is the onset of Silk damping
beyond the 3rd peak.

The three-year spectrum is compared to the first-year spectrum in
Figure~\ref{fig:tt_3yr_1yr}.  We show the new spectrum in black and the old one
in red.  The best-fit $\Lambda$CDM model, fit to the three-year data, is shown
in grey.  In the top panel, the as-published first-year spectrum is shown.  The
most noticeable differences between the two spectra are, (1) the change at
low-$l$ due to the adoption of the maximum likelihood estimate for $l \le 30$,
(2) the smaller uncertainties in the noise-dominated high-$l$ regime, discussed
further below, and (3) a small but systematic difference in the mid-$l$ range
due to improvements in our determination of the beam window functions
(\S\ref{sec:tt_windows}).  The middle panel shows the ratio of the new spectrum
to the old.  For comparison, the red curve shows the (inverse) ratio of the
three-year and first-year window functions, which differ by up to 2\%.  The
spectrum ratio tracks the window function ratio well up to $l \sim 500$ at which
point the sensitivity of the first-year spectrum starts to diminish.  For $l \le
30$ in this panel, we have substituted the pseudo-$C_l$ based spectrum from the
three-year data into this ratio to show the stability of the underlying sky map
data.  When based on the same estimation method, the two spectra agree to within
$\sim$2\%, despite changes in the gain model and in the details of the Galactic
foreground subtraction, both of which affect the low-$l$ data.  (The gain model
changes {\em are} important for the polarization data, while the
temperature-based foreground model has no direct bearing on polarization.)  The
bottom panel shows the three-year and first-year spectra again, but this time
the first-year data have been deconvolved with the three-year window functions
and we have substituted the maximum likelihood estimate into the first-year
spectrum.  The agreement between the two is now excellent.

As noted above, the power spectrum is now measured with cosmic variance limited
sensitivity to $l=400$, which happens to coincide with the measured position of
the first trough in the acoustic spectrum (Table~\ref{tab:tt_peak}). 
Figure~\ref{fig:tt_1st_pk} shows the measurement of the first acoustic peak with
no binning in $l$.  The black trace shows the three-year measurement, while the
grey trace shows the first-year result for comparison.  The background error
band gives the 1$\sigma$ uncertainty per $l$ for the three-year data, including
the effects of Galaxy masking and a minimal contribution from instrument noise
at the high-$l$ limit.  In the first-year power spectrum there were several
localized features, which have become known, technically, as ``glitches''.  The
most visible was a ``bite'' near the top of the first acoustic peak, from
$l=205$ to 210.  Figure~\ref{fig:tt_1st_pk} shows that the feature still exists
in the three-year data, but it is not nearly as prominent, and it disappears
almost entirely from the binned spectrum.  In this $l$ range, the pixel weights
used to evaluate the pseudo-$C_l$ data changed from ``transitional'' in the
first-year analysis [Appendix~A of \citet{hinshaw/etal:2003}] to uniform in the
three-year analysis.  This reduces the effective level of cosmic variance in
this $l$ range, while increasing the small noise contribution somewhat.  We
conclude that the feature was likely a noise fluctuation superposed on a
moderate signal fluctuation.  

Several other glitches remain in the low-$l$ power spectrum, perhaps including
the low quadrupole.  Several authors have commented on the significance of these
features \citep{efstathiou:2003d, lewis:2003oekstqyclewmgkmwsa,
bielewicz/gorski/banday:2004, slosar/seljak/makarov:2004}, and several more have
used the $C_l$ data to search for features in the underlying primordial
spectrum, $P(k)$ \citep{shafieloo/souradeep:2004, martin/ringeval:2004,
martin/ringeval:2005, tocchini/etal:2006}, see also \citet{spergel/etal:prep}. 
Figure~\ref{fig:tt_chi2} shows that the binned $\chi^2$ per $l$-band is slightly
elevated at low $l$, but not to a compelling degree.  In the absence of an
established theoretical framework in which to interpret these glitches (beyond
the Gaussian, random phase paradigm), they will likely remain curiosities.

\section{BEYOND THE ANGULAR POWER SPECTRUM -- LARGE-SCALE FEATURES}
\label{sec:phase}

The low $l$ CMB modes trace the largest structure observable in the universe
today.  In an inflationary scenario, these modes were the first to leave the
horizon and the most recent to re-enter.  Since they are largely unaffected by
sub-horizon scale acoustic evolution, they also present us with the cleanest
remnant of primordial physics available today.  Unusual behavior in these modes
would be of potentially great and unique importance.

Indeed there do appear to be some intriguing features in the data, but their
significance is difficult to ascertain and has been a topic of much debate.  In
this section we briefly review the claims that have been made to date and
comment on them in light of the three-year \map\ data.  These features include:
low power, especially in the quadrupole moment; alignment of modes, particularly
along an ``axis of evil''; unequal fluctuation power in the northern and
southern sky; a surprisingly low three-point correlation function in the
northern sky; an unusually deep/large cold spot in the southern sky;  and
various ``ringing'' features, ``glitches'', and/or ``bites'' in the power
spectrum.  Of course one expects to encounter low probability features with any
{\it a posteriori} data analysis, and there are no clear cut rules for assigning
the degree of posterior bias to any given observation, so some amount of
judgment is called for.  Ultimately, the most scientifically compelling
development would be the introduction of a new model that explains a number of
currently disparate phenomena in cosmology (such as the behavior of the low $l$
modes and the nature of the dark energy) while also making testable predictions
of new phenomena.

\subsection{Summary of First-Year Results}
\label{sec:phase_yr1}

The 2-point correlation function computed from the first-year \map\ data showed
a notable lack of signal on large angular scales \citep{spergel/etal:2003}. 
Similar behavior was also seen in the \cobe-DMR data \citep{hinshaw/etal:1996b},
making unidentified experimental systematic effects an unlikely cause.  Because
the 2-point function is the Legendre transform of the angular power spectrum,
the large-scale behavior of $C(\theta)$ is dominated by the lowest $l$ modes,
especially the quadrupole.  So this signature is at least partially a reflection
of the low quadrupole amplitude.  

To study this further, \citet{spergel/etal:2003} characterized the feature by an
integral, $S = \int_{-1}^{+1/2} C^2(\theta) d\cos\theta$, which measures the
power in $C(\theta)$ for $\theta > 60^{\circ}$.  In a Monte Carlo simulation,
only 0.15\% of the realizations produced a value of $S$ as small as did the
\map\ data.  It remains to assess the degree to which Galactic errors and
posterior bias (the selection of the $60^{\circ}$ integration limit) affect this
result.  \citet{niarchou/etal:2003} concluded that the evidence for measurement
error was at least as likely as the evidence for a cutoff in the primordial
power spectrum, but that both in both cases the evidence was weak.

\citet{tegmark/deoliveira-costa/hamilton:2003} produced a high resolution
full-sky CMB map by using a variant of the foreground cleaning approach
developed by \citet{tegmark/efstathiou:1996}. They cautiously quote a quadrupole
amplitude of $202$ \uKsq, evaluated over the full-sky.  They further remark that
the quadrupole and octupole phases are notably aligned with each other and that
the octupole is unusually ``planar'' with most of its power aligned
approximately with the Galactic plane. \citet{deoliveira-costa/etal:2004a}
estimate the {\em a priori} chance of observing both the low quadrupole, the
$l=2,3$ alignment, and the $l=3$ planarity to be $\sim$1 in 24000.  Note that
this estimate does not formally attempt to account for Galactic modeling
uncertainty, which will tend to reduce the significance of the noted features.
\citet{bielewicz/gorski/banday:2004} confirmed the above result, and they
further conclude that the $l=3$ properties are stable with respect to the
applied mask and the level of foreground correction, but that the quadrupole is
much less so.

\citet{schwarz/etal:2004} claimed that the quadrupole and octopole are even more
correlated (99.97\% CL), with the quadrupole plane and the three octopole planes
``remarkably aligned.'' Further, they claimed that three of these planes  are
orthogonal to the ecliptic and the normals to these planes are aligned with the 
direction of the cosmological dipole and with the equinoxes. This had led to
speculation that the low-$l$ signal is not cosmological in origin. 
\citet{copi/huterer/starkman:2003} use ``multipole vectors'' to characterize the
geometry of the $l$ modes.  They conclude that the ``oriented area of planes
defined by these vectors $\ldots$ is inconsistent with the isotropic Gaussian
hypothesis at the 99.4\% level for the ILC map.''  See also
\citet{katz/weeks:2004}.  

\begin{figure}
\figurenum{14}
\epsscale{0.8}
\plotone{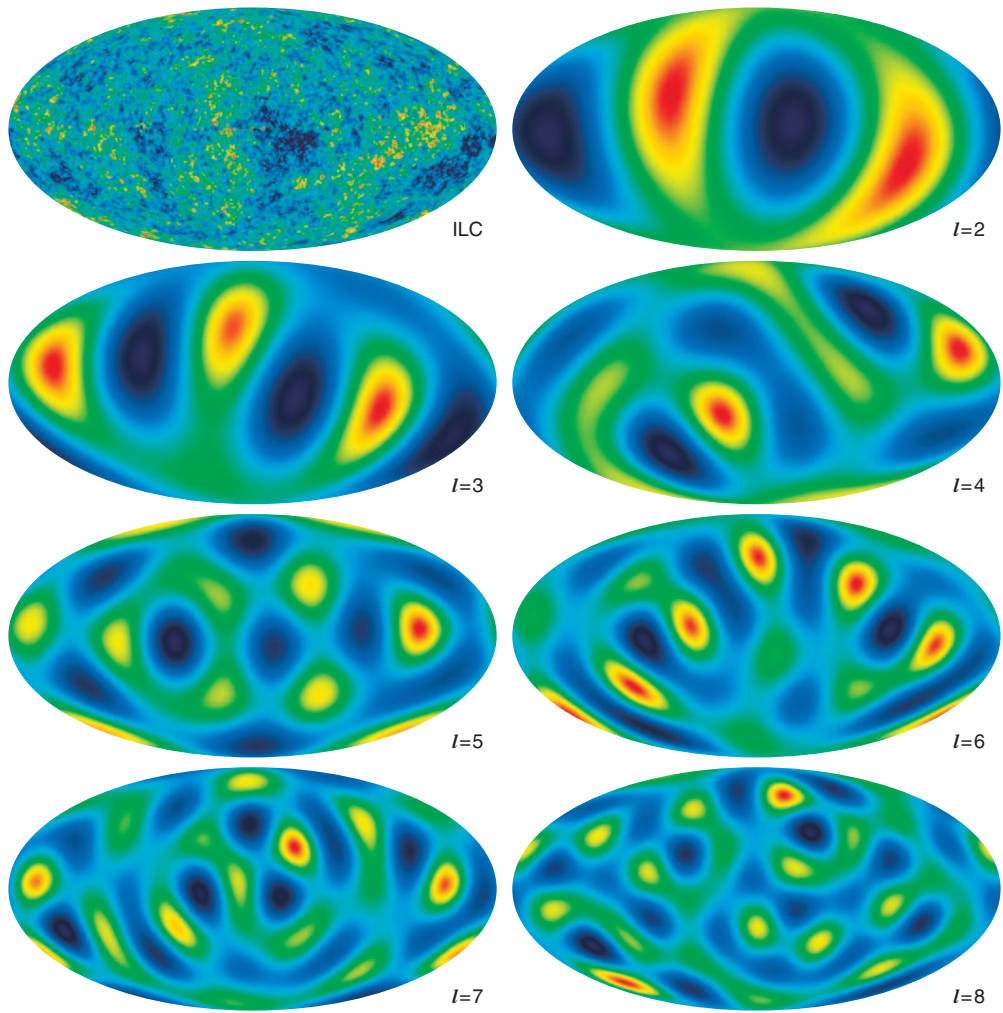}
\caption{Maps of power spectrum modes $l=2-8$ computed from full-sky fits to the
ILC map, shown at top left.  Many authors note peculiar patterns in the phase of
these modes, and many claim that the behavior is inconsistent with Gaussian
random-phase fluctuations, as predicted by inflation.  For example, the $l=5$
mode appears strikingly symmetric (a non-random distribution of power in $m$),
while the $l=2$ and $3$ modes appear unusually aligned.  The significance of
these {\it a posteriori} observations is being actively debated.  See
\S\ref{sec:phase} for a more detailed discussion.}
\label{fig:ilc_2_8}
\end{figure}

\begin{figure}
\figurenum{15}
\epsscale{1.0}
\plotone{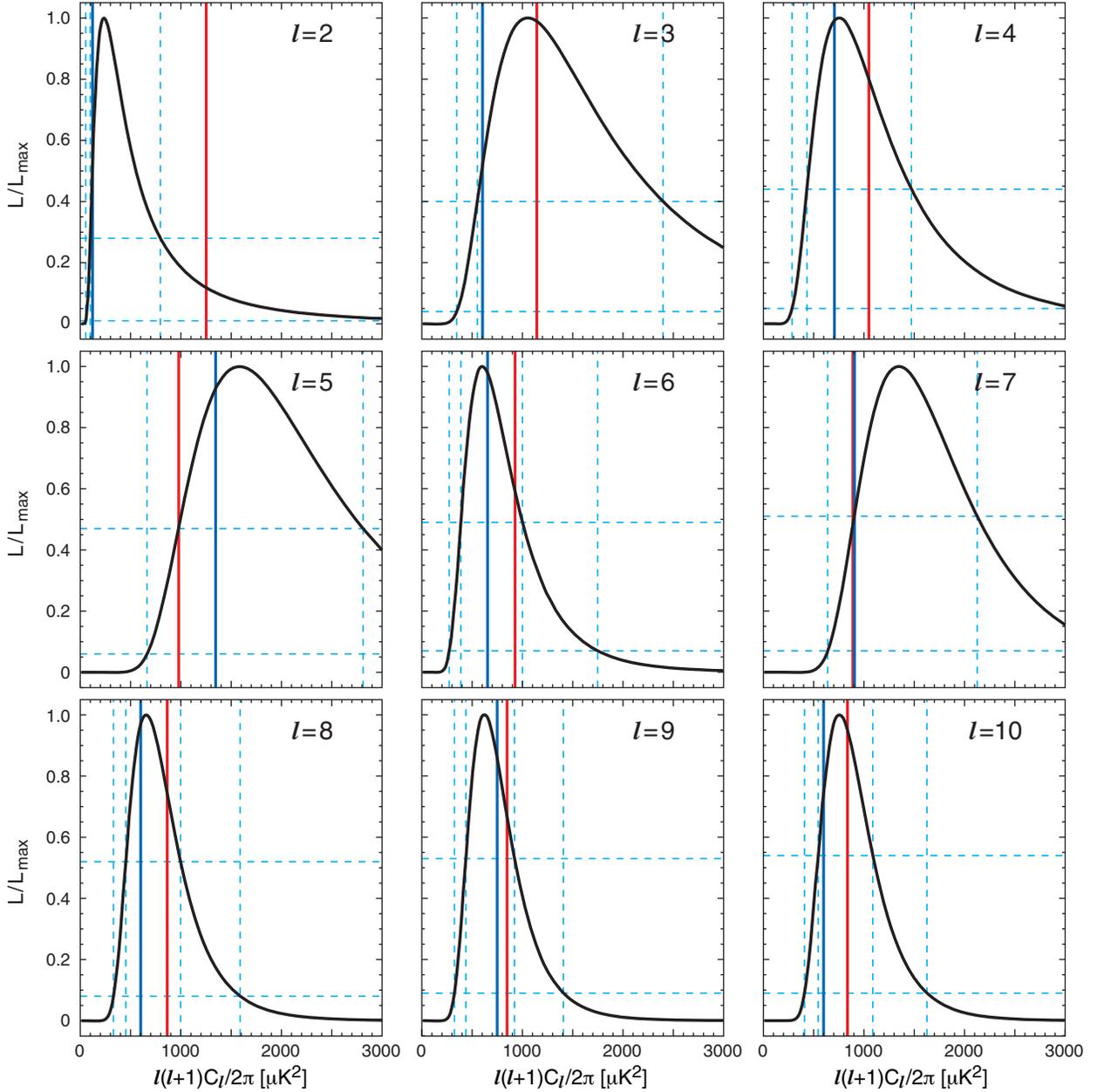}
\caption{The posterior likelihood of $l(l+1)C_l/2\pi$ given the \map\ data, for
$l=2-10$.  The curves are computed using data from the ILC map evaluated outside
the Kp2 sky cut.  The vertical blue lines show the values inferred using the
pseudo-$C_l$ estimate.  These values are well within the posterior likelihood
distribution, but there is a tendency for them to be lower than the peak,
especially for $l=2,3,7$.  The vertical red lines show the predicted power
spectrum from the best-fit $\Lambda$CDM model (fit to \map\ data).  These points
are well within the posterior likelihood in all cases; the faint dashed lines
indicate the 68\% and 95\% confidence regions of the distribution (see
\S\ref{sec:tt_low_l}).  Note that the curves approach a Gaussian distribution as
$l$ increases.}
\label{fig:lowl_tt_like}
\end{figure}

\begin{figure}
\figurenum{16}
\plotone{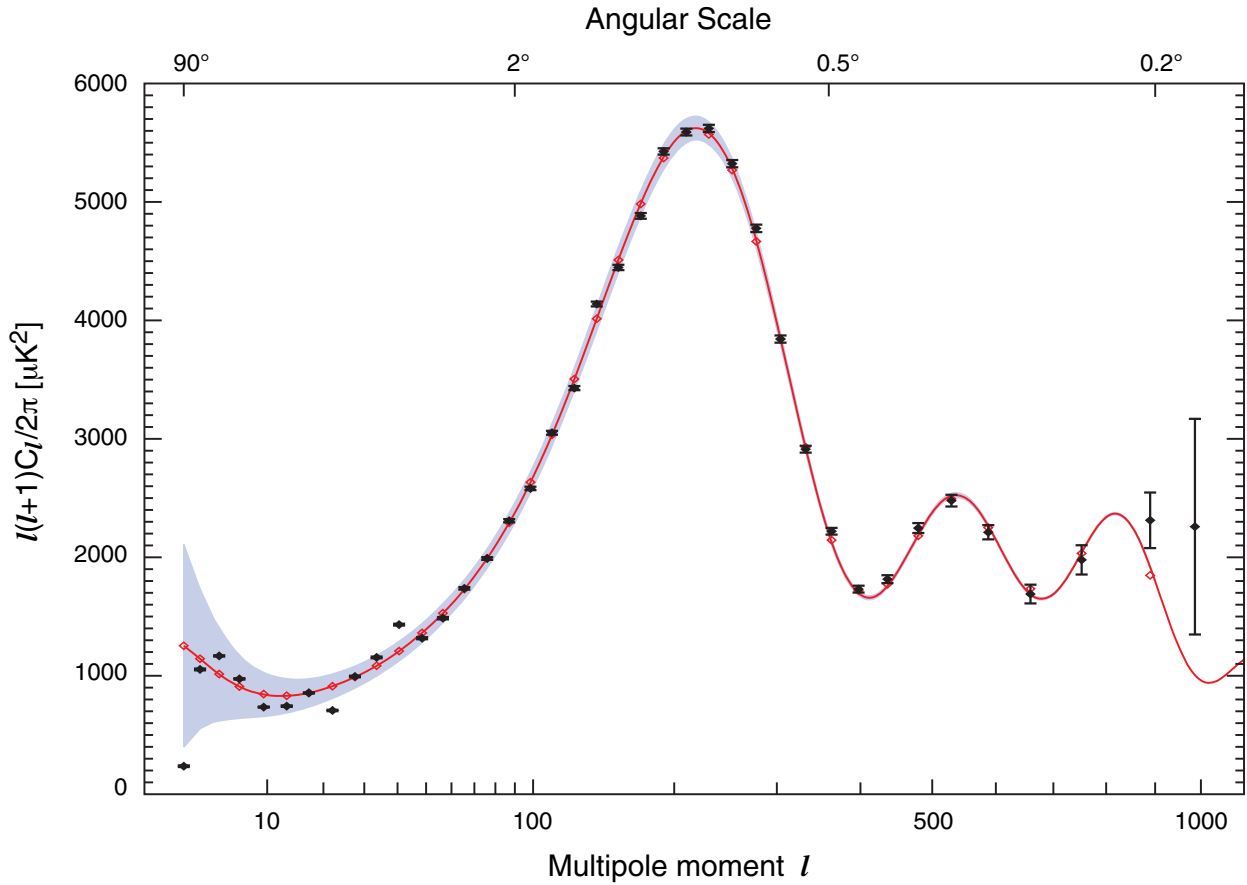}
\caption{The binned three-year angular power spectrum (in black) from
$l=2-1000$, where it provides a cosmic variance limited measurement of the first
acoustic peak, a robust measurement of the second peak, and clear evidence for
rise to the third peak.  The points are plotted with noise errors only (see
text).  Note that these errors decrease linearly with continued observing time. 
The red curve is the best-fit $\Lambda$CDM model, fit to \map\ data only
\citep{spergel/etal:prep}, and the band is the binned 1$\sigma$ cosmic variance
error.  The red diamonds show the model points when binned in the same way as
the data.}
\label{fig:tt_final}
\end{figure}

\begin{figure}
\figurenum{17}
\epsscale{1.0}
\plotone{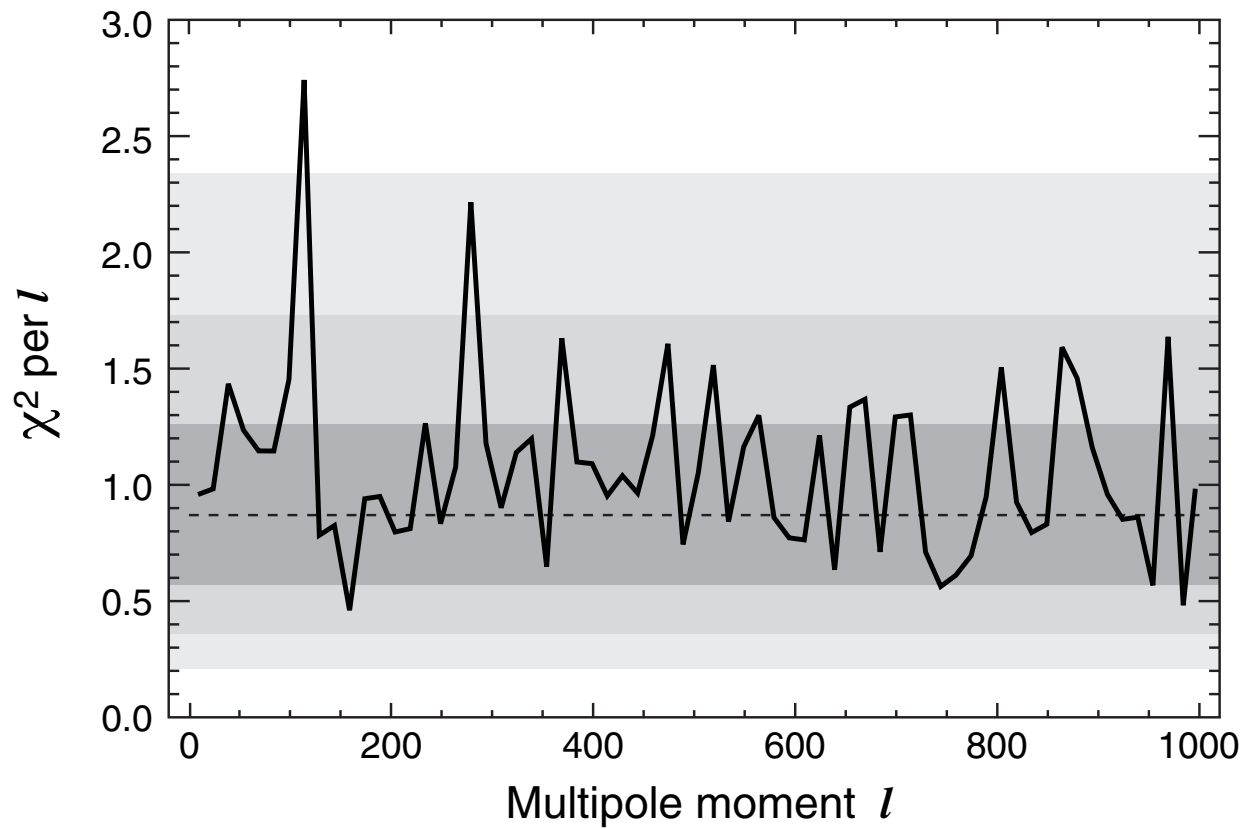}
\caption{$\chi^2$ vs. $l$ for the full power spectrum relative to the best-fit
$\Lambda$CDM model, fit to \map\ data only.  The $\chi^2$ per $l$ has been
averaged in $l$-bands of width $\Delta l = 15$.  The dark to light grey shading
indicates the 1, 2, and 3$\sigma$ confidence intervals for this distribution. 
The dashed line indicates the mode.}
\label{fig:tt_chi2}
\end{figure}

\begin{figure}
\figurenum{18}
\plotone{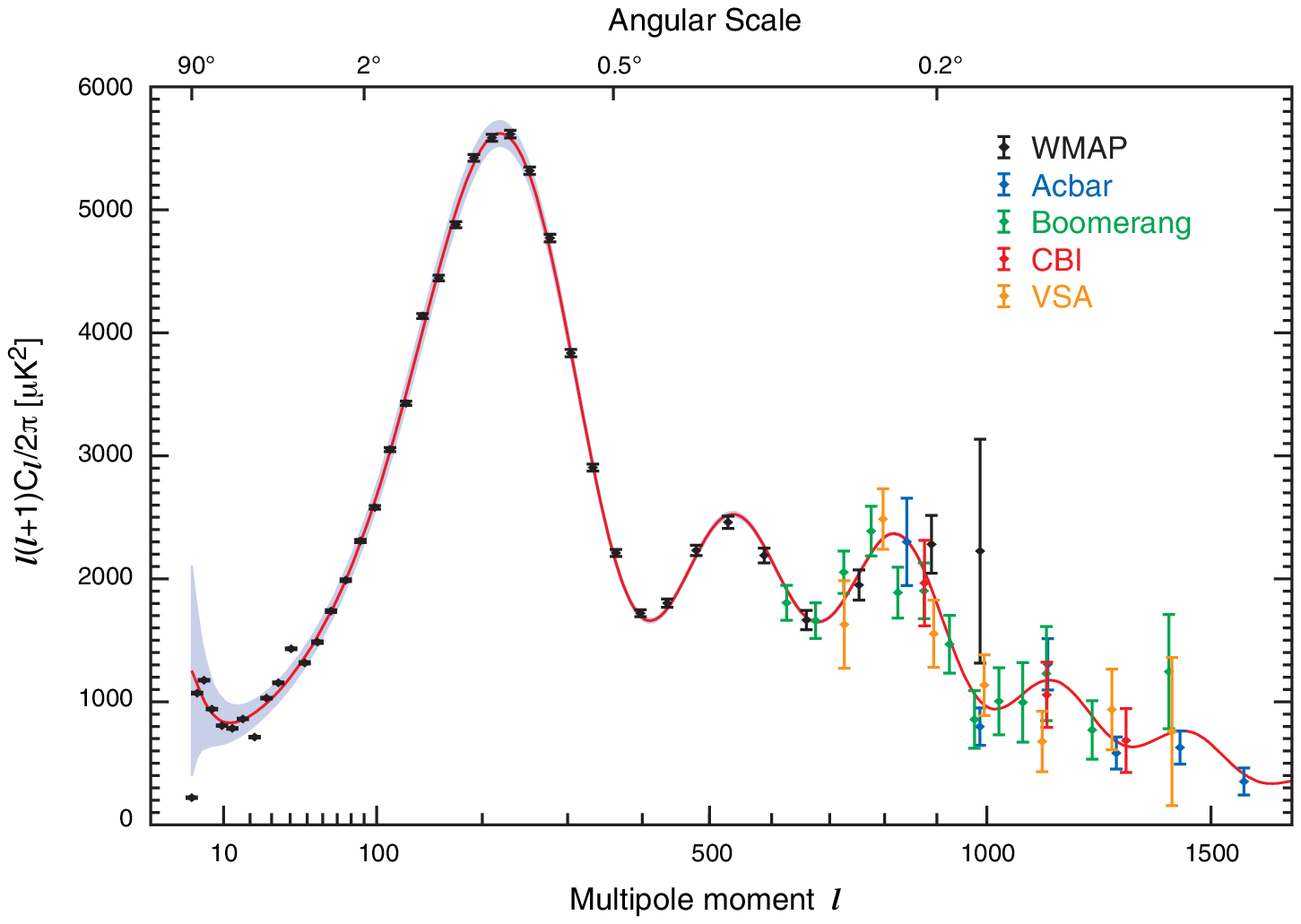}
\caption{The \map\ three-year power spectrum (in black) compared to other recent
measurements of the CMB angular power spectrum, including Boomerang
\citep{jones/etal:2005}, Acbar \citep{kuo/etal:2004}, CBI
\citep{readhead/etal:2004}, and VSA \citep{dickinson/etal:2004}.  For clarity,
the $l<600$ data from Boomerang and VSA are omitted; as the measurements are
consistent with \map, but with lower weight.  These data impressively confirm
the turnover in the 3rd acoustic peak and probe the onset of Silk damping.  With
improved sensitivity on sub-degree scales, the \map\ data are becoming an
increasingly important calibration source for high-resolution experiments.}
\label{fig:tt_final_ext}
\end{figure}

\begin{figure}
\figurenum{19}
\epsscale{0.4}
\plotone{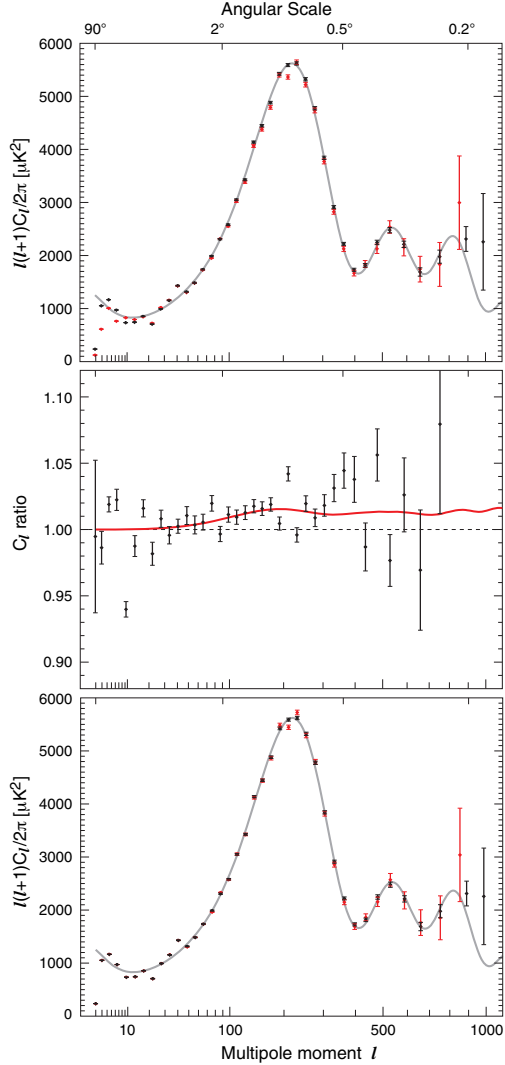}
\caption{Comparison of the three-year angular power spectrum with the first-year
result. ({\it top}) The three-year combined spectrum, in black, is shown with
the first-year spectrum, as published, in red.  The best-fit three-year
$\Lambda$CDM model is shown in grey for comparison.  For $l<30$, the difference
is due to a change in estimation methodology: the three-year spectrum is based
on a maximum likelihood estimate, while the first-year results is based on a
pseudo-$C_l$ estimate.  For $l>100$ the difference is due primarily to (i) an
improved determination of the \map\ beam response, and (ii) the improved
sensitivity of the three-year data.  ({\it middle}) Ratio of the three-year
spectrum to the first-year spectrum.  For $l<30$ we plot the ratio of the two
pseudo-$C_l$-based spectra to show the consistency of the underlying data.  The
red curve is the ratio of the first-year window function to the three-year
window function. ({\it bottom}) Same as top panel except that the first-year
spectrum has been multiplied by the window function ratio depicted in the middle
panel, and the maximum likelihood estimate has been substituted for $l<30$.}
\label{fig:tt_3yr_1yr}
\end{figure}

\begin{figure}
\figurenum{20}
\epsscale{1.0}
\plotone{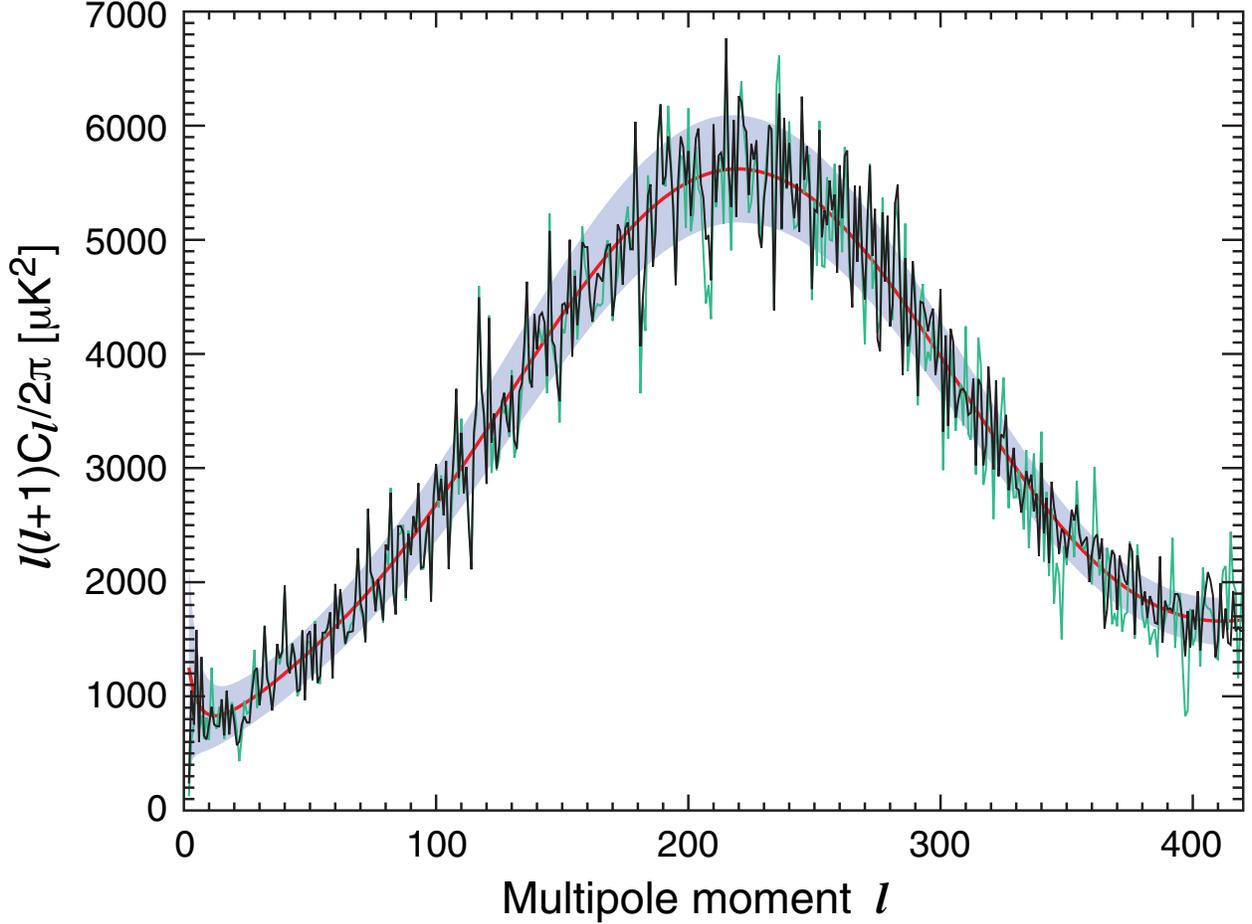}
\caption{The unbinned three-year angular power spectrum (in black) from
$l=2-400$, where it provides a cosmic variance limited measurement of the first
acoustic peak.  The first-year spectrum is shown in green for comparison.  The
red curve with the grey band is the best-fit $\Lambda$CDM model and 1$\sigma$
error band per $l$.  The width of the band is dominated by cosmic variance to
$l=400$.  One feature that was singled out in the first-year spectrum was the
``bite'' at $l \sim 208$.  The feature is still visible in the three-year
spectrum, but not prominently.  We believe this feature was predominantly a
noise fluctuation in the first-year data.}
\label{fig:tt_1st_pk}
\end{figure}

\citet{eriksen/etal:2004d}, in their study of the ILC method, confirm that the
quadrupole and octopole are strongly aligned.  They also note that the $l=5$
mode is ``spherically symmetric'' at  $\sim3\sigma$, and the $l=6$ mode is
planar at $\sim 2\sigma$ confidence (see Figure~\ref{fig:ilc_2_8}).  But they
add that the first-year ILC map is probably not clean enough to use for
cosmological analyses of this type.  \citet{land/magueijo:2005} point out that
the $l=3$ and 5 modes are aligned in both direction and azimuth, thereby
``rejecting statistical isotropy with a probability in excess of 99.9\%.''

\citet{hansen/etal:2004b} fit cosmological parameters to the \map\ data
separately in the northern and southern hemispheres in three coordinate systems.
They conclude that, ``it may be necessary to question the assumption of
cosmological isotropy.''  \citet{eriksen/etal:2004a}, evaluate the ratio of
low-$l$ power between two hemispheres and conclude that only 0.3\% of simulated
skies have as low a ratio as observed, even when allowing the (simulated) data
to define the direction.  \citet{hansen/banday/gorski:2004} reach a similar
conclusion and note that it is ``hard to explain in terms of residual
foregrounds and known systematic effects.''

\begin{figure}
\figurenum{21}
\epsscale{0.8}
\plotone{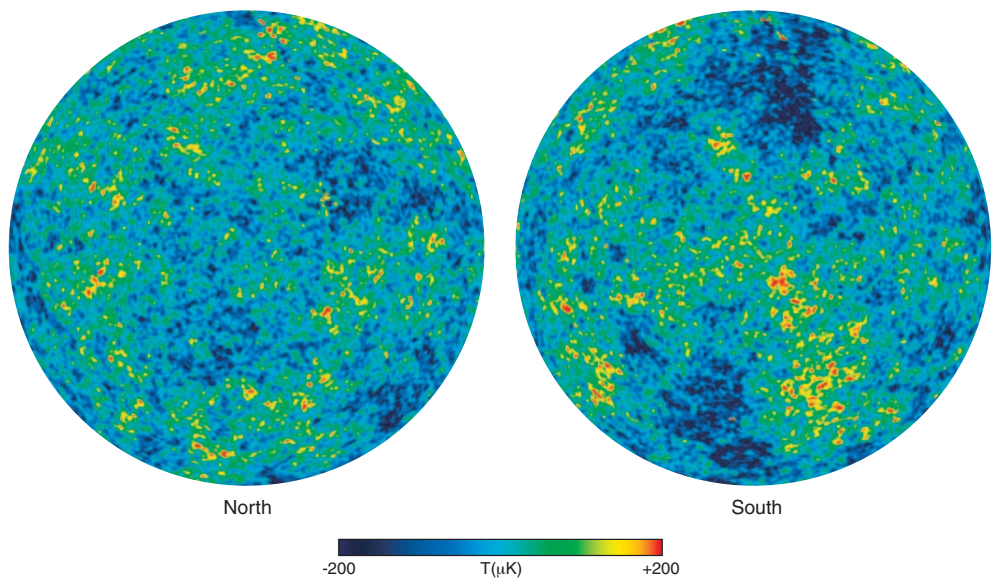}
\caption{Full-sky maps in ecliptic coordinates smoothed to $1^{\circ}$
resolution, shown in Lambert Azimuthal Equal Area projection. {\it top}:
three-year ILC map,  {\it bottom}: three-year W-band map with the Kp2 Galaxy
mask superposed, but no other foreground removal applied.  Note the qualitative
differences in large-scale power between the two hemispheres, as has been noted
by several authors (see text).  The statistical significance of this difference
continues to be an area of active study.}
\label{fig:ilc_ecl}
\end{figure}

Faced with so many apparently unlikely features in the data, our foremost
priority is to re-evaluate potential sources of systematic error in the maps.
Indeed, a by-product of the exhaustive three-year polarization analysis is
enhanced confidence in the accuracy of the temperature signal.  
Figure~\ref{fig:maps_3yr_1yr} shows that, even after all of the processing
changes outlined in \S\ref{sec:change}, the three-year maps are consistent with
the first-year maps up to a small quadrupole difference
(Table~\ref{tab:3yr_1yr_diff}) that is well within the first-year error budget
\citep{hinshaw/etal:2003b}.  Furthermore, the improvements applied to the ILC
processing to reduce Galactic foreground residuals, did not visibly alter  the
low-$l$ phase structure in the three-year ILC map (Figure~\ref{fig:ilc_2_8}),
nor the north-south asymmetry that is plainly visible in
Figure~\ref{fig:ilc_ecl}.  We have made no attempt to evaluate the above-noted
statistics using the three-year maps, so the degree to which they persist
remains to be seen.

\section{SUMMARY AND CONCLUSIONS}
\label{sec:conclude}

For the three-year \map\ data release, we have made improvements in nearly every
aspect of the data processing pipeline, many of which were driven by the need to
successfully characterize the instrument noise to the level required to measure
the polarization signal, $\sim$0.1 $\mu$K.  The improvements are spelled out in
detail here and in the companion papers by \citet{jarosik/etal:prep},
\citet{page/etal:prep}, and \citet{spergel/etal:prep}.  The key points follow.

\begin{figure}
\figurenum{22}
\epsscale{1.0}
\plotone{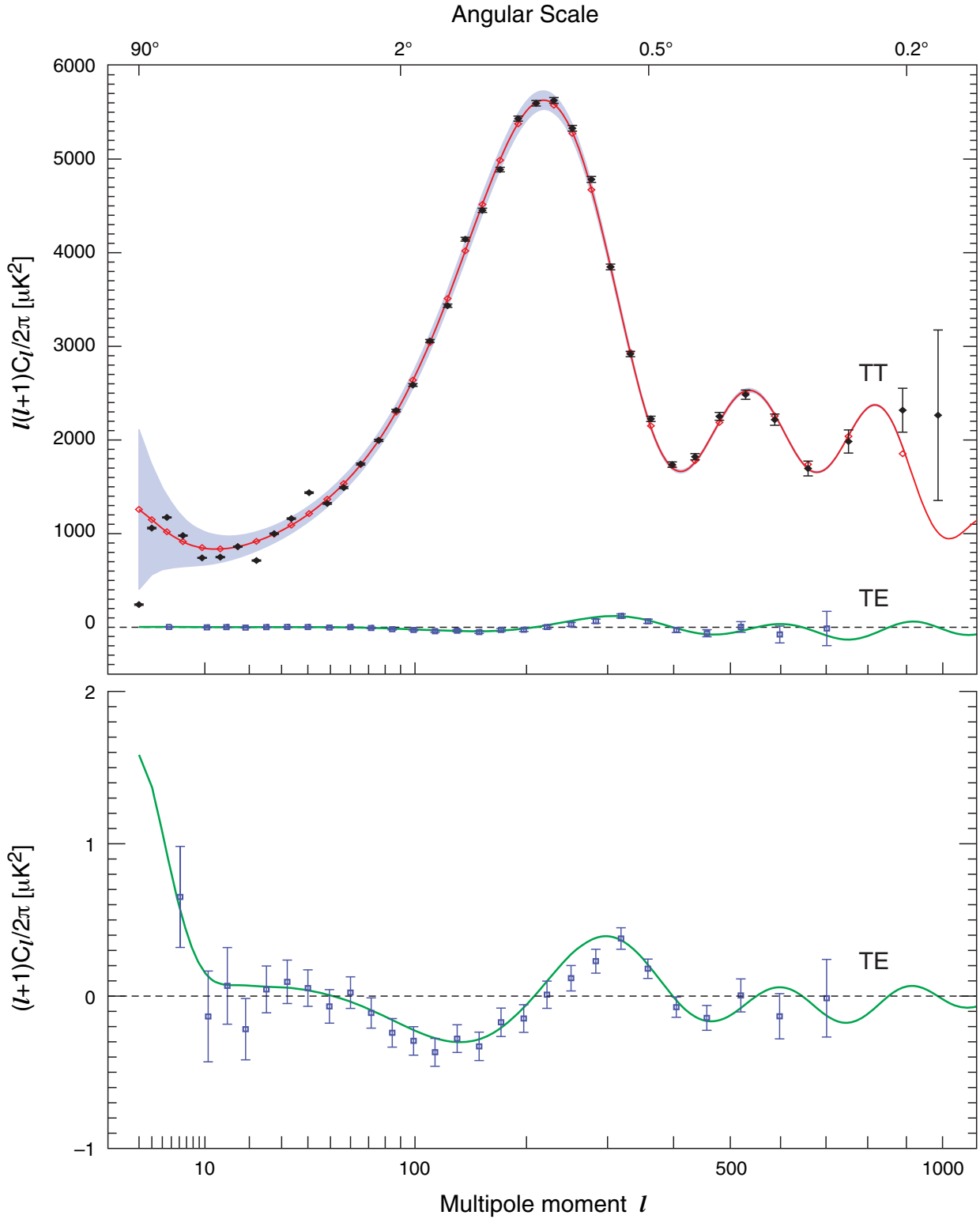}
\caption{Angular power spectra $C_l^{TT}$ \& $C_l^{TE}$ from the three-year
\map\ data.  {\it top}: The TT data are as shown in Figure~\ref{fig:tt_final}. 
The TE data are shown in units of $l(l+1)C_l/2\pi$, on the same scale as the TT
signal for comparison.  {\it bottom}: The TE data, in units of $(l+1)C_l/2\pi$. 
This updates Figure~12 of \citet{bennett/etal:2003b}.}
\label{fig:tt_te}
\end{figure}

\begin{enumerate}

\item Improved models of the instrument gain and beam response are now accurate
to better than 1\% and do not presently limit the scientific conclusions that
can be drawn from the data.  Polarization-specific effects such as spurious
pickup from radiometer bandpass mismatch have also been accounted for in the
data processing, and do not limit the data.

\item The map-making procedure has been overhauled to produce genuine maximum
likelihood maps of the temperature and polarization signal.  In concert with
this, we have produced code to evaluate the corresponding pixel-pixel weight
matrix (inverse covariance matrix) at pixel resolution r4.  This was required to
adequately characterize the noise for the polarization analysis
\citep{jarosik/etal:prep}.

\item The three-year temperature maps are consistent with the first-year maps
(\S\ref{sec:maps}) and, to a good approximation, have 3 times lower variance. A
by-product of the exhaustive polarization analysis is enhanced confidence in the
accuracy of the temperature signal.

\item We have updated the MEM, ILC, and template-based temperature foreground
emission models and assessed their uncertainties.  We have developed new models
of the polarized foreground emission that allow us to extract the cosmological
reionization signal and that pave the way for future studies of CMB polarization
\citep{page/etal:prep}.

\item Our analysis of the temperature power spectrum is improved at low $l$ by
employing a maximum likelihood estimate.  At high $l$, the spectrum is $>\,$3
times more sensitive due to the additional years of data and the use of r10 sky
maps to reduce the effects of pixel smoothing.  We have also placed new limits
on systematic effects in the power spectrum due to beam ellipticity.

\item We have updated the \map\ point source catalog.  The contribution of
unresolved sources to the power spectrum has been updated and subtracted.

\item We have developed new methods to evaluate polarization power spectra from
sky map data.  Our approach accounts for the following key issues: (1) sky cuts,
(2) low signal amplitude, (3) correlated noise from 1/f and scan-related
effects, (4) other polarization-specific effects, such as baseline sensitivity,
and bandpass mismatch.  This effort was developed in conjunction with the
processing of polarization sky maps \citep{page/etal:prep}.

\item Figure~\ref{fig:tt_te}, which updates Figure~12 from
\citet{bennett/etal:2003b}, shows the three-year TT and TE power spectra
(\S\ref{sec:tt_full} and \citet{page/etal:prep}).  The TT measurement is cosmic
variance limited up to $l=400$, and has a signal-to-noise ratio greater than one
up to $l=1000$, in $l$ bands of width 3\%.  The high-$l$ TE signal is consistent
with the first-year result, while for $l<6$ the signal is reduced. A detailed
comparison of the three-year TE signal to the first-year is shown in Figure~24
of  \citet{page/etal:prep}.  The implications for the inferred optical depth,
$\tau$,  are shown in their Figure~26, which includes a detailed comparison of
the three-year results with the first-year.

\item We have refined the evaluation of the likelihood function used to infer
cosmological parameters.  The new approach uses pixel-space inputs for the
low-$l$ temperature and polarization data and correctly accounts for the joint
probability of the temperature and polarization signal \citep{page/etal:prep}.
We use this to infer cosmological parameters from the three-year data.  Many
parameter degeneracies that existed in the first-year results are greatly
reduced reduced by having new EE data and more sensitive high-$l$ TT data.
Figure~1 of \citet{spergel/etal:prep} illustrates this for the parameter pairs
($n_s$,$\tau$) and ($n_s$,$\Omega_b h^2$).  A cosmological model with only 6
parameters still provides an excellent fit to \map\ and other cosmological data.

\item With new constraints on $\tau$ from the EE data, we get correspondingly
tighter constraints on the scalar spectral index, $n_s$.  Using \map\ data only,
we now find $n_s = 0.958 \pm 0.016$, while combining \map\ data with other data
gives $n_s = 0.947 \pm 0.015$ \citep{spergel/etal:prep}.  Joint limits on $n_s$
and the tensor to scalar ratio, $r$, are shown in Figure~14 of
\citet{spergel/etal:prep}.

\item The \map\ observatory continues to operate flawlessly, and many results
that are currently measured at 2-3 $\sigma$ confidence, such as the apparent
deviation from scale invariance, hints of a running spectral index, and the
details of the reionization history, will be significantly clarified by
additional years of data.

\end{enumerate}

\section{DATA PRODUCTS}
\label{sec:data}

All of the three-year \map\ data products are being made available through the
Legacy Archive for Microwave Background Data Analysis (LAMBDA), NASA's CMB
Thematic Data Center.  The low-level products include the time-ordered data
(calibrated and raw), the functions used to filter the time-ordered data, and
the beam response data.  The processed temperature and polarization maps in both
single-year and three-year forms, with and without foreground subtraction, are
supplied at pixel resolutions r4, r9, and r10.  Full pixel-pixel inverse
covariance matrices are supplied for analyzing the r4 polarization maps.  The
processed foreground products include the MEM component maps, the ILC map, and
the temperature and polarization templates used for foreground subtraction.  The
analyzed CMB products include the angular power spectra, the likelihood used in
the three-year parameter analysis, and a complete web-based set of parameter
results for every combination of data set and cosmological model that was run
for the three-year analysis.  For each of the model/data combinations, we also
supply the best-fit model spectra and the full Markov chains.  The products are
described in detail in the \map\ Explanatory Supplement
\citep{limon/etal:prep}.  The LAMBDA URL is \verb"http://lambda.gsfc.nasa.gov/".

\acknowledgments

We thank Jim Condon, Kevin Huffenberger, Dominik Schwarz, Sergei Trushkin, and
the anonymous referee for comments that helped to improve the accuracy and
clarity of the material in this paper.  We especially thank Hans Kristian
Eriksen for helpful comments and for providing us with very thorough cross
checks of our low-$l$ likelihood results. This work led us to adopt a full
maximum likelihood evaluation of the spectrum for multipoles up to $l=30$.

The \map\ mission is made possible by support from the Science Mission
Directorate at NASA Headquarters and by the hard and capable work of scores of 
scientists, engineers, technicians, machinists, data analysts, budget analysts, 
managers, administrative staff, and reviewers.  This research was additionally
supported by NASA grants LTSA 03-000-0090 and ATP NNG04GK55G. EK acknowledges
support from an Alfred P. Sloan Research Fellowship.  HVP is supported by NASA
through Hubble Fellowship grant HF-01177.01-A awarded by the Space Telescope
Science Institute, which is operated by the Association of Universities for
Research in Astronomy, Inc., for NASA, under contract NAS 5-26555.  LV is
supported by: NASA ADP03-0000-0092 and NASA ADP04-0000-0093.  We acknowledge use
of the HEALPix and CAMB packages.

\appendix

\section{IMPLEMENTATION OF BEAM AND POINT SOURCE ERRORS}
\label{app:beam_src_err}

In this Appendix we discuss our simplified handling of beam deconvolution
uncertainties and point source subtraction errors.  Since both of these errors
primarily affect the high-$l$ spectrum, we adopt the approximation that the
likelihood is Gaussian and develop methodology for computing the change in the
likelihood induced by these errors.

\subsection{Likelihood Decomposition}

At all but the lowest multipoles, $l$, the likelihood of the data, ${\bf
d}={\hat C}_l$, given a theoretical model, ${\bf m}=C_l$, is well approximated
by treating the power spectrum as Guassian distributed,
\beq
{\cal L} \equiv -2\ln L({\bf d}|{\bf m}) 
 = \sum_{ll'}({\hat C}_l-C_l) \Sigma^{-1}_{ll'} ({\hat C}_{l'}-C_{l'})
 + \ln\det {\bf \Sigma}
\eeq
where ${\bf \Sigma}$ is the covariance matrix of the data, $\Sigma_{ll'} =
\EV{\Delta\hat C_l \Delta\hat C_{l'}}$.  As discussed in
\citet{hinshaw/etal:2003}, the full covariance matrix may be decomposed into
its constituent contributions
\beq
{\bf \Sigma} = {\bf \Sigma}_{\rm cv} + {\bf \Sigma}_{\rm noise}
             + {\bf \Sigma}_{\rm mask} + {\bf \Sigma}_{\rm beam}
             + {\bf \Sigma}_{\rm src}.
\eeq
Since we are only interested in the final two terms in this Appendix, we
introduce a simplified notation whereby
\beq
{\bf \Sigma} \equiv {\bf \Sigma}_0 + {\bf \Sigma}_1,
\eeq
where ${\bf \Sigma}_0$ consists of the first three terms: cosmic variance,
instrument noise and mode coupling due to the sky mask; while ${\bf \Sigma}_1$
consists of the last two terms, the beam decovolution and point source
subtraction uncertainties.  With this decomposition of the covariance matrix,
the likelihood becomes
\beq
{\cal L} = \sum_{ll'} 
 ({\hat C}_l-C_l) ({\bf \Sigma}_0 + {\bf \Sigma}_1)^{-1}_{ll'} ({\hat C}_{l'}-C_{l'})
 + \ln\det ({\bf \Sigma}_0 + {\bf \Sigma}_1).
\eeq

The key point in the new treatment is that the beam and point source
uncertainties have a very restricted form in $l$ space.  The source model has
the form given in equation~(\ref{eq:source_model}), in which we treat $A$ as
uncertain, while the beam deconvolution uncertainty may be described by only a
handful of modes. As discussed in detail below, this allows us to decompose
${\bf \Sigma}_1$ in the form
\beq
{\bf \Sigma}_1 \approx {\bf U U}^T,
\eeq
where ${\bf U}$ is an $N_l \times M$ matrix with $M \ll N_l$ (typically $M \sim
10$).  This approximation allows us to efficiently compute ${\bf \Sigma}^{-1}$
using the Sherman-Morrison-Woodbury formula,
\beqa
({\bf \Sigma}_0 + {\bf \Sigma}_1)^{-1} & \approx & ({\bf \Sigma}_0 + {\bf U U}^T)^{-1} \\
& = & {\bf \Sigma}_0^{-1} - {\bf \Sigma}_0^{-1} {\bf U} 
      \left({\bf I} + {\bf U}^T {\bf \Sigma}_0^{-1} {\bf U} \right)^{-1} 
      {\bf U}^T {\bf \Sigma}_0^{-1}.
\eeqa
With this result the likelihood decomposes into ${\cal L} = {\cal L}_0 + {\cal
L}_1$, where
\beq
{\cal L}_0 = \sum_{ll'} 
 ({\hat C}_l-C_l) ({\bf \Sigma}_0)^{-1}_{ll'} ({\hat C}_{l'}-C_{l'})
 + \ln\det {\bf \Sigma}_0,
\eeq
and
\beq
{\cal L}_1 = -\sum_{ll'} ({\hat C}_l-C_l)
 \left[{\bf \Sigma}_0^{-1} {\bf U} 
      \left({\bf I} + {\bf U}^T {\bf \Sigma}_0^{-1} {\bf U} \right)^{-1} 
      {\bf U}^T {\bf \Sigma}_0^{-1} \right]_{ll'} ({\hat C}_{l'}-C_{l'})
 + \ln\det({\bf I} + {\bf U}^T {\bf \Sigma}_0^{-1}{\bf U}).
\eeq
Note that we have rewritten the $\ln\det$ term,
\beq
\ln\det({\bf I} + {\bf \Sigma}_0^{-1}{\bf U U}^T)
 = \ln\det({\bf I} + {\bf U}^T {\bf \Sigma}_0^{-1}{\bf U}),
\eeq
to reduce the dimensionality of the matrix in the determinant from $N_l \times
N_l$ to $M \times M$.  As a further simplification, we take ${\bf \Sigma}_0$ to
be diagonal when evaluating ${\cal L}_1$.  We have checked that this has a
negligible effect on the accuracy of the expression.

\subsection{Beam Deconvolution Uncertainty}

The detailed form of the covariance matrix due to beam uncertainties was
discussed in \citet{hinshaw/etal:2003}, but we review the salient features here.
For a pair of cross-power spectra, $C_l^{\bf i}$ and $C_l^{\bf j}$, the beam
covariance has the form
\beq
({\bf \Sigma}_{\rm b})^{\bf ij}_{ll'} = C_l ({\bf S}_{\rm b})^{\bf ij}_{ll'} C_{l'},
\eeq
where
\beq
({\bf S}_{\rm b})^{\bf ij}_{ll'}
 = B^{i}_{ll'}(\delta_{ij}+\delta_{ij'}) + B^{i'}_{ll'}(\delta_{i'j}+\delta_{i'j'}).
\eeq
Here $i, i'$ denote the pair of DA's in the spectrum $C_l^{\bf i}$, and
similarly for $j, j'$.  $B^{i}_{ll'}$ is the fractional covariance matrix,
$B^{i}_{ll'} = \EV{u^i_l u^i_{l'}}$, where $u^i_l = \Delta b^i_l/b^i_l$ is the
fractional error in the beam transfer function $b^i_l$.  The complete beam
covariance matrix for the final combined spectrum is 
\beq
({\bf \Sigma}_{\rm b})_{ll'} = C_l ({\bf S}_{\rm b})_{ll'} C_{l'} = \sum_{\bf ij} 
 C_l w^{\bf i}_l ({\bf S}_{\rm b})^{\bf ij}_{ll'} w^{\bf j}_{l'} C_{l'},
\eeq
where the $w^{\bf i}_l$ are the weights used to form the combined spectrum,
${\hat C}_l = \sum_{\bf i} w^{\bf i}_l C^{\bf i}_l$ (see \S\ref{sec:tt_high_l}).

The beam covariance matrix is dominated by a small number of modes. We take its
singular value decomposition, ${\bf S}_b = {\bf P w Q}^T$, where ${\bf P}$ and
${\bf Q}$ are orthogonal matrices (${\bf P}={\bf Q}$, since ${\bf S}_b$ is
symmetric), and ${\bf w} = {\rm diag}(w_1,\dots,w_{N_l})$ is diagonal with
singular values $w_1 > \dots > w_{N_l}$.  Only 10 modes have $w_i/w_1 >
10^{-3}$, and only 18 have $w_i/w_1 > 10^{-6}$.  Thus we can approximate
${\bf \Sigma}_b$ as
\beq
({\bf \Sigma}_b)_{ll'} \approx \sum_{i=1}^M
C_l \, P_{li} w_i P^T_{il'} \, C_{l'},
\eeq
where $M \ll N_l$ is the number of modes used in the approximation. Then,
defining an $N_l \times M$ matrix $U_{lk} = C_l P_{li} w^{1/2}_{i}$, we have
${\bf \Sigma}_b \approx {\bf UU}^T$.  For the three-year analysis we use $M=9$;
tests with additional modes changed the TT likelihood by $\Delta \chi^2 < 0.01$.

\subsection{Point Source Subtraction Uncertainty}

The amplitude of the point source correction is uncertain at the 20\% level. To
include this uncertainty in the likelihood evaluation we simply add another mode
to the matrix ${\bf U}$ defined above.  Specifically, $U_{l(M+1)} = \sigma_A \,
C^{\rm src}_l$, where $C^{\rm src}_l = \sum_{\bf i} w^{\bf i}_l C^{{\bf i},{\rm
src}}_l$ is the point source model defined in equation~(\ref{eq:source_model}),
and $\sigma_A=0.21$ is the fractional uncertainty of the point source
amplitude $A$.

\section{BEAM ASYMMETRIES}
\label{app:beam_asymm}

In \citet{jarosik/etal:prep} we outline our approach to determining the beam
transfer functions from flight observations of Jupiter.  This method implicitly
assumes that the effective beam response is equivalent to the azimuthal average
of the response, which is appropriate in the limit that each sky map pixel is
observed with equal weight over a full range of azimuth.  In practice this is a
good approximation near the ecliptic poles, but is less so near the ecliptic
plane. In this Appendix we quantify how \map's beam asymmetry distorts our
estimate of the angular power spectrum by taking into account: (1) the intrinsic
shape of the beam response in spacecraft-fixed coordinates, (2) the map-making
algorithm, including the spacecraft scan strategy, and (3) the form of the
estimator used to measure the cross-power spectra.

\citet{wu/etal:2001} treated the MAXIMA beams using an effective symmetric beam.
\citet{wandelt/gorski:2001} developed a fast method of convolving two arbitrary
band-limited functions on the sphere by transforming it into a 3-dimensional
Fourier transform.

\subsection{Formalism}

The calibrated, differential time-ordered data, ${\bf d}$, is related to the sky
map, ${\bf t}$, via
\beq
{\bf d} = {\bf Mt} + {\bf n},
\label{eq:dmtn}
\eeq
where ${\bf M}$ is the mapping function and ${\bf n}$ is the noise which we
assume to have zero mean, $\EV{{\bf n}}=0$ \citep{hinshaw/etal:2003b}. The
maximum likelihood estimate for ${\bf t}$ is $\hat{\bf t} = {\bf Wd}$, where 
${\bf W} = ({\bf M}^T{\bf N}^{-1}{\bf M})^{-1}{\bf M}^T{\bf N}^{-1}$, 
and ${\bf N} = \EV{{\bf nn}^T}$ is the noise covariance matrix of the
time-ordered data. Using the fact that ${\bf WM} = {\bf I}$, where ${\bf I}$ is
the identity matrix, it follows that  $\hat{\bf t} = {\bf Wd} = {\bf t} + 
{\bf Wn}$.  The maximum likelihood map is unbiased because the sky map noise, 
${\bf Wn}$, has zero mean.

For a differential instrument like \map\ that observes temperature differences
between two sides A and B, the mapping function ${\bf M}$ can be decomposed into
two pieces, ${\bf M}_A - {\bf M}_B$, which denote the A- and B-side
contributions, respectively. For a given side, the matrix elements  
${\bf M}_{ip}$ have the form $g_i(p)\Omega_p$, where $g_i(p)$ is the beam 
response in pixel $p$ for the $ith$ observation, and $\Omega_p$ is the pixel 
solid angle.  The normalization is such that $\int d\Omega \, g_i(p) 
\rightarrow \sum_p g_i(p) \Omega_p = 1$.

In practice, the solution $\hat{\bf t} = {\bf Wd}$ is too costly to evaluate
exactly given the properties of ${\bf M}$ and ${\bf N}^{-1}$.  Rather, we adopt
approximate forms for these quantities, ${\bf M}'$ and ${\bf N}'^{-1}$, and 
estimate the sky map using ${\bf t}' = {\bf W}'{\bf d}$, where 
${\bf W}' = ({\bf M}'^T{\bf N}'^{-1}{\bf M}')^{-1}{\bf M}'^T{\bf N}'^{-1}$.
The resulting sky map is biased relative to the true sky according to 
\beq
{\bf t}' = {\bf W}'{\bf Mt} \equiv {\bf Jt},
\eeq
where ${\bf J} \equiv {\bf W}'{\bf M}$. (The distorted sky map noise, ${\bf
W}'{\bf n}$, still has zero mean.) The map-making algorithm used by \map\
assumes that the beams are circularly symmetric with infinite resolution. That
is, each row (observation) of ${\bf M}'$ contains a $+1$ in the column (pixel)
seen by the A-side beam and a $-1$ in the column (pixel) seen by the B-side
beam. In addition, since the \map\ instrument noise is nearly white (${\bf N}
\approx \sigma_0^2 {\bf I}$) the algorithm solves for ${\bf t}'$ using
${\bf W}' = ({\bf M}'^T{\bf M}')^{-1}{\bf M}'^T$.  The matrix 
${\bf M}'^T{\bf M}'$ is diagonally dominant with diagonal elements 
$N_p = N^A_p+N^B_p$ equal to the number of times pixel $p$ has been observed 
by either the A- or B-side beam. The action of ${\bf W}'$ on ${\bf d}$ is thus 
approximately
\beq
{\bf W}'_{pi} \approx \frac{1}{N_p}(\delta^A_{pi} - \delta^B_{pi}),
\label{eq:wpi}
\eeq
where $\delta^A_{pi}$ is one if the A-side beam is in pixel $p$ and zero
otherwise, and likewise for $\delta^B_{pi}$. Combining \refeqnt{eq:gip} and
\refeqnt{eq:wpi}, we obtain the following expression for the distortion matrix
${\bf J}$
\beq
{\bf J}_{pp'} \approx \frac{\Omega_p}{N_p} 
\left( \sum_{i|A\in p}g^A_i(p') + \sum_{i|B\in p}g^B_i(p') \right)
\equiv {\bf J}^A_{pp'} + {\bf J}^B_{pp'}.
\eeq
Here ``$i|A\in p$'' denotes summation over all observations $i$ where the A-side
beam observes pixel $p$.  This algorithm produces a sky map in which the
effective beam response is an average of the A- and B-side responses convolved
with the scan pattern at each pixel.

At small angular scales where beam asymmetries may become important, a
pseudo-$C_l$ method is used to estimate the \map\ angular power spectrum from
these maps  \citep{hauser/peebles:1973, wandelt/hivon/gorski:2001,
hivon/etal:2002, hinshaw/etal:2003}. This method estimates the power spectrum
$C_l$ by first transforming a weighted map ${\bf Qt}'$, where ${\bf Q}$ is a
weight matrix used to mask regions of strong foreground contamination and/or
high instrument noise.  This produces the pseudo-$C_l$ spectrum, $\tilde C_l$,
which is then corrected for the statistical bias induced by ${\bf Q}$ under the
assumption that the underlying signal is statistically isotropic.

In the present context we wish to correct the power spectrum for the biases
induced by both ${\bf Q}$ and ${\bf J}$. We transform the distorted map into the
spherical harmonic basis, ${\bf a}' = \Omega_p {\bf Y}^\dagger {\bf Qt}'  =
\Omega_p {\bf Y}^\dagger {\bf QJt} = \Omega_p {\bf Y}^\dagger {\bf QJYa}$, where
${\bf a}$ is the (full-sky) spherical harmonic transform of  
${\bf t}$.\footnote{Here ${\bf Y}$ is the pixelized spherical harmonic transform
matrix with elements ${\bf Y}_{p(lm)} = Y_{lm}(p)$, where $p$ is the pixel
index. It obeys ${\bf YY}^\dagger = \delta_{pp'}/\Omega_p$ and ${\bf Y}^\dagger
{\bf Y} = \delta_{(lm)(l'm')}/\Omega_p$, where $\Omega_p$ is the pixel solid
angle.} The covariance matrix of ${\bf a}'$ is then related to the underlying
signal covariance by $\EV{{\bf a}'{\bf a}'^\dagger} = \JJ{\bf S}\JJ^\dagger$,
where  $\JJ = \Omega_p {\bf Y}^\dagger {\bf QJY}$ and ${\bf S}$ is the
undistorted CMB  signal matrix ${\bf S} = \EV{{\bf aa}^\dagger} = C_l
\delta_{(lm)(l'm')}$. The distorted pseudo-$C_l$  spectrum is thus related to
the true power spectrum, in ensemble-average, by
\beq
\EV{\tilde{C}'_l} = \frac{1}{2l+1} \sum_{m} \EV{|a'_{lm}|^2} 
\equiv \sum_{l'} G_{ll'} \EV{C_{l'}},
\eeq
where
\beq
G_{ll'} \equiv \frac{1}{2l+1} \sum_{mm'} {|\JJ_{(lm)(l'm')}|^2}
\eeq
is the mode coupling matrix. For cross-power spectra $\EV{{\bf a}_1
{\bf a}_2^\dagger}$, $|\JJ|^2$ is replaced by $\JJ_1\JJ^*_2$ where $\JJ_1$ is
evaluated with the weights and beam properties appropriate to channel $1$, and
so on.  The unbiased power spectrum estimate is then $\sum_{l'}G^{-1}_{ll'}
\tilde{C}'_{l'}$.

We now turn to the explicit evaluation of the distortion matrix $\JJ$. First, we
assume that the time dependence of the beam response on the sky is due only to
changes in the spacecraft orientation. That is, $g_i(p') = g(p_i,\alpha_i;p')$
where $p_i$ is the pixel observed by the beam centroid during the $i$th
observation, and $\alpha_i$ is the azimuth angle of the observation at that
time. If we expand the spacecraft-fixed beam response in spherical harmonics,
\beq
g(\hat{z},0;p) = \sum_{lm} g_{lm} Y_{lm}(p),
\eeq
then the response on the sky may be expressed as
\beq
g_i(p') = \sum_{lmm'} D^l_{mm'}(p_i,\alpha_i) g_{lm'} Y_{lm}(p'),
\label{eq:gip}
\eeq
where the $D^l_{mm'}$ are the Wigner D-functions.

Since $\JJ$ is a linear function of ${\bf J}$, it  follows that $\JJ({\bf
J}^A+{\bf J}^B) = \JJ^A+\JJ^B$.  Furthermore, if ${\bf Q}$ is diagonal, we can
write ${\bf J}^S$ (with S=A,B) as
\beq
{\bf J}^S_{pp'} = \frac{\Omega_p}{N_p} \sum_{i|S\in p}g^S_i(p') 
     = \Omega_p \int{\frac{d\alpha}{2\pi} w^S(p,\alpha) g^S(p,\alpha;p')},
\eeq
where
\beq
w^S(p,\alpha) = {\bf Q}_{pp} \frac{2\pi}{N_{p}} 
\sum_{i|S\in p}{\delta(\alpha-\alpha_i)}.
\label{eq:scanstrategy}
\eeq
The Wigner $D$-functions form an orthogonal basis for functions defined over the
the 3-dimensional rotation group, SO(3), thus we can expand $w^S(p,\alpha)$ as
\beq
w^S(p,\alpha)
= \sum_{lmm'}{w^{S,l}_{mm'} D^l_{mm'}(p,\alpha)},
\eeq
where the sum is over all integral values of $l,m,m'$. Then, using the product
rule
\beq
D^{l_1}_{m_1m_1'}(p,\alpha) D^{l_2}_{m_2m_2'}(p,\alpha)
  = \sum_{lmm'}{ C^{lm}_{l_1m_1l_2m_2} C^{lm'}_{l_1m_1'l_2m_2'} D^l_{mm'}(p,\alpha) },
\label{eq:dlmmproductrule}
\eeq
where the $C^{lm}_{l_1m_1l_2m_2}$ are the Clebsch-Gordon coefficients, along
with the fact that
\beq
\int{\frac{d\alpha}{2\pi} \, D^l_{mm'}(p,\alpha)}
    = \sqrt{\frac{4\pi}{2l+1}} Y^*_{lm}(p) \, \delta_{m'0},
\eeq
we find
\beq
{\bf Q}_{pp}{\bf J}^S_{pp'} = \Omega_p \sum_{lm,l_1m_1m_1',l_2m_2m_2'}
  b^S_{l_1m_1'} w^{S,l_2}_{m_2m_2'} C^{lm}_{l_1m_1l_2m_2} C^{l0}_{l_1m_1'l_2m_2'}
  \sqrt{\frac{2l_1+1}{2l+1}} Y^*_{lm}(p) Y_{l_1m_1}(p').
\label{eq:almprime}
\eeq
Here we have defined $b_{lm} \equiv \sqrt{4\pi/(2l+1)}\,g_{lm}$, and we note
that when the beams are circular $b_{lm} = b_l\delta_{m0}$, where $b_l$ is the
beam transfer function defined in \citet{page/etal:2003b}.  Now, recalling that
$\JJ = \Omega_p {\bf Y}^\dagger {\bf QJY}$, we have
\beq
\JJ^S_{(lm)(l_1m_1)} = \sum_{m_1',l_2m_2m_2'}
  b^{S*}_{l_1m_1'} w^{S,l_2*}_{m_2m_2'} C^{lm}_{l_1m_1l_2m_2} C^{l0}_{l_1m_1'l_2m_2'}
  \sqrt{\frac{2l_1+1}{2l+1}}.
\label{eq:XiX}
\eeq
Note that the sums over $m_2$ and $m_2'$ effectively drop out of
\refeqnt{eq:XiX}, since the Clebsch-Gordon coefficients are zero unless
$m_1+m_2=m$ and $m_1'+m_2'=0$.

Calculating $G_{ll'}$ is now straightforward; for cross-power spectra,
$\EV{{\bf a}_1 {\bf a}_2^\dagger}$, we can expand it as
\beq
G_{ll'} = G^{A_1A_2}_{ll'} + G^{A_1B_2}_{ll'} + G^{B_1A_2}_{ll'} + G^{B_1B_2}_{ll'},
\eeq
where $A_1$ refers to side A of channel 1, and so forth.  Then, using
\beq
\sum_{mm_1}{C^{lm}_{l_1m_1l_2m_2} C^{lm}_{l_1m_1l_2'm_2'}} =
\delta_{l_2l_2'}\delta_{m_2m_2'}(2l+1)/(2l_2+1),
\eeq
we can write
\beq
G^{S_1S_2}_{ll'} = \sum_{l_1m_1}{\frac{({\bf I}^{S_1})^{ll'}_{l_1m_1} 
                   ({\bf I}^{S_2*})^{ll'}_{l_1m_1}} {2l_1+1} },
\label{eq:Gll}
\eeq
where
\beq
({\bf I}^{S})^{ll'}_{l_1m_1} \equiv \sum_{m'm_1'}
 b^S_{l'm'} w^{S,l_1}_{m_1m_1'} C^{l0}_{l'm'l_1m_1'} \sqrt{\frac{2l'+1}{2l+1}}.
\label{eq:Illlm}
\eeq

In the case of uniform sky coverage, \refeqnt{eq:Gll} reduces to a familiar
limit. Specifically, if all pixels are observed for the same length of time with
a uniform distribution of azimuth angles, then $w(p,\alpha) = 1$, which gives
$w^{l}_{mm'} = \delta_{l0}\delta_{m0}\delta_{m'0}$.  This result, along with the
identity $C^{lm}_{l'm'00} = \delta_{ll'}\delta_{mm'}$, gives $G_{ll'} =
b^2_{l0}\,\delta_{ll'}$. Thus, even if the beams are not circularly symmetric,
the full azimuthal coverage effectively symmetrizes the response.

If we retain full azimuthal coverage but introduce a mask, $w(p,\alpha) = q(p)$,
we have
\beqa
w^{l}_{mm'} &=& \frac{2l+1}{8\pi^2} 
   \int{d\Omega \, d\alpha \, q(p) \, D^{l*}_{mm'}(p,\alpha)}
\label{eq:wlmm} \\
            &=& \sqrt{\frac{2l+1}{4\pi}} \, q_{lm} \, \delta_{m'0},
\eeqa
where the $q_{lm}$ are the spherical harmonic coefficients of $q(p)$. Then,
expressing the Clebsch-Gordon coefficients in terms of the Wigner-$3j$ symbols,
\beq
C^{lm}_{l'm'l''m''} =
  (-1)^{m+l'+l''} \sqrt{2l+1} \wjjj{l'}{l''}{l}{m'}{m''}{-m},
\eeq
we can write
\beq
I^{ll'}_{l''m''} = b_{l'0} \, q_{l''m''} \, (-1)^{l'+l''} \wjjj{l'}{l''}{l}{0}{0}{0}
  \sqrt{\frac{(2l'+1)(2l''+1)}{4\pi}},
\eeq
which gives the standard result for the coupling matrix \citep{hivon/etal:2002}
\beq
G_{ll'} = \sum_{l''m''}{\frac{2l'+1}{4\pi} \, b^2_{l'0} \, |q_{l''m''}|^2
          \wjjj{l'}{l''}{l}{0}{0}{0}^2}.
\label{eq:simple-Gll}
\eeq
Note that the beam window function is naturally incorporated in this expression,
because of the way the mapping matrix was defined.  This result also generalizes
to the case of non-uniform noise per pixel by allowing the mask function,
$q(p)$, to become a weight map.  Finally, if the beams are circularly symmetric,
we can drop the requirement that pixels be observed with uniform azimuthal
weight and derive the same result.

\subsection{Results}

Evaluating \refeqnt{eq:Gll} requires $\order(l^5_{\rm max})$ operations and is
thus impractical when $l_{\rm max} \sim 10^3$.  However, if we ignore sky cuts
and apply uniform sky weight, ${\bf Q} = {\bf I}$, we can exploit an approximate
symmetry of the \map\ scan pattern to significantly speed up the calculation. 
Specifically, the \map\ scan strategy is approximately independent of ecliptic
longitude, implying $w^S(\theta,\phi,\alpha) \approx w^S(\theta,0,\alpha)$ where
$\theta$ and $\phi$ are ecliptic latitude and longitude, respectively.  This
eliminates the sum over $m_1$ in \refeqnt{eq:Gll}.

The weight coefficients are
\beq
w^{S,l}_{mm'} = \frac{2l+1}{8\pi^2} \int{d\Omega \, d\alpha \, 
  w^S(\theta,\phi,\alpha) D^{l*}_{mm'}(\theta,\phi,\alpha)},
\eeq
and since
\beq
D^{l*}_{0m'}(\theta,\phi,\alpha) 
  = \sqrt{\frac{4\pi}{2l+1}}\, Y^*_{l,-m'}(\theta,\alpha),
\eeq
the weight coefficients simplify to
\beq
w^{S,l}_{mm'} = \delta_{m0} \, \sqrt{\frac{2l+1}{4\pi}} {\tilde w}^S_{l,-m'}
\eeq
where
\beq
{\tilde w}^S_{lm'} = \int{d(\cos\theta) \, d\alpha \, w^S(\theta,\alpha) \,
  Y^*_{lm}(\theta,\alpha)}
\eeq
is a standard spherical harmonic transform.  Then
\beq
({\bf I}^S)^{ll'}_{l''m''} = \delta_{m''0}
 \sqrt{\frac{(2l'+1)(2l''+1)}{4\pi}} \,
 \sum_{m'}{ b^S_{l'm'} {\tilde w}^S_{l'' m'}
 \wjjj{l'}{l''}{l}{m'}{-m'}{0} (-1)^{l'+l''} },
\eeq
and
\beq
G^{S_1S_2}_{ll'} = \frac{2l'+1}{4\pi} \sum_{l'' m'm''}
 { b^{S_1}_{l'm'} \, {\tilde w}^{S_1}_{l''m'} \,
   b^{S_2*}_{l'm''} \, {\tilde w}^{S_2*}_{l''m''}
  \wjjj{l'}{l''}{l}{m'}{-m'}{0} \, 
  \wjjj{l'}{l''}{l}{m''}{-m''}{0} }.
\label{eq:GAB}
\eeq
If the sums over $m',m''$ are truncated at $m_{\rm max}$, this requires ${\cal
O}(m_{\rm max}l^3_{\rm max})$ operations. For elliptical beams, $b_{lm}$ falls
off rapidly with $m$, becoming negligible for $m \ga 6$.

\begin{figure}
\figurenum{23}
\epsscale{1.0}
\plotone{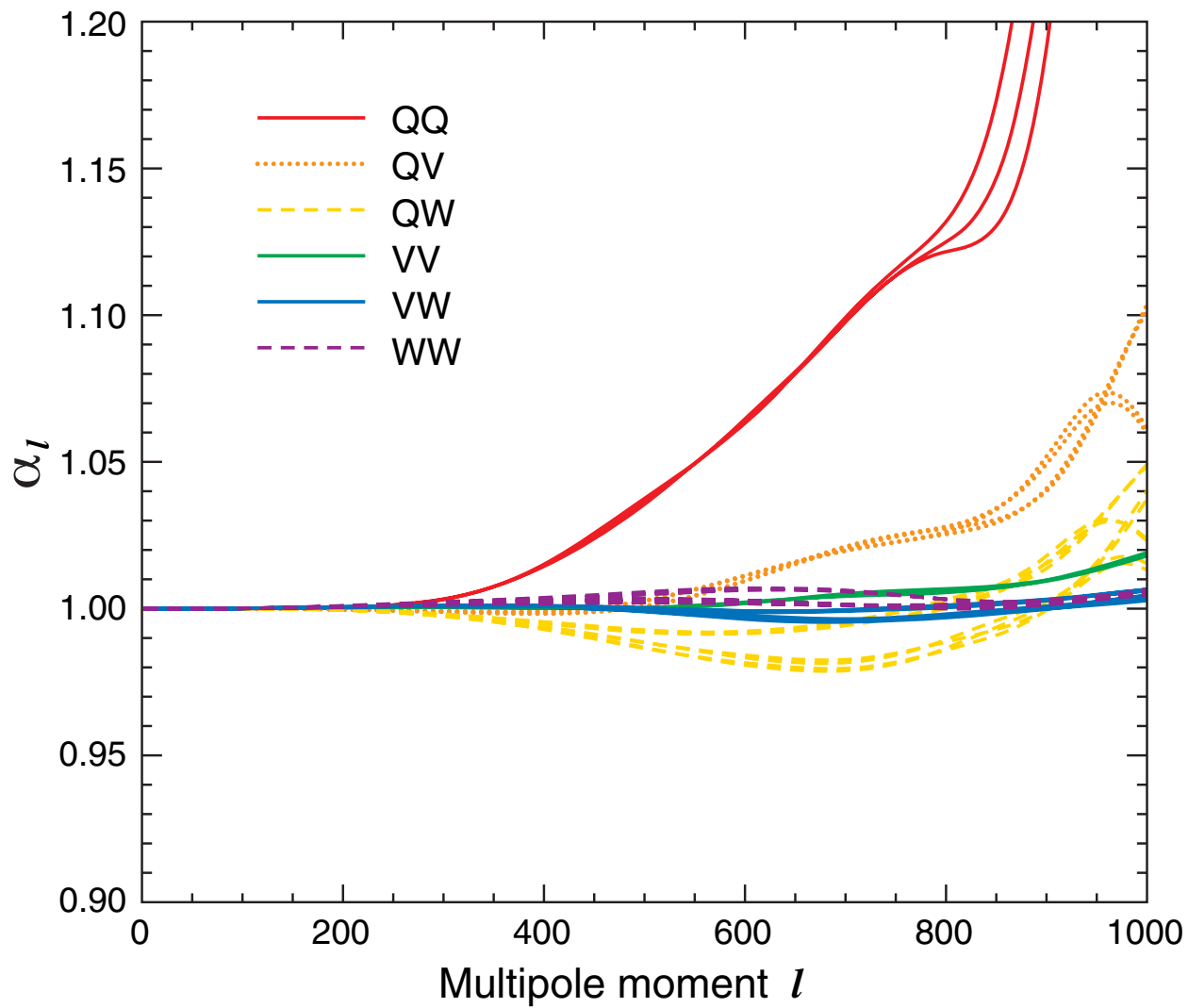}
\caption{An estimate of the multiplicative error introduced in the angular power
spectrum, $C_l$, due to the effects of beam asymmetry.  See
\S\ref{sec:tt_windows} and Appendix~\ref{app:beam_asymm} for details on how this
estimate was obtained.  Note that the final power spectrum does not include data
from Q band.}
\label{fig:beam_asym}
\end{figure}

To quantify how beam asymmetry distorts the power spectrum, we compute the ratio
\beq
\alpha_l \equiv \frac{1}{b^i_l b^{i'}_l C^{\rm fid}_l} 
                \sum_{l'} G^{\bf i}_{ll'} C^{\rm fid}_{l'},
\eeq
where $C^{\rm fid}_l$ is a fiducial power spectrum, $b^i_l$ is the symmetrized
beam transfer function noted above, and $G^{\bf i}_{ll'}$ is the coupling matrix
appropriate to cross-power spectrum ${\bf i}$ (this index notation is defined in
\S\ref{sec:tt_instrument}).  The matrices $G_{ll'}$ were computed using the
three-year sky coverage with beam transforms derived from the hybrid beam maps
\citep{jarosik/etal:prep}, and with $l_{\rm max}=1500$, $m_{\rm max}=16$. The
asymmetry corrections for each of the Q-, V-, and W-band auto- and cross-power
spectra are shown in \reffigt{fig:beam_asym}. The QQ spectrum show the largest
effect ($6.3\%$ at $l=600$), because the Q-band beams are relatively elliptical.
For example, the FWHM along the major and minor axes of the Q2A beam are
$0\ddeg60$ and $0\ddeg42$, respectively. The rise of $\alpha_l$ at high-$l$
indicates that the symmetric window underestimates the power in the beam, and
thus overestimates the power spectrum.  Because of the magnitude of this effect
in the Q-band data, and because of the large foreground correction needed at
low-$l$, Q-band was excluded from the final combined power spectrum.  The
corrections to the V- and W-band spectra are all $\la1\%$ for $l<1000$.

\section{MAXIMUM LIKELIHOOD NOTES}
\label{app:tt_ml}

\subsection{The General Case Without Noise}

The log of the likelihood function has the form
\beq
-2 \ln L({\bf d}|{\bf m}) = \chi^2 + \ln \det {\bf C},
\eeq
where ${\bf d}$ is the data, which is assumed to be drawn from a multivariate
Gaussian distribution, ${\bf m}$ is a set of model parameters, and ${\bf C}$ is
the covariance matrix of the data.  In the limit of no noise and full sky
coverage, it is convenient to represent the data in terms of the spherical
harmonic coefficients, ${\bf d} = {\bf a} \equiv a_{lm}$, in which case
\beq
{\bf C} = {\rm diag}[C_0,C_1,\ldots,C_2,\ldots]
\eeq
and
\beq
\chi^2 = {\bf a}^T\cdot {\bf C}^{-1} \cdot {\bf a} 
       = \sum_{lm} \frac{|a_{lm}|^2}{C_l}
\eeq
where $C_l$ is the power spectrum that encodes the variance of the $a_{lm}$
coefficients
\beq
\left< a_{lm} a_{lm}^*\right> = C_l \delta_{ll'}\delta_{mm'}.
\eeq
It is straightforward to show that the maximum likelihood estimate of the power
spectrum, $C^{\rm ML}_l$, is given by
\beq
C^{\rm ML}_l = \sum_{m=-l}^{l} \frac{|a_{lm}|^2}{2l+1}.
\label{ml_full}
\eeq

To set up the case of incomplete sky coverage, consider a new representation of
the data which is related to ${\bf a}$ by a simple linear transformation
\beq
{\bf d} = {\bf L} \cdot {\bf a},
\label{dLa}
\eeq
where ${\bf L}$ is an $N_d \times (l_{\rm max}+1)^2$ matrix, $N_d$ is the number
of points in ${\bf d}$ and $l_{\rm max}$ is the maximum spherical harmonic order
being analyzed.  For example, if ${\bf d} = {\bf t}$ is a sky map, the matrix
${\bf L} = {\bf Y}$ is the matrix of spherical harmonics evaluated at each pixel
$i$.  (We need not assume that ${\bf t}$ covers the full sky.)

The likelihood of ${\bf d}$, given the model spectrum, $C_l$, is
\beq
-2 \ln L({\bf d}|{\bf m}) = {\bf d}^T \cdot \tilde{\bf C}^{-1} \cdot {\bf d} + \ln \det \tilde{\bf C},
\eeq
where $\tilde{\bf C}$ is is the covariance matrix of ${\bf d}$ which has the
form
\beq
\tilde{\bf C} = \left< {\bf d}{\bf d}^T \right>
              = {\bf L} \cdot \left< {\bf a}{\bf a}^T \right> \cdot {\bf L}^T
              = {\bf L} \cdot {\bf C} \cdot {\bf L}^T,
\eeq
with
\beq
\tilde{\bf C}^{-1} = ({\bf L}^T)^{-1} \cdot {\bf C}^{-1} \cdot {\bf L}^{-1}.
\eeq
The determinant of $\tilde{\bf C}$ factors as
\beq
\ln \det \tilde{\bf C} = \ln (\det{\bf L}\det{\bf C}\det{\bf L}^T)
                       = \ln \det{\bf C} + \ln (\det{\bf L})^2,
\eeq
and since ${\bf C}$ is diagonal and ${\bf L}$ is independent of $C_l$, this 
reduces to
\beq
\ln \det \tilde{\bf C} = \sum_{lm} \ln C_l + {\rm const.}
\eeq
Thus we can rewrite the likelihood as
\beq
-2 \ln L({\bf d}|{\bf m}) = ({\bf L}^{-1} {\bf d})^T \cdot {\bf C}^{-1} \cdot ({\bf L}^{-1} {\bf d}) + \ln \det{\bf C} + {\rm const.}
\eeq
This form is a trivial recasting of the original form of $L$ in terms of  ${\bf
L}^{-1} {\bf d}$.  By the same argument that led to equation~(\ref{ml_full}) we
have the following expression for the maximum likelihood estimate of $C_l$
\beq
C^{\rm ML}_l = \sum_{m=-l}^{l} \frac{|({\bf L}^{-1} {\bf d})_{lm}|^2}{2l+1}.
\label{ml_cut}
\eeq
So, in the limit of no noise, a complete maximum likelihood estimate of the low
$l$ spectrum requires only a single matrix inversion to evaluate  ${\bf
L}^{-1}$.  This enables very rapid computation of the likelihood and it isolates
the source of systematic errors to the form of ${\bf L}$ and its inverse.

\subsection{The Pseudo-$a_{lm}$ Case}

One representation of the data are the pseudo-$a_{lm}$ coefficients,
$\tilde{a}_{lm}$, which are obtained by transforming the data on the cut sky
\beq
\tilde{a}_{lm} = \sum_i w_i \, Y^*_{lm}(\hat{n}_i) \, t_i,
\eeq
where the sum is over all pixels $i$, $w_i$ is the weight per pixel, which
includes any mask applied, $Y^*_{lm}(\hat{n}_i)$ is the spherical harmonic in
the direction of pixel $i$, and $t_i$ is the temperature in pixel $i$.  If we
expand the temperature in spherical harmonics, we can express the
$\tilde{a}_{lm}$ in terms of the true $a_{lm}$ coefficients as
\beq
\tilde{a}_{lm} = \sum_i w_i \, Y^*_{lm}(\hat{n}_i) \, \sum_{l'm'} a_{l'm'} Y_{l'm'}(\hat{n}_i)
               = \sum_{l'm'} M_{(lm)(lm)'} a_{l'm'},
\eeq
where the coupling matrix is defined as
\beq
M_{(lm)(lm)'} \equiv \sum_i w_i \, Y^*_{lm}(\hat{n}_i) Y_{l'm'}(\hat{n}_i),
\label{def_M}
\eeq
which depends only on the weight array, ${\bf w}$.  (If one expands ${\bf w}$ in
spherical harmonics, the coupling matrix can be expressed in terms of the Wigner
3-$j$ symbols, but this form is not especially useful in this context.) We
express the pseudo-$a_{lm}$ in matrix notation as $\tilde{\bf a} = {\bf
M}\cdot{\bf a}$, which has the same form as equation~(\ref{dLa}).   It follows
that the maximum likelihood solution for $C_l$ is then
\beq
C^{\rm ML}_l = \sum_{m=-l}^{l} \frac{|({\bf M}^{-1} \tilde{\bf a})_{lm}|^2}{2l+1}.
\label{ml_palm}
\eeq

In practice, the coupling matrix depends on both $l_{\rm max}$ and $l'_{\rm
max}$, the harmonic orders to which the true and pseudo-$a_{lm}$ are analyzed,
respectively. In general ${\bf M}$ need not be square.  It is instructive to
look at the structure of ${\bf M}$ using SVD
\beq
\tilde{\bf a} = {\bf M}\cdot{\bf a}
              = {\bf u}{\bf w}{\bf v}^T \cdot {\bf a}.
\eeq
Since ${\bf u}$ and ${\bf v}$ are orthogonal, the values of ${\bf w}$ encode the
level of deprojection that occurs when solving for ${\bf a} = {\bf M}^{-1}
\cdot\tilde{\bf a}$. Specifically, since
\beq
{\bf M}^{-1} = {\bf v}{\bf w}^{-1}{\bf u}^T,
\eeq
it follows that pseudo-$a_{lm}$ modes corresponding to small values of ${\bf w}$
will be greatly amplified in the recovery of the true $a_{lm}$.  Thus small
errors in the $\tilde{a_{lm}}$ can be also be greatly magnified.  In the case of
full sky, ${\bf w}=1$, while for the Kp2 cut sky and $l_{\rm max} = l'_{\rm max}
= 10$, we find $0.05 < {\bf w} < 1$. Note also that any error in the
$\tilde{a}_{lm}$ will bias the estimate of $C_l$ obtained with this method,
since the estimator for $C^{\rm ML}_l$ is quadratic.

\subsection{The Pixel-Space Case}

As noted above, the data may be represented as a (cut) sky map, ${\bf t}$, which
can be expanded in spherical harmonics, ${\bf t} = {\bf Y}\cdot{\bf a}$. Here
${\bf Y}$ is an $N_{\rm pix} \times (l_{\rm max}+1)^2$ matrix of spherical
harmonic functions evaluated at each pixel $i$ that survives the sky cut.

The likelihood of ${\bf t}$, given the model spectrum, $C_l$, is
\beq
-2 \ln L({\bf t}|{\bf m}) = {\bf t}^T \cdot \hat{\bf C}^{-1} \cdot {\bf t} + \ln \det \hat{\bf C},
\eeq
where $\hat{\bf C}$ is is the covariance matrix of ${\bf t}$ which has the
form
\beq
\hat{\bf C} = \left< {\bf t}{\bf t}^T \right>
            = {\bf Y} \cdot \left< {\bf a}{\bf a}^T \right> \cdot {\bf Y}^T
            = {\bf Y} \cdot {\bf C} \cdot {\bf Y}^T.
\eeq
This may be expressed in the more familiar form of the 2-point correlation
function, $C(\theta_{ij})$
\beq
\hat{{\bf C}}_{ij} = C(\theta_{ij}) 
                   = \sum_l \frac{2l+1}{4\pi} C_l P_l(\cos\theta_{ij})
                   = \sum_{lm} C_l \, Y^*_{lm}(\hat{n}_i) Y_{lm}(\hat{n}_j).
\eeq
where the last equality follows from the addition theorem for spherical
harmonics.  The inverse of $\hat{\bf C}$ has the form
\beq
\hat{\bf C}^{-1} = ({\bf Y}^T)^{-1} \cdot {\bf C}^{-1} \cdot {\bf Y}^{-1},
\eeq
and the determinant of $\hat{\bf C}$ factors as
\beq
\ln \det \hat{\bf C} = \ln (\det{\bf Y}\det{\bf C}\det{\bf Y}^T)
                     = \ln \det{\bf C} + \ln (\det{\bf Y})^2
                     = \ln \det{\bf C} + {\rm const.}
\eeq
Thus
\beq
-2 \ln L({\bf t}|{\bf m}) = ({\bf Y}^{-1} {\bf t})^T \cdot {\bf C}^{-1} \cdot ({\bf Y}^{-1} {\bf t}) + \ln \det{\bf C} + {\rm const,}
\eeq
and
\beq
C^{\rm ML}_l = \sum_{m=-l}^{l} \frac{|({\bf Y}^{-1} {\bf t})_{lm}|^2}{2l+1}.
\label{ml_pix}
\eeq

What is the nature of ${\bf Y}^{-1}$?  In the limit of full sky coverage the
spherical harmonics form an orthonormal basis,
\beq
\sum_{i} Y^*_{lm}(\hat{n}_i) Y_{l'm'}(\hat{n}_i) = \delta_{ll'}\delta_{mm'},
\eeq
or ${\bf Y}^T \cdot {\bf Y} = {\bf I}$, where ${\bf I}$ is the identity matrix. 
In this limit ${\bf Y}$ is orthogonal, ${\bf Y}^{-1} = {\bf Y}^T$, and ${\bf
Y}^{-1}{\bf t} = {\bf Y}^T{\bf t}$ is just the spherical harmonic transform of
${\bf t}$.  

More generally, in the presence of a sky cut, we can relate the pixel space
formalism to the pseudo-$a_{lm}$ formalism as follows.  The pseudo-$a_{lm}$
coefficients may be written as
\beq
\tilde{\bf a} = {\bf Y}^T \cdot {\bf t}
              = {\bf Y}^T {\bf Y} \cdot {\bf a}
              = {\bf M} \cdot {\bf a},
\eeq
where ${\bf M} = {\bf Y}^T {\bf Y}$ is the coupling matrix defined in
equation~(\ref{def_M}).  Note that we are now implicitly assuming that the sky
mask, ${\bf w}$, is sharp in the sense that it consists of only 1's or 0's.  The
case of an apodized mask needs to be studied further.  It follows that
\beq
{\bf a} = {\bf M}^{-1} \cdot \tilde{\bf a}
        = ({\bf Y}^T {\bf Y})^{-1}{\bf Y}^T \cdot {\bf t}
        = {\bf Y}^{-1} ({\bf Y}^T)^{-1} {\bf Y}^T \cdot {\bf t}
        = {\bf Y}^{-1} \cdot {\bf t},
\label{pix_palm}
\eeq
thus ${\bf Y}^{-1} = {\bf M}^{-1}{\bf Y}^T$ (this reduces to ${\bf Y}^{-1} =
{\bf Y}^T$ in the limit of full sky coverage). The relations in
equation~(\ref{pix_palm}) establish the equivalence of equation~(\ref{ml_palm})
and equation~(\ref{ml_pix}) in the case of a sharp mask.  Note also that
equation~(\ref{ml_pix}) is the same expression one would get by least squares
fitting the $a_{lm}$ coefficients on the cut sky (with uniform weight outside
the cut) and summing them to get the power spectrum.

Since ${\bf Y}^{-1} = {\bf M}^{-1}{\bf Y}^T$, the essential numerical features
of the maximum likelihood estimates are encoded in the coupling matrix ${\bf
M}$.  Specifically, one must ensure that ${\bf M}^{-1}$ has sufficient range to
adequately capture the harmonic content in the pseudo-$a_{lm}$ data.

\begin{deluxetable}{cccccccccc}
\tablecaption{Differencing Assembly (DA) Properties \label{tab:radiometers}}
\tablewidth{0pt}
\tablehead{
\colhead{DA} &
\colhead{$\lambda$\tablenotemark{a}} &
\colhead{$\nu$\tablenotemark{a}} &
\colhead{$g(\nu)$\tablenotemark{b}} &
\colhead{$\theta_{\rm FWHM}$\tablenotemark{c}} &
\colhead{$\sigma_0$(I)\tablenotemark{d}} &
\colhead{$\sigma_0$(Q,U)\tablenotemark{d}} &
\colhead{$\nu_{\rm s}$\tablenotemark{e}} &
\colhead{$\nu_{\rm ff}$\tablenotemark{e}} &
\colhead{$\nu_{\rm d}$\tablenotemark{e}} \\
\colhead{} &
\colhead{(mm)} &
\colhead{(GHz)} &
\colhead{} &
\colhead{($^\circ$)} &
\colhead{(mK)} &
\colhead{(mK)} &
\colhead{(GHz)} &
\colhead{(GHz)} &
\colhead{(GHz)}}
\startdata
 K1 & 13.17 & 22.77 & 1.0135 & 0.807 & 1.439 & 1.455 & 22.47 & 22.52 & 22.78 \\
Ka1 & 9.079 & 33.02 & 1.0285 & 0.624 & 1.464 & 1.483 & 32.71 & 32.76 & 33.02 \\
 Q1 & 7.342 & 40.83 & 1.0440 & 0.480 & 2.245 & 2.269 & 40.47 & 40.53 & 40.85 \\
 Q2 & 7.382 & 40.61 & 1.0435 & 0.475 & 2.135 & 2.156 & 40.27 & 40.32 & 40.62 \\
 V1 & 4.974 & 60.27 & 1.0980 & 0.324 & 3.304 & 3.330 & 59.65 & 59.74 & 60.29 \\
 V2 & 4.895 & 61.24 & 1.1010 & 0.328 & 2.946 & 2.970 & 60.60 & 60.70 & 61.27 \\
 W1 & 3.207 & 93.49 & 1.2480 & 0.213 & 5.883 & 5.918 & 92.68 & 92.82 & 93.59 \\
 W2 & 3.191 & 93.96 & 1.2505 & 0.196 & 6.532 & 6.571 & 93.34 & 93.44 & 94.03 \\
\enddata
\tablenotetext{a}{Effective wavelength and frequency for a thermodynamic spectrum.}
\tablenotetext{b}{Conversion from antenna temperature to CMB thermodynamic temperature;\\
$\Delta T = g(\nu)\Delta T_A$; $g(\nu)=[(e^x-1)^2/x^2e^x]$; $x=h\nu/kT_0$;
$T_0=2.725$ K \citep{mather/etal:1999}.}
\tablenotetext{c}{Full-width-at-half-maximum from radial profile of A- and
B-side average beams.  Note: beams are not Gaussian.}
\tablenotetext{d}{Noise per observation for resolution 9 and 10 $I$, $Q$, \& $U$
maps, to $\sim$0.1\% uncertainty. $\sigma(p)=\sigma_0 N_{\rm obs}^{-1/2}(p)$.}
\tablenotetext{e}{Effective frequency for synchrotron (s), free-free (ff), and
dust (d) emission, assuming spectral indices of $\beta = -2.9, -2.1, +2.0$,
respectively, in antenna temperature units.}
\end{deluxetable}

\begin{deluxetable}{lccccc}
\tablecaption{Data Flagging Summary \label{tab:baddata}}
\tablewidth{0pt}
\tablehead{
\colhead{Category} & \colhead{K-band} & \colhead{Ka-band} & \colhead{Q-band} &
\colhead{V-band} & \colhead{W-band}}
\startdata
\sidehead{Lost or rejected data}
Lost\tablenotemark{a} (\%)     & 0.43 & 0.43 & 0.43 & 0.43 & 0.43 \\
Thermal disturbance\tablenotemark{b} (\%)
                               & 0.51 & 0.51 & 0.51 & 0.51 & 0.51 \\
Gain/baseline step (\%)        & 0.02 & 0.04 & 0.05 & 0.00 & 0.06 \\
                               &------&------&------&------&------\\
 Total lost or rejected (\%)   & 0.96 & 0.98 & 0.99 & 0.94 & 1.00 \\
\sidehead{Data not used in maps}
Planet in beam (\%)            & 0.11 & 0.11 & 0.11 & 0.11 & 0.11 \\
\enddata
\tablenotetext{a}{Primarily due to one solar storm induced safehold.}
\tablenotetext{b}{Primarily due to station-keeping maneuvers at $L_2$.}
\end{deluxetable}

\begin{deluxetable}{lrrr}
\tablecaption{Change in low-$l$ Power \label{tab:3yr_1yr_diff}}
\tablewidth{0pt}
\tablehead{
\colhead{Band} &
\colhead{$l=1$\tablenotemark{a}} &
\colhead{$l=2$\tablenotemark{b}} &
\colhead{$l=3$\tablenotemark{b}} \\
\colhead{} &
\colhead{($\mu$K)} &
\colhead{($\mu$K$^2$)} &
\colhead{($\mu$K$^2$)} }
\startdata
 K & 10.1 & 38.8 &  9.1 \\
Ka &  7.3 &  2.7 &  3.4 \\
 Q &  6.1 &  7.1 & 12.3 \\
 V &  5.1 &  7.1 &  2.2 \\
 W &  7.0 &  5.8 &  1.5 \\
\enddata
\tablenotetext{a}{$l=1$ - Amplitude in the difference map, in $\mu$K.}
\tablenotetext{b}{$l>1$ - Power in the difference map, $l(l+1)\,C_l / 2\pi$, in $\mu$K$^2$.}
\end{deluxetable}

\begin{deluxetable}{cccccc}
\tablecaption{Galactic Template Fit Coefficients \label{tab:gal_template}}
\tablewidth{0pt}
\tablehead{
\colhead{DA} & \colhead{$b_1$} & \colhead{$b_2$\tablenotemark{a}} & \colhead{$b_3$\tablenotemark{b}}
 & \colhead{$\beta_s$\tablenotemark{c}} & $h_{\rm ff}$ \\}
\startdata
Q1 & $0.243$ & $1.020$ & $0.195$ & $-3.15$ & 6.28 \\
Q2 & $0.242$ & $1.046$ & $0.193$ & $-3.19$ & 6.28 \\ 
V1 & $0.054$ & $0.700$ & $0.448$ & $-3.50$ & 6.57 \\
V2 & $0.052$ & $0.678$ & $0.463$ & $-3.48$ & 6.57 \\
W1 & $0.000$ & $0.405$ & $1.226$ & \nodata & 6.72 \\
W2 & $0.000$ & $0.400$ & $1.240$ & \nodata & 6.72 \\
W3 & $0.003$ & $0.397$ & $1.207$ & \nodata & 6.71 \\
W4 & $0.000$ & $0.402$ & $1.234$ & \nodata & 6.72 \\
\enddata
\tablenotetext{a}{Units of $\mu$K R$^{-1}$.  Coefficients constrained to follow a
free-free spectrum.  See text.}
\tablenotetext{b}{Dimensionless amplitude relative to FDS Model 8 evaluated at
94 GHz. Coefficients forced to follow $\beta_{\rm d}=+2.0$.  See text.}
\tablenotetext{c}{Synchrotron spectral index computed from
equation~(\ref{eq:beta_synch}).  The W-band fits imply $\beta_{\rm s}
(\nu_K,\nu_W) < -4$.}
\end{deluxetable}

\begin{deluxetable}{lrrrrr}
\tablecaption{ILC Weights by Region \label{tab:ilc_wt}}
\tablewidth{0pt}
\tablehead{\colhead{Region\tablenotemark{a}}
& \colhead{K-band} & \colhead{Ka-band} & \colhead{Q-band} 
& \colhead{V-band} & \colhead{W-band}}
\startdata
$ 0$ & $ 0.1559$ & $-0.8880$ & $ 0.0297$ & $2.0446$ & $-0.3423$ \\
$ 1$ & $-0.0862$ & $-0.4737$ & $ 0.7809$ & $0.7631$ & $ 0.0159$ \\
$ 2$ & $ 0.0358$ & $-0.4543$ & $-0.1173$ & $1.7245$ & $-0.1887$ \\
$ 3$ & $-0.0807$ & $ 0.0230$ & $-0.3483$ & $1.3943$ & $ 0.0118$ \\
$ 4$ & $-0.0781$ & $ 0.0816$ & $-0.3991$ & $0.9667$ & $ 0.4289$ \\
$ 5$ & $ 0.1839$ & $-0.7466$ & $-0.3923$ & $2.4184$ & $-0.4635$ \\
$ 6$ & $-0.0910$ & $ 0.1644$ & $-0.4983$ & $0.9821$ & $ 0.4428$ \\
$ 7$ & $ 0.0718$ & $-0.4792$ & $-0.2503$ & $1.9406$ & $-0.2829$ \\
$ 8$ & $ 0.1829$ & $-0.5618$ & $-0.8002$ & $2.8464$ & $-0.6674$ \\
$ 9$ & $-0.0250$ & $-0.3195$ & $-0.0728$ & $1.4570$ & $-0.0397$ \\
$10$ & $ 0.1740$ & $-0.9532$ & $ 0.0073$ & $2.7037$ & $-0.9318$ \\
$11$ & $ 0.2412$ & $-1.0328$ & $-0.2142$ & $2.5579$ & $-0.5521$ \\
\enddata
\tablenotetext{a}{See Figure~\ref{fig:ilc_bias} for the region definitions.  
See \S\ref{sec:gal_ilc} for notes on the handling of regions 0 and 1.}
\end{deluxetable}

\begin{deluxetable}{llrrr}
\tablecaption{Low-$l$ Multipole Moments ($\mu$K)\tablenotemark{a} \label{tab:tt_123}}
\tablewidth{0pt}
\tablehead{ & & \colhead{$l=1$\tablenotemark{b}} &
\colhead{$l=2$\tablenotemark{c}} & \colhead{$l=3$\tablenotemark{d}}}
\startdata
$\tilde{a}_{l+3}$ & &         \nodata     &         \nodata & $-11.24\pm1.48$ \\
$\tilde{a}_{l+2}$ & &         \nodata     & $-14.41\pm3.13$ & $ 22.03\pm0.37$ \\
$\tilde{a}_{l+1}$ & & $ -239.3\pm\phn1.2$ & $ -0.05\pm2.62$ & $-13.05\pm1.55$ \\
$\tilde{a}_{l0}$  & & $ 2505.0\pm11.1$    & $ 11.48\pm3.59$ & $ -5.99\pm0.46$ \\
$\tilde{a}_{l-1}$ & & $-2223.6\pm12.5$    & $  4.86\pm2.62$ & $  2.45\pm0.83$ \\
$\tilde{a}_{l-2}$ & &         \nodata     & $-18.80\pm2.88$ & $  0.70\pm0.26$ \\
$\tilde{a}_{l-3}$ & &         \nodata     &         \nodata & $ 33.46\pm1.35$ \\
\enddata
\tablenotetext{a}{Phase convention: the coefficients of the real valued
harmonics reported here, $\tilde{a}_{lm}$, are related to the complex
coefficients by $\tilde{a}_{lm} = \sqrt{2}\,{\rm Im}\,a_{lm}$ for $l<0$,
$\tilde{a}_{lm} = a_{lm}$ for $l=0$, and $\tilde{a}_{lm} = \sqrt{2}\,{\rm
Re}\,a_{lm}$ for $l>0$.  Also $\sum_m |\tilde{a}_{lm}|^2 = \sum_m |a_{lm}|^2$.}
\tablenotetext{b}{The dipole components in Galactic rectilinear coordinates are
($x$,$y$,$z$) = ($\tilde{a}_{1+1}$,$\tilde{a}_{1-1}$,$\tilde{a}_{10}$); in
Galactic polar coordinates the components are ($d$,$l$,$b$) = ($3.358\pm0.017$
mK, $263.86\pm0.04^{\circ}$, $48.24\pm0.10^{\circ}$).  The uncertainty in the
dipole magnitude is dominated by the 0.5\% calibration uncertainty.}
\tablenotetext{c}{Uncertainties from a quadrature sum of calibration error
\citep{jarosik/etal:prep} and Galaxy subtraction error (\S\ref{sec:gal_ilc}).}
\tablenotetext{d}{Uncertainties from Galaxy subtraction (\S\ref{sec:gal_ilc}).}
\end{deluxetable}

\begin{deluxetable}{lrrrrr}
\tablecaption{Maximum Likelihood Spectrum ($l=2-10$) \label{tab:tt_low_l_like}}
\tablewidth{0pt}
\tablehead{\colhead{$l$} & \colhead{$\Delta T_l^2$} & \colhead{$-$95\%} 
& \colhead{$-$68\%} & \colhead{$+$68\%} & \colhead{$+$95\%}}
\startdata
$ 2$ & $ 236$\phn & $ 54$\phn & $ 99$\phn & $ 796$\phn & $3827$\phn \\
$ 3$ & $1053$\phn & $347$\phn & $552$\phn & $2398$\phn & $6036$\phn \\
$ 4$ & $ 752$\phn & $285$\phn & $434$\phn & $1472$\phn & $3131$\phn \\
$ 5$ & $1582$\phn & $664$\phn & $977$\phn & $2811$\phn & $5338$\phn \\
$ 6$ & $ 601$\phn & $272$\phn & $388$\phn & $1000$\phn & $1747$\phn \\
$ 7$ & $1345$\phn & $639$\phn & $904$\phn & $2126$\phn & $3591$\phn \\
$ 8$ & $ 655$\phn & $330$\phn & $453$\phn & $ 998$\phn & $1589$\phn \\
$ 9$ & $ 621$\phn & $324$\phn & $438$\phn & $ 922$\phn & $1407$\phn \\
$10$ & $ 755$\phn & $409$\phn & $545$\phn & $1088$\phn & $1626$\phn \\
 \enddata
\end{deluxetable}

\begin{deluxetable}{lcc}
\tablecaption{\map\ Power Spectrum Peak and Trough Data \label{tab:tt_peak}}
\tablewidth{0pt}
\tablehead{\colhead{Quantity} & \colhead{$l$} & \colhead{$\Delta T_l^2$} \\
\colhead{} & \colhead{} & \colhead{($\mu$K$^2$)}}
\startdata
First peak     & $220.8 \pm  0.7$ & $5624 \pm 30$ \\
First trough   & $412.4 \pm  1.9$ & $1716 \pm 28$ \\
Second peak    & $530.9 \pm  3.8$ & $2485 \pm 44$ \\
Second trough  & $675.2 \pm 11.1$ & $1688 \pm 81$ \\
\enddata
\end{deluxetable}

\clearpage
\setlength{\hoffset}{-20mm}
\begin{deluxetable}{ccccccccccc}
\tablecaption{\label{tbl:sources} WMAP Source Catalog}
\tabletypesize{\footnotesize}
\tablewidth{0pt}
\tablehead{
 \colhead{RA [hms]}
&\colhead{Dec [dm]}
&\colhead{ID}
&\colhead{K [Jy]}
&\colhead{Ka [Jy]}
&\colhead{Q [Jy]}
&\colhead{V [Jy]}
&\colhead{W [Jy]}
&\colhead{$\alpha$}
&\colhead{5 GHz ID}
&
}
\startdata
 00 06 05
&$-$06 22
&060
&$2.4\pm0.08$
&$2.2\pm0.1$
&$2.3\pm0.2$
&$2.3\pm0.3$
&\nodata
&$-0.0\pm0.2$
&PMN J0006-0623
&
\\
 00 12 56
&$-$39 53
&202
&$1.4\pm0.06$
&$1.3\pm0.1$
&$1.0\pm0.1$
&$1.5\pm0.3$
&\nodata
&$-0.2\pm0.3$
&PMN J0013-3954
&
\\
 00 19 32
&\phs26 02
&   
&$1.1\pm0.08$
&$1.1\pm0.2$
&$0.9\pm0.2$
&$0.8\pm0.2$
&\nodata
&$-0.3\pm0.5$
&GB6 J0019+2602
&
\\
 00 19 33
&\phs20 19
&   
&$1.1\pm0.07$
&$1.1\pm0.1$
&$0.9\pm0.1$
&$1.9\pm0.4$
&\nodata
&$0.2\pm0.4$
&GB6 J0019+2021
&
\\
 00 25 26
&$-$26 04
&   
&$1.0\pm0.08$
&$0.8\pm0.1$
&\nodata
&\nodata
&\nodata
&$-0.4\pm1$
&PMN J0025-2602
&\tablenotemark{a}
\\
 00 26 03
&$-$35 11
&   
&$1.0\pm0.08$
&$1.0\pm0.1$
&$1.2\pm0.1$
&$1.2\pm0.3$
&\nodata
&$0.3\pm0.4$
&PMN J0026-3512
&
\\
 00 29 35
&\phs05 55
&   
&$1.0\pm0.07$
&$1.4\pm0.1$
&$1.2\pm0.1$
&\nodata
&\nodata
&$0.4\pm0.4$
&GB6 J0029+0554B
&\tablenotemark{a}
\\
 00 42 57
&\phs52 08
&   
&$0.9\pm0.05$
&$0.6\pm0.08$
&\nodata
&\nodata
&\nodata
&$-0.9\pm0.8$
&GB6 J0043+5203
&
\\
\enddata

\tablenotetext{a}{Indicates the source has multiple possible identifications.}
\tablenotetext{b}{Source J0322-3711 (Fornax A) is extended, and the fluxes
listed were obtained by aperture photometry.}
\tablenotetext{c}{Source J0519-0540 is a blend of the Lynds Bright Nebulae LBN
207.65-23.11 and LBN 207.29-22.66.}
\tablenotetext{d}{Source J1357+7644 is outside of the declination range of the
GB6 and PMN catalogs.  Identified as QSO NVSSJ135755+764320 by Trushkin (2006,
private communication).}
\tablenotetext{e}{Source J1633+8227 is outside of the declination range of the
GB6 and PMN catalogs. It was identified as NGC 6251 by \citet{trushkin:2003}.}
\tablenotetext{f}{Source J1657+4749 is identified as QSO GB6J1658+4737 by
Trushkin (2006, private communication). Offset from the \map\ position is 11.6
arcminutes.}
\tablenotetext{g}{Source J2356+4952 is identified as GB6 J2355+4950 by Trushkin
(2003). Offset from the \map\ position is 11.9 arcminutes.}
\end{deluxetable}

\clearpage
\setlength{\hoffset}{0mm}

\clearpage

\clearpage

\clearpage

\clearpage

\clearpage

\clearpage

\clearpage

\clearpage

\clearpage

\clearpage

\clearpage

\clearpage

\clearpage

\clearpage

\clearpage

\clearpage

\clearpage

\clearpage

\clearpage

\clearpage

\end{document}